# CARBON-ENHANCED METAL-POOR STAR FREQUENCIES IN THE GALAXY: CORRECTIONS FOR THE EFFECT OF EVOLUTIONARY STATUS ON CARBON ABUNDANCES

Vinicius M. Placco[1], Anna Frebel[2], Timothy C. Beers[3], Richard J. Stancliffe[4]

## ABSTRACT

We revisit the observed frequencies of Carbon-Enhanced Metal-Poor (CEMP) stars as a function of the metallicity in the Galaxy, using data from the literature with available high-resolution spectroscopy. Our analysis excludes stars exhibiting clear over-abundances of neutron-capture elements, and takes into account the expected depletion of surface carbon abundance that occurs due to CN processing on the upper red-giant branch. This allows for the recovery of the initial carbon abundance of these stars, and thus for an accurate assessment of the frequencies of carbon-enhanced stars. The correction procedure we develope is based on stellar-evolution models, and depends on the surface gravity, $\log g$, of a given star. Our analysis indicates that, for stars with [Fe/H]$\leq -2.0$, 20% exhibit [C/Fe]$\geq +0.7$. This fraction increases to 43% for [Fe/H]$\leq -3.0$ and 81% for [Fe/H]$\leq -4.0$, which is higher than have been previously inferred without taking the carbon-abundance correction into account. These CEMP-star frequencies provide important inputs for Galactic and stellar chemical-evolution models, as they constrain the evolution of carbon at early times and the possible formation channels for the CEMP-no stars. We also have developed a public online tool with which carbon corrections using our procedure can be easily obtained.

## 1. INTRODUCTION

A number of recent studies have shown that Carbon-Enhanced Metal-Poor (CEMP) stars are one of the most important objects for constraining the formation and evolution of the first stellar populations in the Galaxy and the Universe (e.g., Carollo et al. 2012, 2014; Norris et al. 2013b; Cooke & Madau 2014). These stars belong to the broader class of very metal-poor (VMP – [Fe/H][5] $< -2.0$, e.g., Beers & Christlieb 2005; Frebel & Norris 2013) stars, which have been vigorously searched for and analyzed over the past quarter century. The definition of a CEMP star has been refined over the years, as more high-resolution spectroscopic data has become available, making it possible to distinguish between possible scenarios for their formation. The initial classification by Beers & Christlieb (2005) distinguishes CEMP stars as objects with [C/Fe] abundance ratios (or "carbon abundances", also sometimes refered to as "carbonicity") at least ten times the solar value ([C/Fe] $> +1.0$). However, subsequent analysis has indicated that a more suitable division appears at [C/Fe] $\geq +0.7$ (Aoki et al. 2007; Carollo et al. 2012; Norris et al. 2013a).

Generally, CEMP stars in the Galaxy occur over a broad range in both metallicity and carbonicity. The metallicity, or strictly speaking, the [Fe/H] abundance ratio, is commonly used as a proxy for chemical-evolution timescales in the Galaxy. Iron has a very distinctive nucleosynthesis channel (e.g. Woosley & Weaver 1995), and well-traces the enrichment of the interstellar medium through the course of the Galactic evolution. Carbon, on the other hand, has several different formation channels (e.g., Norris et al. 2013b) in the early universe, making it unsuitable for tracing evolutionary timescales. Hence, quantifying the occurance of CEMP stars as a function of metallicity maps out the evolution of carbon in the early universe.

Broadly speaking, the carbon-enhancement phenomenon can be either *extrinsic* or *intrinsic* to a given star:

*Extrinsic* enrichment accounts for the observed abundance patterns of the CEMP-*s* ([Ba/Fe] $> +1.0$ and [Ba/Eu] $> +0.5$) and CEMP-*r/s* ($0.0\leq$[Ba/Fe] $\leq +0.5$) stars (but see Hollek et al. 2014). This pattern is thought to be the result of mass transfer across a binary system, coming from an evolved star that has passed through the asymptotic giant-branch (AGB; e.g., Herwig 2005) evolutionary stage. Radial-velocity monitoring (e.g., Lucatello et al. 2005) confirms the binarity of the majority of these CEMP stars, and extensive studies have been conducted to compare the observed abundance patterns with theoretical models (e.g., Bisterzo et al. 2011; Placco et al. 2013; Hollek et al. 2014).

*Intrinsic* enrichment is thought to be the main formation channel for the CEMP-no ([Ba/Fe] $< 0.0$) and CEMP-*r* ([Eu/Fe] $> +1.0$) subclasses of stars. Given that their metallicities are almost exclusively below [Fe/H] $= -2.7$ (Aoki et al. 2007), such stars most likely formed from chemically primitive gas clouds. In the case of the CEMP-no stars, there appears to exist a distinct carbon abundance regime where they are found ($\log_\epsilon$(C) $\sim$ 6.5; Spite et al. 2013). Contrary to the extrinsically enriched CEMP-s stars, the CEMP-no stars must have formed from carbon-enhanced natal gas clouds. Norris et al. (2013b), and references therein, suggest a number of scenarios for the early production of carbon, and thus the origins of CEMP-no stars. Among these are massive, zero/low-metallicity stars, with/without rotation (Meynet et al. 2006, 2010), and mixing and fallback Type II supernovae, often referred to as "faint supernovae" (Umeda & Nomoto 2005; Tominaga et al. 2007).

[1] Gemini Observatory, Hilo, HI 96720, USA
[2] Department of Physics and Kavli Institute for Astrophysics and Space Research, Massachusetts Institute of Technology, Cambridge, MA 02139, USA
[3] Department of Physics and JINA Center for the Evolution of the Elements, University of Notre Dame, Notre Dame, IN 46556, USA
[4] Argelander-Institut für Astronomie der Universität Bonn, 53121 Bonn, Germany
[5] [A/B] $= log(N_A/N_B)_\star - log(N_A/N_B)_\odot$, where $N$ is the number density of atoms of a given element in the star ($\star$) and the Sun ($\odot$), respectively.

One remarkable example of the CEMP-no subclass is BD+44°493 (Ito et al. 2009, 2013; Placco et al. 2014b), a $V = 9$ star with a light-element abundance pattern (e.g., C, N, O, Na, Mg, Al, etc.) that agrees well with yields from faint-supernovae models (Nomoto et al. 2006). Furthermore, observational evidence suggests that the CEMP-no abundance pattern is dominant at low metallicity, given that five of the six stars known to have [Fe/H]$< -4.5$ are CEMP-no stars (Christlieb et al. 2002; Frebel et al. 2005; Norris et al. 2007; Caffau et al. 2011; Hansen et al. 2014; Keller et al. 2014).

In this work, we employ new stellar-evolution models that quantify the changes in surface carbon abundances of metal-poor stars during stellar evolution on the giant branch. We also provide an online tool that allows the calculations of these carbon corrections for a given set of input stellar parameters. After excluding recognized CEMP-$s$ and CEMP-$r/s$ stars, we obtain appropriate corrections to apply to the observed carbon abundances, as a function of the observed [Fe/H] and $\log g$. Proper treatment of the carbon depletion allows for an assessment of the true (intrinsic) CEMP-no stellar frequencies as a function of metallicity. These frequencies, in turn, provide important constraints on Galactic chemical-evolution (e.g., Kobayashi & Nakasato 2011) and population-synthesis models (e.g., Pols et al. 2012), on the initial mass function (IMF; e.g., Lee et al. 2014), and on the chemical compositions of progenitor stellar populations.

This paper is outlined as follows. Section 2 describes the theoretical models used for determining the carbon corrections, followed by details of the data selection in Section 3. Corrections for carbon abundances based on these models are provided in Section 4, including a discussion on the sources of uncertainties in our analysis. We present the corrected carbon abundances for the literature sample in Section 5. Section 6 presents a corrected determination of the cumulative CEMP-star frequencies as a function of metallicity, based on high-resolution spectroscopic analyses reported in the literature. We discuss our results and their astrophysical implications in Section 7.

## 2. STELLAR-EVOLUTION MODELS

The evolutionary stage of a given star has an impact on the observed carbon (and similarly, nitrogen and oxygen) abundances. During evolution on the upper red-giant branch, carbon from the lower layers of a stellar atmosphere is converted to nitrogen due to the CN cycle, then mixed to the surface of the star, resulting in an increase of the surface nitrogen abundance and reduction in the surface carbon abundance. The amount of carbon depletion depends mostly on the metallicity and the initial stellar carbon and nitrogen abundances. This effect has already been discussed extensively in the literature (Gratton et al. 2000; Spite et al. 2006; Aoki et al. 2007). However, apart from the CEMP-star classification suggested by Aoki et al. (2007), which takes into account the luminosity of a given star (and hence its evolutionary status), no further investigations have been undertaken to consider the impact of carbon depletion on the giant branch when describing the populations of CEMP stars in Galaxy.

Using the STARS stellar-evolution code (Eggleton 1971; Stancliffe & Eldridge 2009), we have computed a grid of $0.8\,M_\odot$ stellar-evolution models with a range of initial compositions. We cover four initial metallicities, namely [Fe/H] = −1.3, −2.3, −3.3, and −4.3. For each of these metallicities, we consider a range of initial [C/Fe] values: [C/Fe] = −0.5, 0.0, +0.5, +0.7, +1.0, +1.5, +2.0, +2.5 and +3.0[6]. For [N/Fe], the models are: [N/Fe] = −0.5, 0.0, +0.5, +0.7, +1.0, and +2.0. In total there are 210 models. Each model is evolved from the pre main-sequence to the tip of the red-giant branch (RGB). It is well-documented that the surface abundances changes occur on the upper part of the RGB (e.g. Gratton et al. 2000) and that some non-convective process is required to account for this. There are many potential mechanisms that can cause this, including (but not limited to) rotation, internal gravity waves, magnetic fields, and thermohaline mixing. For reviews of these mechanisms, we refer the reader to the works of, e.g., Maeder et al. (2013), Mathis et al. (2013), and Stancliffe & Lattanzio (2011).

In this work, to account for extra mixing on the upper giant branch, we follow Stancliffe et al. (2009), who use a diffusive prescription for thermohaline mixing based on the work of Ulrich (1972) and Kippenhahn et al. (1980). This prescription was first shown to reproduce the abundance patterns of red giants by Charbonnel & Zahn (2007), when the one free parameter of the theory[7] is appropriately chosen. It has been subsequently shown that the same parameter choice reproduces the observed abundance trends across a wide range of metallicities, for both carbon-rich and carbon-normal metal-poor field stars (Stancliffe et al. 2009) and globular-cluster stars (Angelou et al. 2011, 2012). However, it should be noted that hydrodynamical simulations of thermohaline mixing do not support the calibration of the free parameter in use by 1D stellar-evolution codes (see, e.g., Denissenkov & Merryfield 2011). In principle, we must remain open to the possibility that thermohaline mixing is *not* the cause of abundance changes on the giant branch, or that the extent of this mixing is *over*- or *under*-estimated (we further discuss this issue in Section 4.1 with respect to our analysis). In addition, we have not accounted for the role potentially played by the other mixing mechanisms mentioned above. Multidimensional hydrodynamic simulations of envelope convection in red giants (e.g., Viallet et al. 2013) may help to establish the relevant physical

[6] There are no available models for [C/Fe] = +3.0 and [Fe/H] = −1.3. Such a substantial carbon enrichment at this metallicity corresponds to a carbon mass fraction of around 0.1, which is implausibly high.

[7] The free parameter is related to the aspect ratio of the salt fingers responsible for the mixing (see Charbonnel & Zahn 2007, for further details).

mechanism(s) at work.

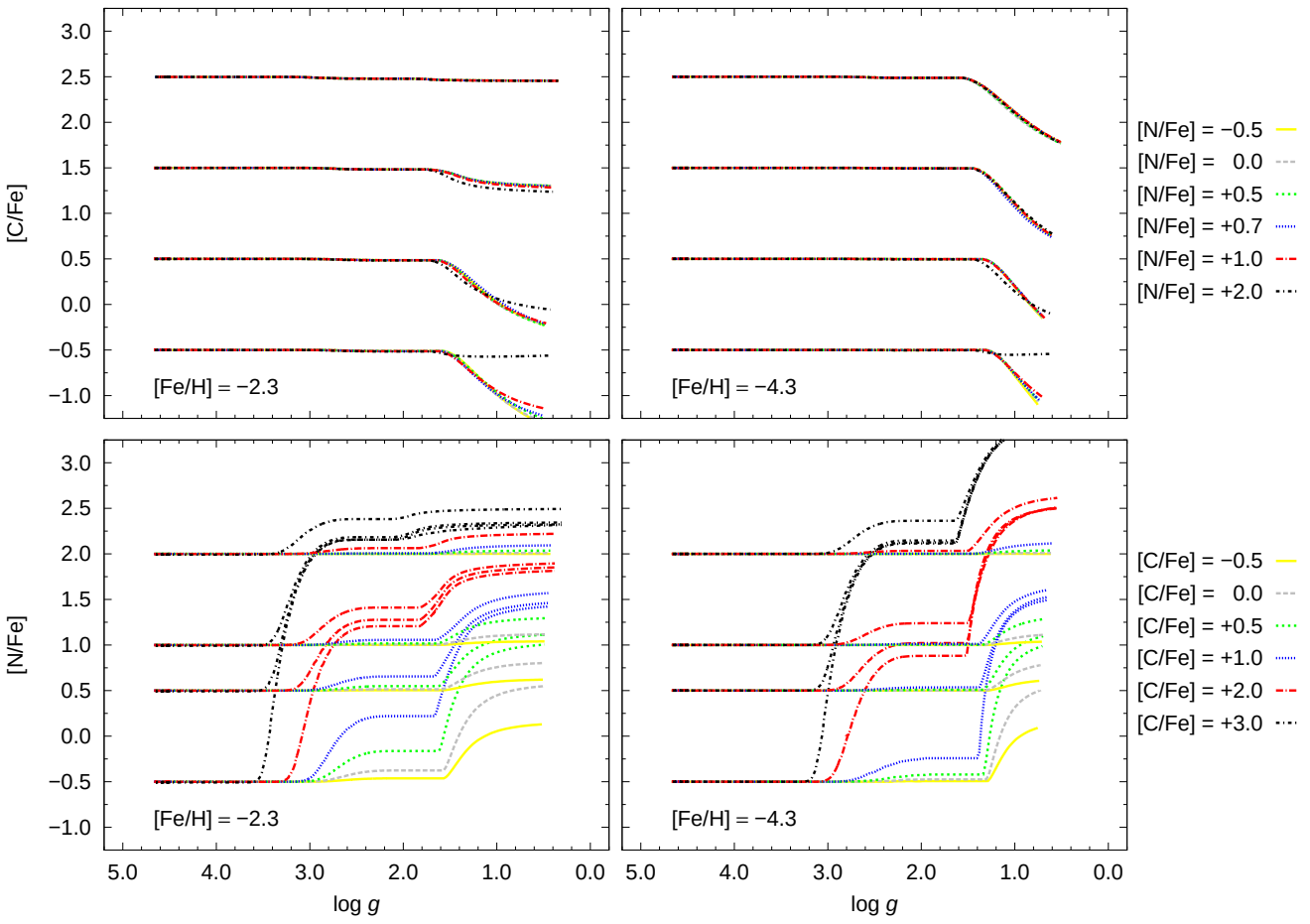

**Figure 1.** [C/Fe] (upper panels) and [N/Fe] (lower panels), as a function of $\log g$, for models with [Fe/H] = −2.3 and [Fe/H] = −4.3. The keys on the right side of the plots list the different initial nitrogen and carbon abundances of each model.

Figure 1 shows the behavior of the carbon (upper panels) and nitrogen (lower panels) abundance ratios, as a function of the surface gravity, for a subset of the models with [Fe/H] =−2.3 and [Fe/H] = −4.3. For [C/Fe], the nitrogen content is largely irrelevant, unless the initial [C/Fe] is very low ([C/Fe]< 0.0). The more C- and N-enhanced models appear to deplete more carbon than the less-enhanced ones. This is expected, since these models behave similarly to more metal-rich models, and spend less time on the RGB (see Stancliffe et al. 2009, for further details). For [N/Fe], there is a clear correlation between initial carbon content and the subsequent nitrogen evolution. More initial C leads to larger amounts of N at both first dredge-up and on the upper RGB. This effect becomes less significant as the initial nitrogen content rises.

Using the models described in this Section, it is also possible to see how the C and N abundance ratios relate to each other during the evolution on the RGB. Figure 2 shows the behavior of [C/Fe]+[N/Fe] (upper panel) and [(C+N)/Fe] (lower panel), as a function of $\log g$, for models with [Fe/H] = −4.3. We caution that, by simply adding [C/Fe] and [N/Fe], as has sometimes been done in previous work, it is not possible to assess the true level of C+N enhancement, because one cannot differentiate between cases with high C or high N. In some regions the C+N measurement is dominated by C, in some it is dominated by N, and in others the two elements provide similar enhancement.

For a proper treatment of the C+N combination, it is thus necessary to employ [(C+N)/Fe] = log [(C+N)/Fe] − log $[(C+N)/Fe]_{\odot}$, i.e., the correct formal definition of [(C+N)/Fe]. As expected, this ratio remains almost flat throughout the evolution, given that the total CN content in the star remains unchanged and proton-burning reactions only influence the relative proportions of the CN nuclei. In addition, there are small variations with metallicity, mainly due to extra mixing having less effect at higher metallicities.

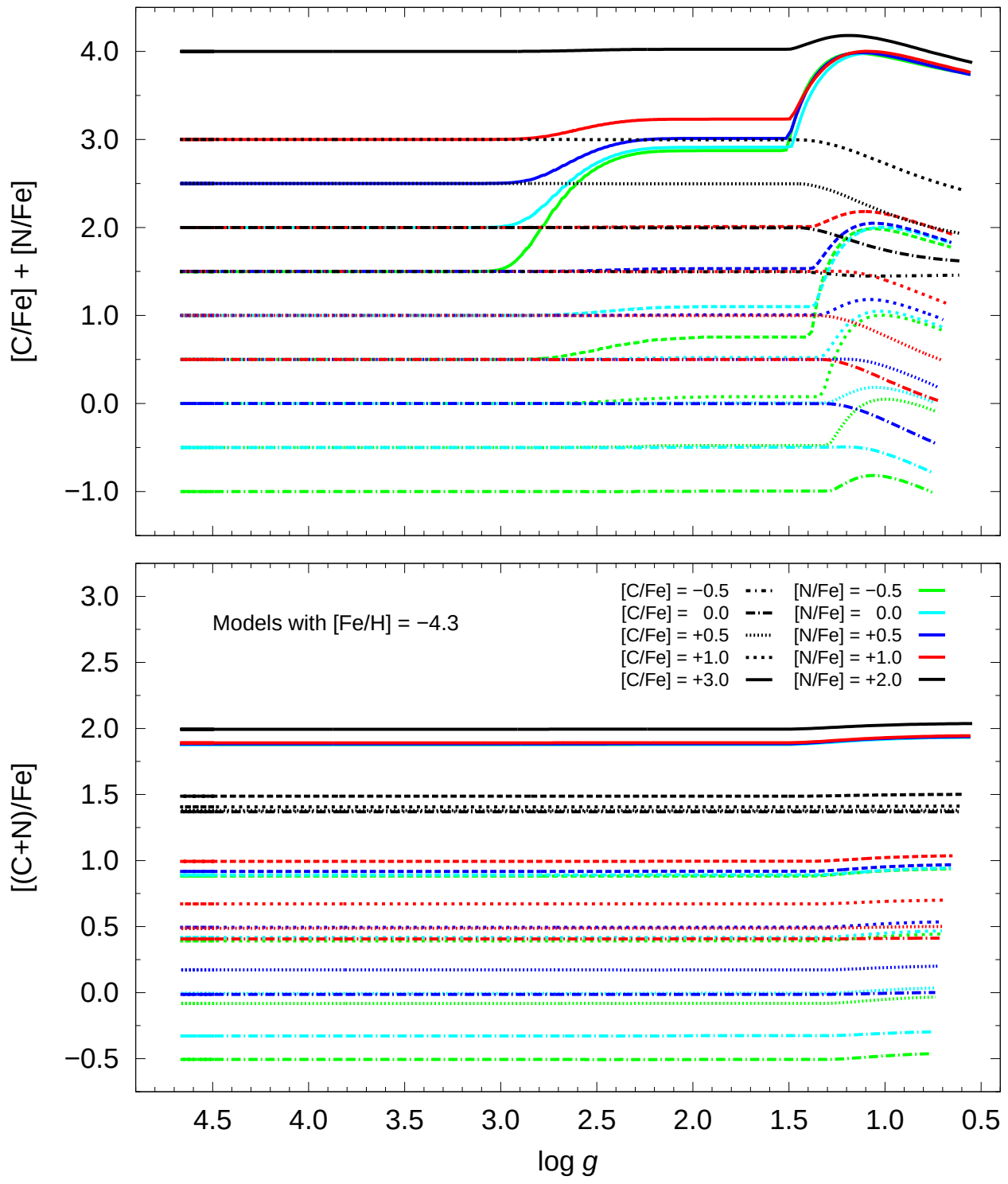

**Figure 2.** [C/Fe]+[N/Fe] (upper panel) and [(C+N)/Fe] (lower panel), as a function of $\log g$, for a series of models with [Fe/H] = −4.3 (see text for definitions). The combinations between different line types and colors give the initial conditions for the 25 models showing in each panel.

## 3. LITERATURE DATA

For the purpose of determining carbon-abundance corrections, based on stellar-evolutionary status, we attempted to collect all available literature data to select a sample of stars with high-resolution spectroscopic metallicities [Fe/H]< −1.0 that have available stellar-atmospheric parameters, along with several critical elemental abundances, including carbon ([C/Fe]), nitrogen ([N/Fe]), strontium ([Sr/Fe]), and barium ([Ba/Fe]) abundance ratios, where available. Our sample is based on the most recent version of the SAGA database (Suda et al. 2008) and the compilation of literature data by Frebel et al. (2010). In addition, we collected data from the literature for studies published after these compilations were assembled. Individual references include: Allen et al. (2012), Akerman et al. (2004), Aoki et al. (2002), Aoki et al. (2006), Aoki et al. (2007), Aoki et al. (2008), Aoki et al. (2013), Barklem et al. (2005), Cohen et al. (2008), Cui et al. (2013), Goswami et al. (2006), Gratton et al. (2000), Hansen et al. (2011), Hansen et al. (2014), Hollek et al. (2011), Hollek et al. (2014), Ito et al. (2013), Johnson et al. (2007), Jonsell et al. (2006), Lai et al. (2007), Lai et al. (2008), Masseron et al. (2010), Mashonkina et al. (2012), McWilliam et al. (1995), Meléndez & Barbuy (2002), Placco et al. (2013), Placco et al. (2014a), Preston et al. (2006), Roederer et al. (2008a), Roederer et al. (2010), Roederer et al. (2014), Simmerer et al. (2004), Sivarani et al. (2006), Sneden et al. (2003), Thompson et al. (2008), Yong et al. (2013), and Zhang et al. (2011). The full sample of literature data contains 863 objects, with a total of 792 stars with [Fe/H] < −1, $\log g$ > 0.0 and [C/Fe] mea-surements. The only exceptions, where upper limits on carbon were used, are: SDSS J102915 (Caffau et al. 2011), CD−38°245, HE 1424−0241 (Yong et al. 2013), and HE 2239−5019 (Hansen et al. 2014). For consistency, we re-scaled all metallicities and abundances to the Asplund et al. (2009) solar photospheric values.

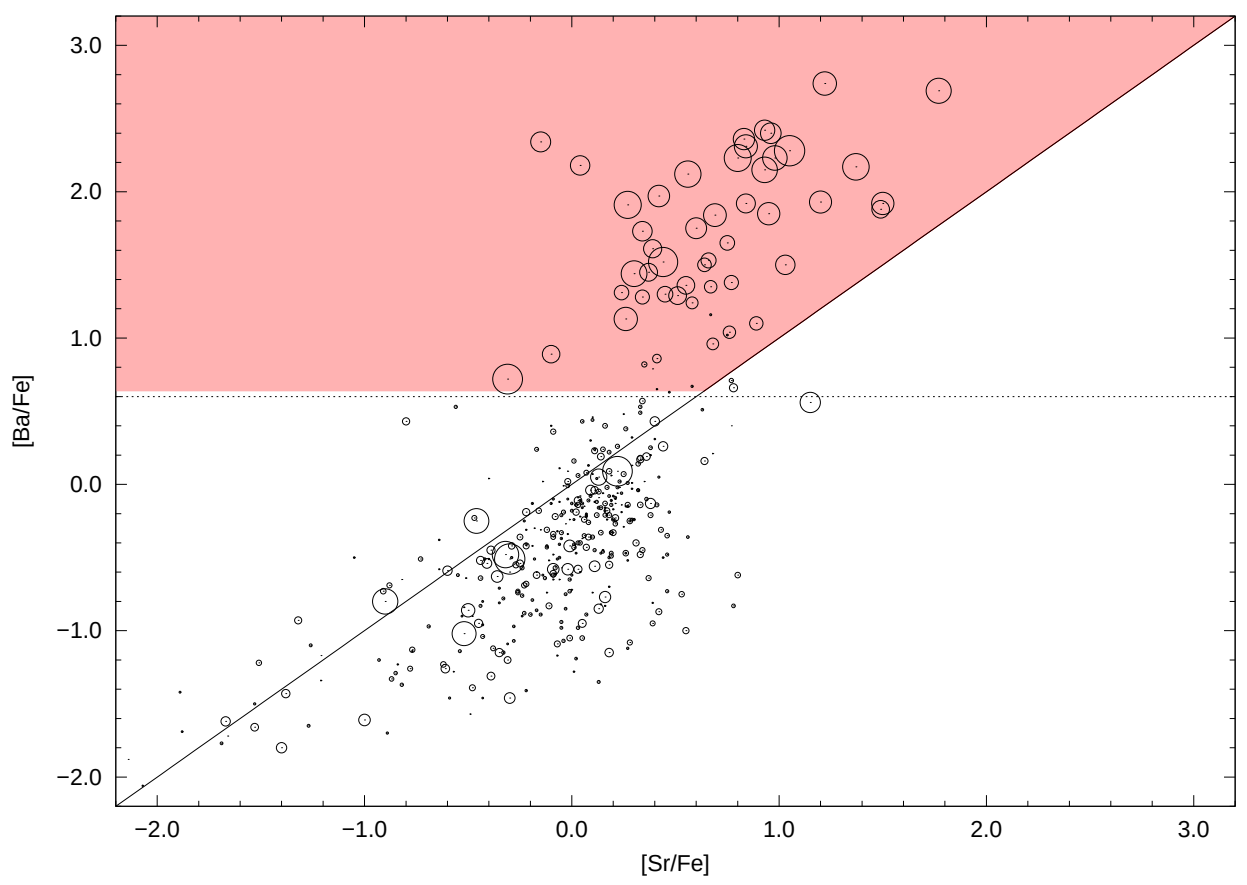

**Figure 3.** [Ba/Fe] vs. [Sr/Fe] distribution for the literature sample. The solid line marks the [Ba/Sr] = 0.0 line, and the dotted line the [Ba/Fe] = +0.6. The symbol size is proportional to the carbon abundance. The shaded area shows the location of the CEMP-$s$ and CEMP-$r/s$ stars that were excluded from the analysis.

To determine the CEMP-star frequencies as a function of metallicity, known CEMP-$s$ and CEMP-$r/s$ stars should be excluded, since these are believed to be enriched at a later time in their evolution by a now-extinct AGB companion. Besides their enhanced carbon, these objects exhibit a distinct signature of $s$-process elements. Figure 3 shows the behavior of the [Ba/Fe] and [Sr/Fe] ratios for the literature data. The size of the points is proportional to the star's [C/Fe], and the red shaded area marks the location of recognized CEMP-$s$ and CEMP-$r/s$ stars. For the purpose of determining carbon corrections and CEMP stellar frequencies, we excluded stars with [Ba/Fe]$>+0.6$ and [Ba/Sr]$>0$ from the subsequent analysis (we studied the effect of changing the criterion to [Ba/Fe]$>+0.8$ on the calculated cumulative CEMP-star frequencies, and results are given in Section 6).

Ideally, only stars with actual [Ba/Fe] and [Sr/Fe] measurements should be used to determine the CEMP-star frequencies. However, it is possible to include stars with upper limits on [Ba/Fe] that indicate [Ba/Fe] $< 0$, and also to assess the level of "contamination" from CEMP-$s$ and CEMP-$r/s$ stars without [Ba/Fe] and [Sr/Fe] measurements. From the 792 stars selected above, 665 exhibit [Ba/Fe]$<+0.6$, upper limits for [Ba/Fe], or no [Ba/Fe] measurements. Within this selected sample, 505 stars have [Fe/H]$\leq -2.0$, which is the metallicity range used for the CEMP-star frequency calculations. There are 87 stars without [Ba/Fe] measurements, and 22 with only upper limits (only 5 upper limits are greater than [Ba/Fe]$>+0.6$[8]). Out of these 66 stars, 40 have either [Sr/Fe]$\geq +0.3$ (the typical lower limit for CEMP-$s$ and -$r/s$ stars; Frebel & Norris 2013) or no Sr abundances measured. Assuming that all 40 stars mentioned above were CEMP-$s$ or -$r/s$, and were mistakenly added to the CEMP-star frequency calculations, they would account for 8% of the total sample (505 stars). However, since the sample has no selection bias on carbon, we would expect a contribution of between 10-20% by CEMP-$s$ or -$r/s$ stars, meaning that no more than $\sim$2% of the selected 505 star sample is contaminated. This fraction could be even lower, since the CEMP-$s$ and CEMP-$r/s$ are more prevalent in the [Fe/H]$>-3.0$ range.

[8] These are: HE 1327−2326 (Frebel et al. 2005), SDSS J2209−0028 (Spite et al. 2013), G 77−61 (Masseron et al. 2012), HE 0107−5240 (Christlieb et al. 2004), and HE 0233−0343 Hansen et al. (2014), which are all well-known CEMP-no stars with [Fe/H]$\leq -4.0$.

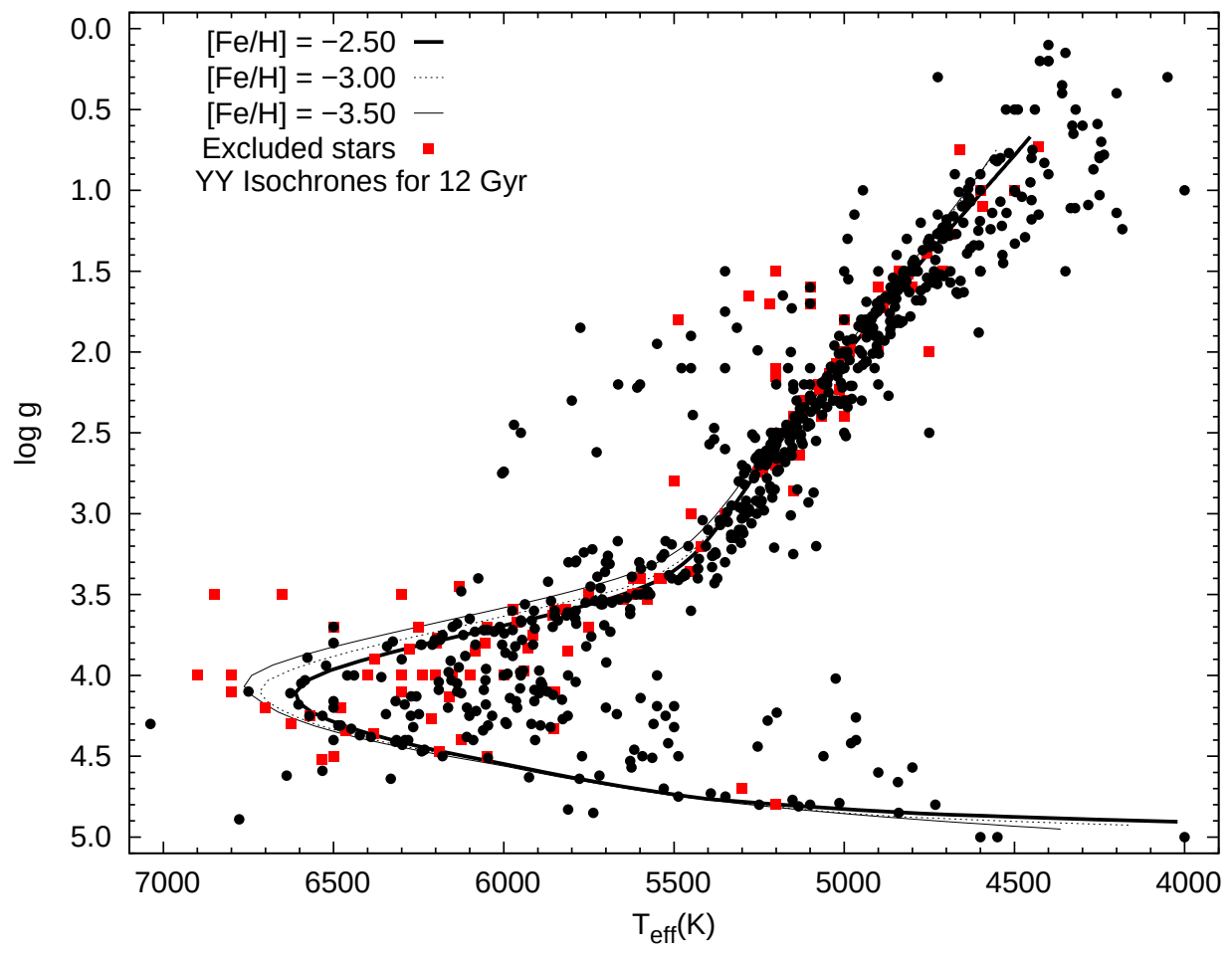

**Figure 4.** H-R diagram for the literature stars. The black filled circles are the dataset used for the determination of the CEMP-star frequencies, and the red filled squares are the excluded CEMP-$s$ and CEMP-$r/s$ stars. Overplotted are the Yale-Yonsei isochrones (Demarque et al. 2004) for ages of 12 Gyr and 3 different values of [Fe/H].

Figure 4 shows the behavior of $T_{\rm eff}$ and $\log g$ for the literature sample, compared with 12 Gyr Yale-Yonsei Isochrones (Demarque et al. 2004) for [Fe/H] $=-3.5$, $-3.0$, and $-2.5$. The CEMP-$s$ and CEMP-$r/s$ stars, which were excluded from the CEMP-star frequency calculations, are also shown. It is possible to see that the bulk of the sample exhibits $\log g < 2.5$, which is the range where the carbon corrections are applied (see Section 4 for further details). In order to establish corrections for [C/Fe] and the frequencies of CEMP stars as a function of metallicity, we used the 505 stars falling outside the shaded area on Figure 3. Figure 5 shows the distribution of carbon abundances, as a function of metallicity, for the literature sample. The symbols are the same as Figure 4. The side panels show the marginal distributions of [C/Fe] and [Fe/H] for the selected stars. As expected, the CEMP-$s$ and CEMP-$r/s$ stars are mostly concentrated at [Fe/H]$>-3.0$ and [C/Fe]$>+1.0$.

Figure 6 shows the carbon abundances, as a function of $\log g$, for the 505 selected stars divided in [Fe/H] bins bracketing the model values. Also shown are the models described in Section 2, assuming an initial nitrogen abundance of [N/Fe]=0.0 (see Section 4.3 for further details). One can see that a number of stars fall outside the $\log g$ range of the theoretical models. In these cases, we used the corrections for the last $\log g$ model value as a constant for all $\log g$ outside the model range, instead of a linear extrapolation that could lead to an over-estimation of the carbon corrections. The following section provides a detailed explanation of this procedure.

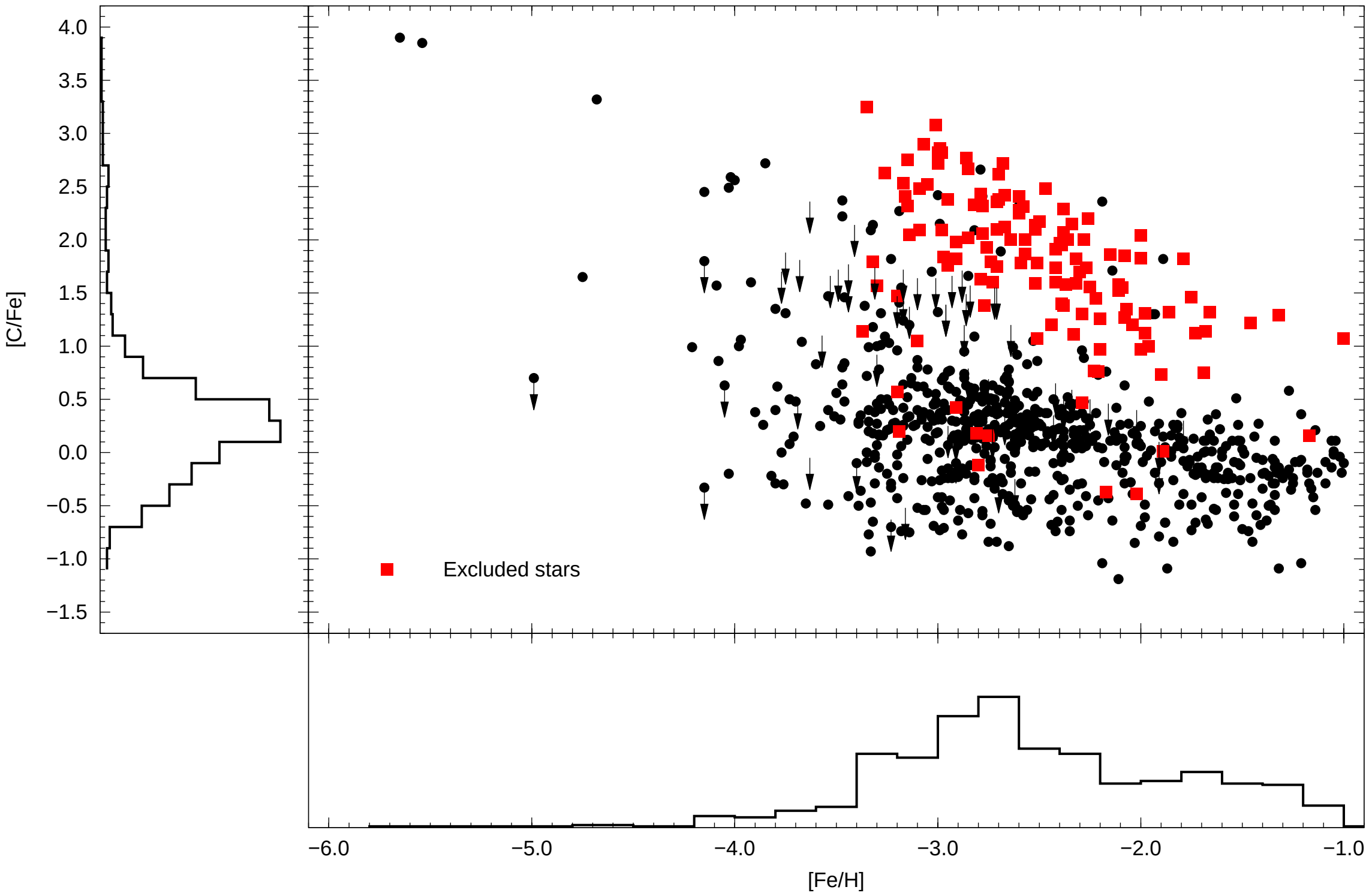

**Figure 5.** [C/Fe], as a function of metallicity, [Fe/H], for the literature stars with available measurements. The black filled dots are the accepted stars, and the red filled squares are CEMP-$s$ and CEMP-$r/s$ stars that were excluded from the analysis. The marginal distributions of each variable for the accepted stars are shown as histograms.

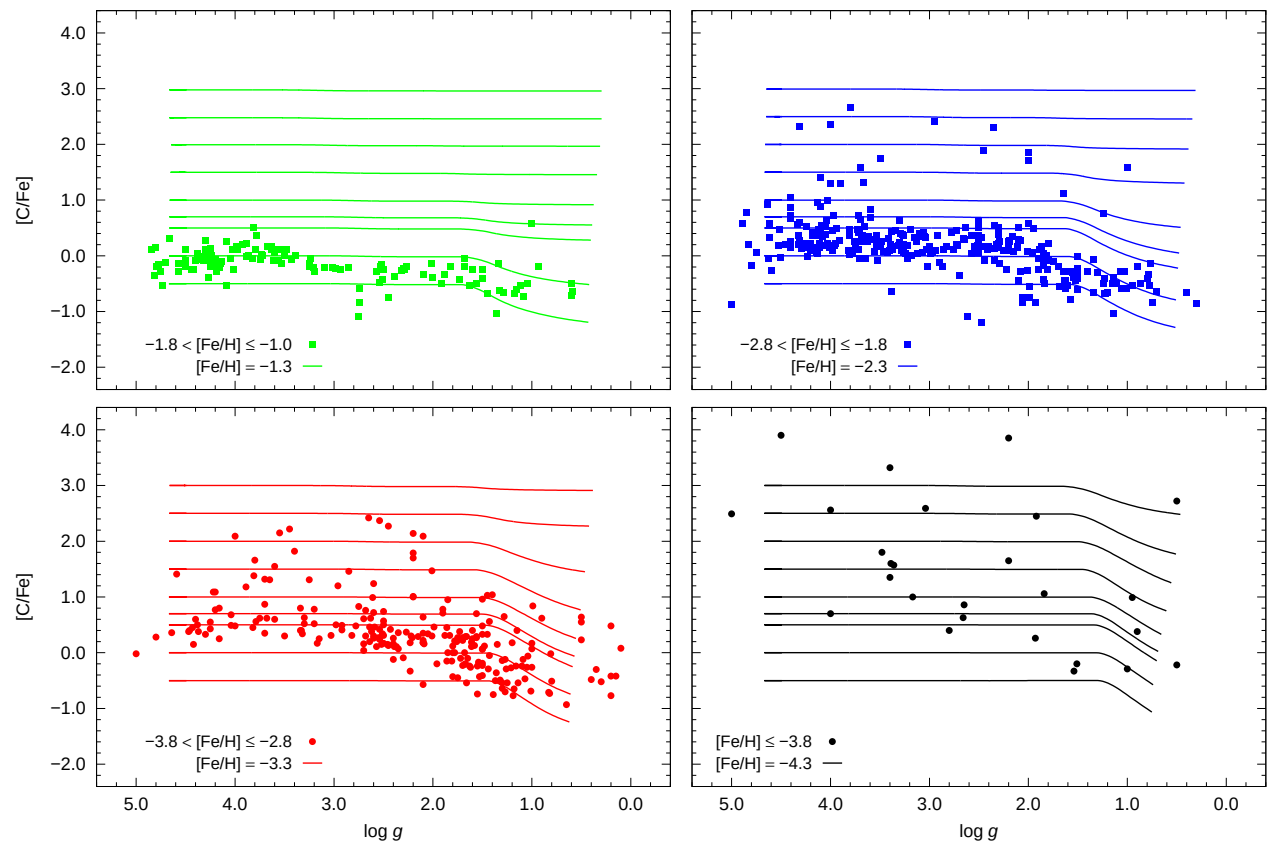

**Figure 6.** Carbonicities, [C/Fe], for the literature data, as a function of $\log g$, divided into four [Fe/H] ranges. The horizontal solid lines are the models with [N/Fe]=0.0.

## 4. CORRECTIONS FOR [C/Fe]

In this section we present our procedure to determine corrections for the observed carbon abundances of CEMP stars, based on their evolutionary status, using the theoretical models described in Section 2. We also discuss possible effects of the uncertainties in the atmospheric parameters and the choice of initial [C/Fe] and [N/Fe] abundances on the derived corrections.

### 4.1. *Further Considerations on the Stellar-Evolutionary Models*

Before proceeding to the determination of the carbon-abundance corrections, it is worth noticing a slight mismatch between the behavior of depletion in the theoretical models and the observations seen in Figure 6, in particular in the top-right panel. The model tracks and the data should show the same decrease in [C/Fe] with decreasing $\log g$ (for $\log g < 2.0$). From the figure it appears, however, that the onset of the mixing mechanism in the models is somewhat "delayed" in $\log g$ space, and only occurs at lower $\log g$ values than the data suggests. This effect is more noticeable for stars with [C/Fe]$< +0.7$. This offset prevents a proper estimate of the amount of depleted carbon. The [C/Fe] corrections would be under-estimated by the models, as would the CEMP-star frequencies. To account for this, we introduce a shift in $\log g$ on the models before calculating the corrections for [C/Fe]. Such a shift should lead to a constant average [C/Fe] as a function of $\log g$ after the abundance corrections are applied.

To test this hypothesis, we calculated the corrections for the carbon abundances for three $\log g$ offsets (using the procedure described below in Section 4.2): (i) original model $\log g$ only; (ii) model $\log g$ + 0.3 dex and; (iii) model $\log g$ + 0.5 dex. Results are shown in Figure 7. One can see that corrections based solely on the original models (Panel b) are not sufficient to recover the depleted carbon on the upper-RGB ($\log g$ <2), whereas the shifted models, with an early mixing onset, are able to keep the [C/Fe] values constant over the entire $\log g$ range. Even though the corrections for case (ii) improve the behavior of the corrected carbon abundances, there is still a decrease in the distribution for $\log g$ <2 (Panel c). Hence, we proceed with the +0.5 dex correction, shown in Panel (d). The effect of these corrections on the [C/Fe] averages are described in Section 5. We also quantify the effect of the $\log g$ shift on the CEMP-star frequencies, as described in Section 6.

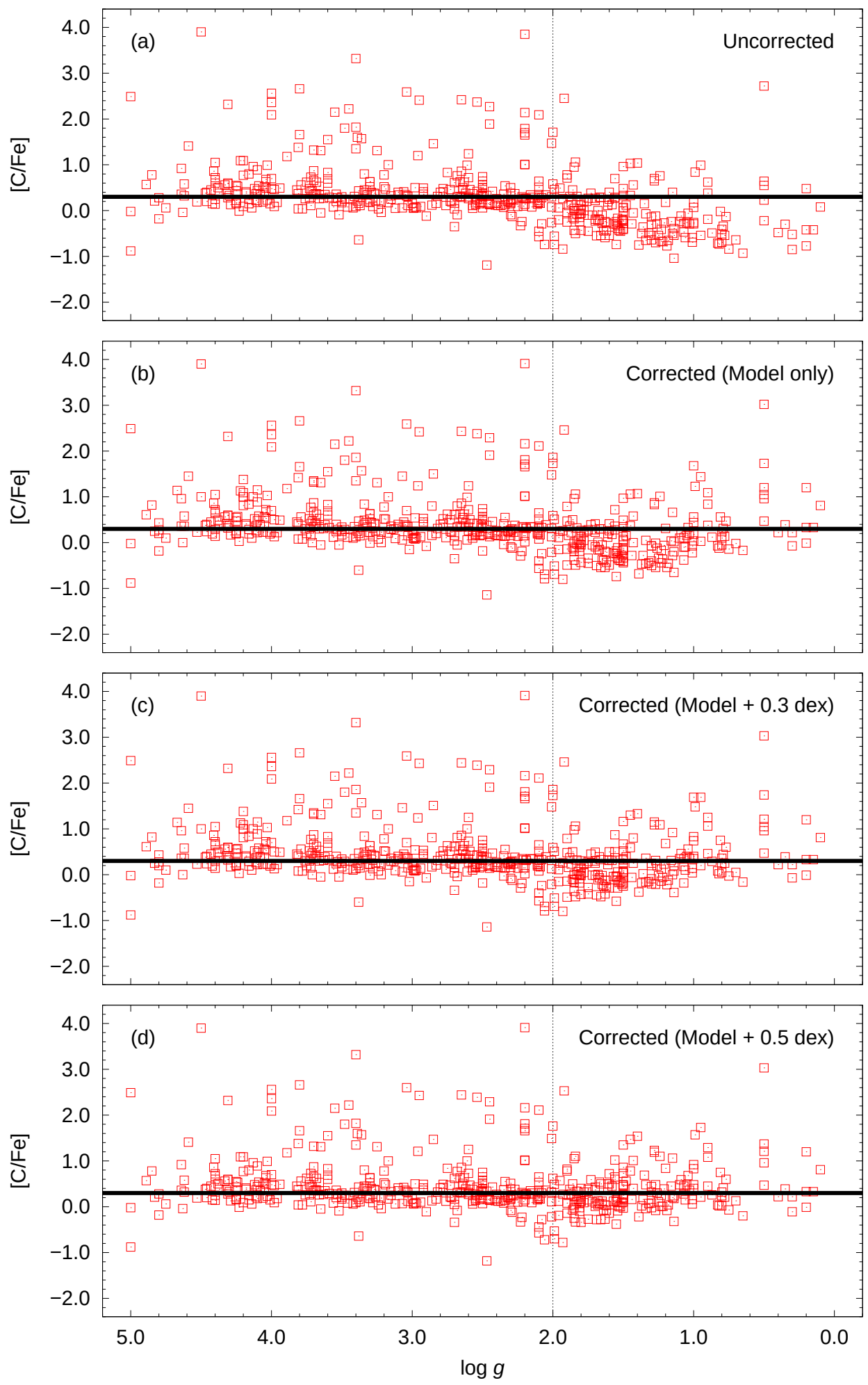

**Figure 7.** [C/Fe] as a function of $\log g$ for the literature sample. Panel (a): Observed carbon abundances. Panel (b): [C/Fe] corrected with the original models only. Panel (c): [C/Fe] corrected with model $\log g$ + 0.3 dex shift. Panel (d): [C/Fe] corrected with model $\log g$ + 0.5 dex shift. The solid horizontal line represents a constant [C/Fe]=+0.3 value to guide the eye. The vertical dashed line is a reference line at $\log g$ = 2.0.

### 4.2. *Procedure*

Since the initial carbon abundance of a given star (at least one that has evolved past the first dredge-up) is a-priori unknown, we cannot match the observed abundance with the *initial* abundance of the model as a first approximation. To deal with this issue (we discuss further implications in Section 4.3), we developed a simple procedure to estimate the amount of carbon depletion for a given set of stellar parameters and carbon abundance, without any assumption on the initial carbon abundance.

For a given set of [Fe/H], [C/Fe], and $\log g$, we first identify the two closest model metallicities and, for each of these, we find the two models with the closest ($\log g$,[C/Fe]) values to the input. Then, for each of the four chosen models, a correction is determined by the difference between the initial [C/Fe] of the model and the [C/Fe] value for the given $\log g$. The final [C/Fe] correction for the input value is then given by a linear interpolation (in [Fe/H] and [C/Fe]) of the four model corrections. This process is repeated for each initial [N/Fe] value. In Section 4.3 we discuss the [N/Fe] model choices and uncertainties in detail.

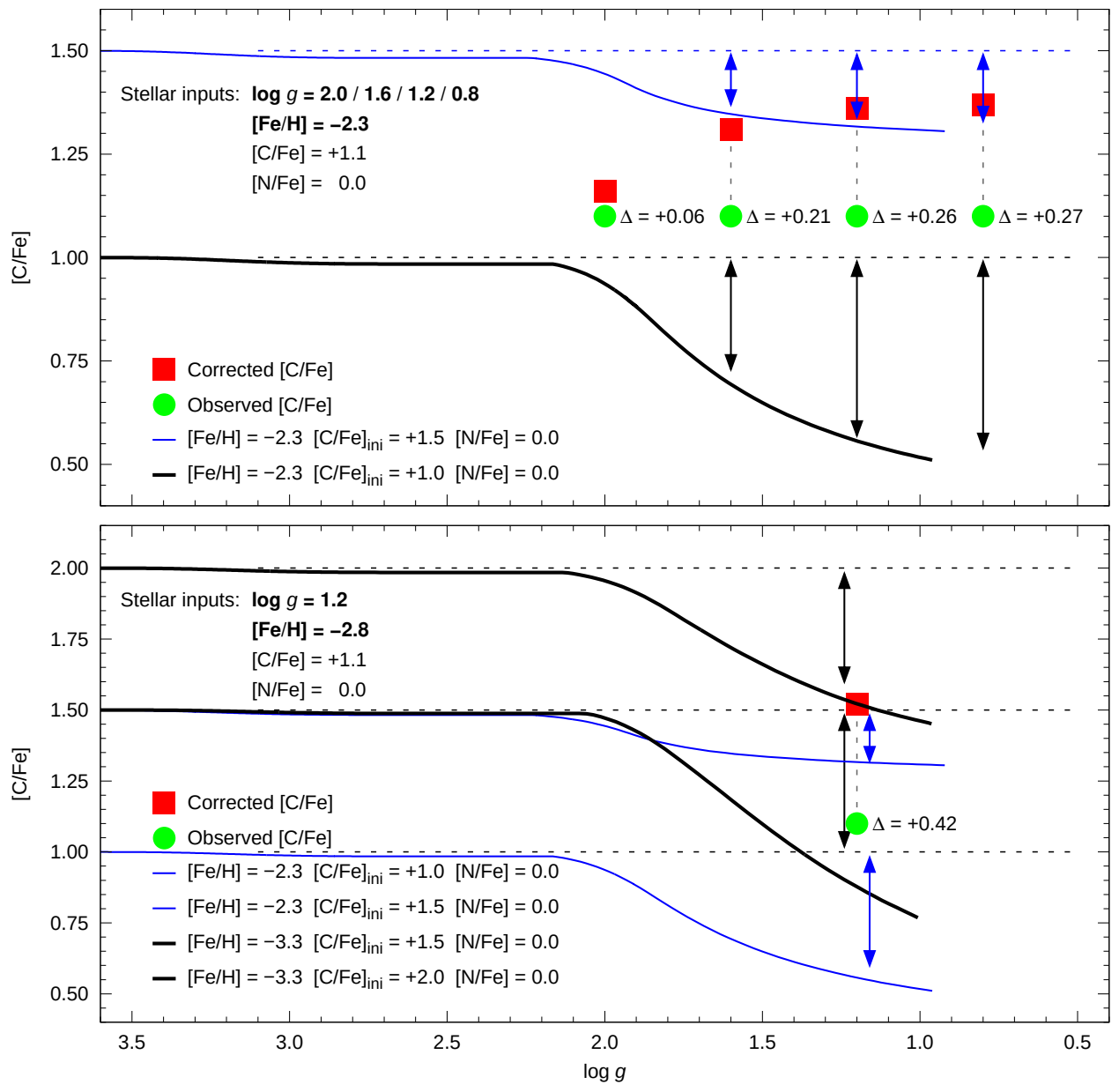

**Figure 8.** Procedure for the determination of carbon-abundance corrections. Top panel: Observed (green circles) and corrected (red squares) carbon abundances for four different $\log g$ values. The horizontal dashed lines show the initial [C/Fe] model values. The final corrections, Δ, are determined from a linear interpolation of the corrections of each model (vertical arrows). Since the [Fe/H] input value coincides with the model values, no interpolation in [Fe/H] is made. Bottom panel: Complete interpolation procedure for $\log g$ = 1.2, and [Fe/H]=−2.8. The four vertical arrows represent the corrections for each model. The final correction is a linear interpolation in both [Fe/H] and $\log g$.

As an example, consider the following input observed parameters: $\log g$=1.3, [Fe/H]=−3.0, and [C/Fe]=+1.0 (considering a fixed [N/Fe]=0.0). The two closest model metallicities are [Fe/H]=−2.3 and −3.3. For $\log g$=0.8 and [C/Fe]=+1.0, the closest *initial* model carbon abundances are [C/Fe]=+1.0 (M1) and +1.5 (M2) for the [Fe/H]=−2.3 model, and [C/Fe]=+1.5 (M3) and +2.0 (M4) for the [Fe/H]=−3.0 model. For each of the four models, a correction is determined: $\Delta_{M1}$ = +0.42 dex, $\Delta_{M2}$ = +0.18 dex, $\Delta_{M3}$ = +0.56 dex, and $\Delta_{M4}$ = +0.44 dex. The final interpolated correction is Δ = +0.47 dex.

Figure 8 illustrates this procedure. The top panel shows four different $\log g$ values, with fixed [Fe/H]=−2.3, [C/Fe]=+1.1, and [N/Fe]=0.0. For clarity, we chose an input [Fe/H] value that matches one of the models, so the interpolation is performed only between models with different initial carbon abundances. The horizontal dashed lines show the initial [C/Fe] model values, and the vertical arrows represent the amount of carbon depleted for a given $\log g$ value, which corresponds to the carbon correction. The double-headed arrows represent the correc-tions for each selected model, and the Δ values are the final interpolated corrections for each log $g$ value.

The bottom panel of Figure 8 shows the complete procedure for determining the carbon correction, for a star with [Fe/H] = −2.8, [C/Fe] = +1.1, [N/Fe]= 0.0, and log $g$=1.2. The solid lines are the four models chosen for the interpolation, and the dashed lines show the initial carbon abundance of the models. The final correction is given by a linear interpolation of the four individual corrections, in the [Fe/H] vs. [C/Fe] plane. It is worth noting from the bottom panel of Figure 8 that, depending on the metallicity, the initial [C/Fe] choices change. In this example, the [Fe/H] = −2.3 models have initial [C/Fe] of +1.0 and +1.5, while the [Fe/H]=−3.3 models have initial [C/Fe] of +1.5 and +2.0. This is just a reflection of the fact that, for a given log $g$ and [C/Fe], the amount of carbon depletion increases with decreasing metallicity.

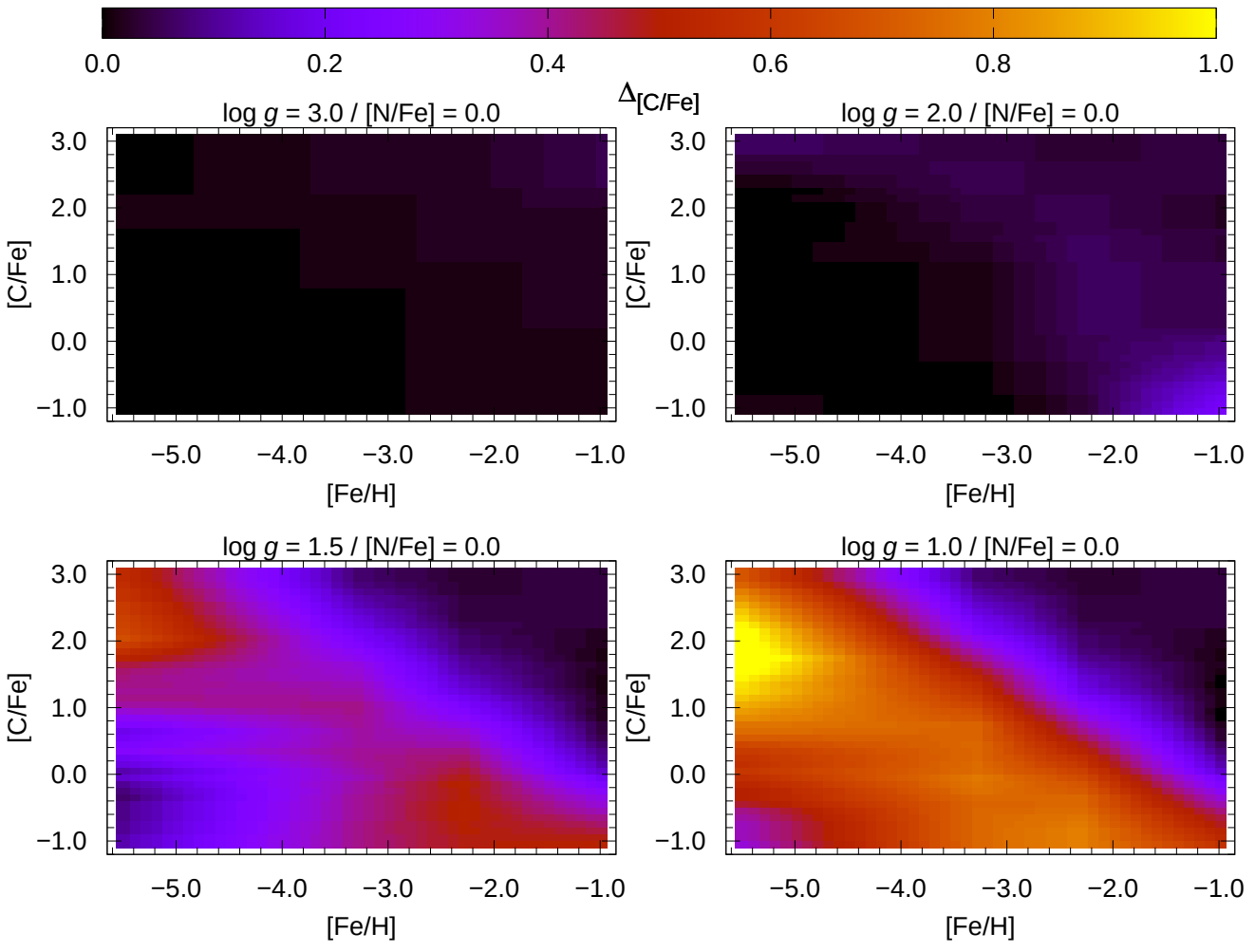

**Figure 9.** [C/Fe] correction map for log $g$ = 3.0, 2.0, 1.5, and 1.0.

By choosing the interpolation instead of fixed bins in [Fe/H] and [C/Fe], the corrections exhibit a smooth transition throughout the parameter space. Figure 9 shows a map of the calculated corrections in the [Fe/H] vs. [C/Fe] plane, for four different log $g$ values and [N/Fe] = 0.0. For log $g$ = 1.0, the corrections can be as high as +1.0 dex for [C/Fe] = +2.0 and [Fe/H] < −5.0. In contrast, the corrections do not exceed +0.25 dex for log $g$ = 2.0, and are almost non-existent (Δ < +0.05 dex) for log $g$ = 3.0. This is physically reasonable – carbon is little affected by the action of first dredge-up, and is only substantially depleted on the upper part of the RGB. We have developed and made available an online tool[9], which allows the user to calculate the carbon corrections for a given set of stellar parameters.

## 4.3. *Uncertainties*

The two main factors that can affect the determination of the carbon-abundance corrections are the choice of an appropriate model (based on [Fe/H], initial [C/Fe], and [N/Fe]), and the uncertainties associated with the input log $g$ values. We discuss these issues below.

[9] http://staff.gemini.edu/∼vplacco/carbon-cor.html

### 4.3.1. *Choice of Model*

The model choice itself results in two sources of uncertainties: (i) the observational uncertainties associated with the carbon abundances and metallicity determinations, and (ii) the choice of the correct initial [N/Fe], when the observed value is not available. The interpolation procedure described above somewhat minimizes these effects, but below we provide estimates that can be used as guidelines on the uncertainties of the carbon corrections.

In order to quantify how the corrections would change, given the uncertainties associated with the observational determinations of [Fe/H] and [C/Fe], we calculated the carbon corrections (assuming fixed log $g$ = 3.0/2.0/1.5/1.0 and [N/Fe] = 0.0), for a series of [Fe/H] and [C/Fe] combinations, in steps of 0.25 dex, with [Fe/H] ranging from −4.0 to −2.0, and [C/Fe] from 0.0 to +2.0. The size step of 0.25 dex is similar to the total uncertainty associated with measurements of [Fe/H] and [C/Fe].

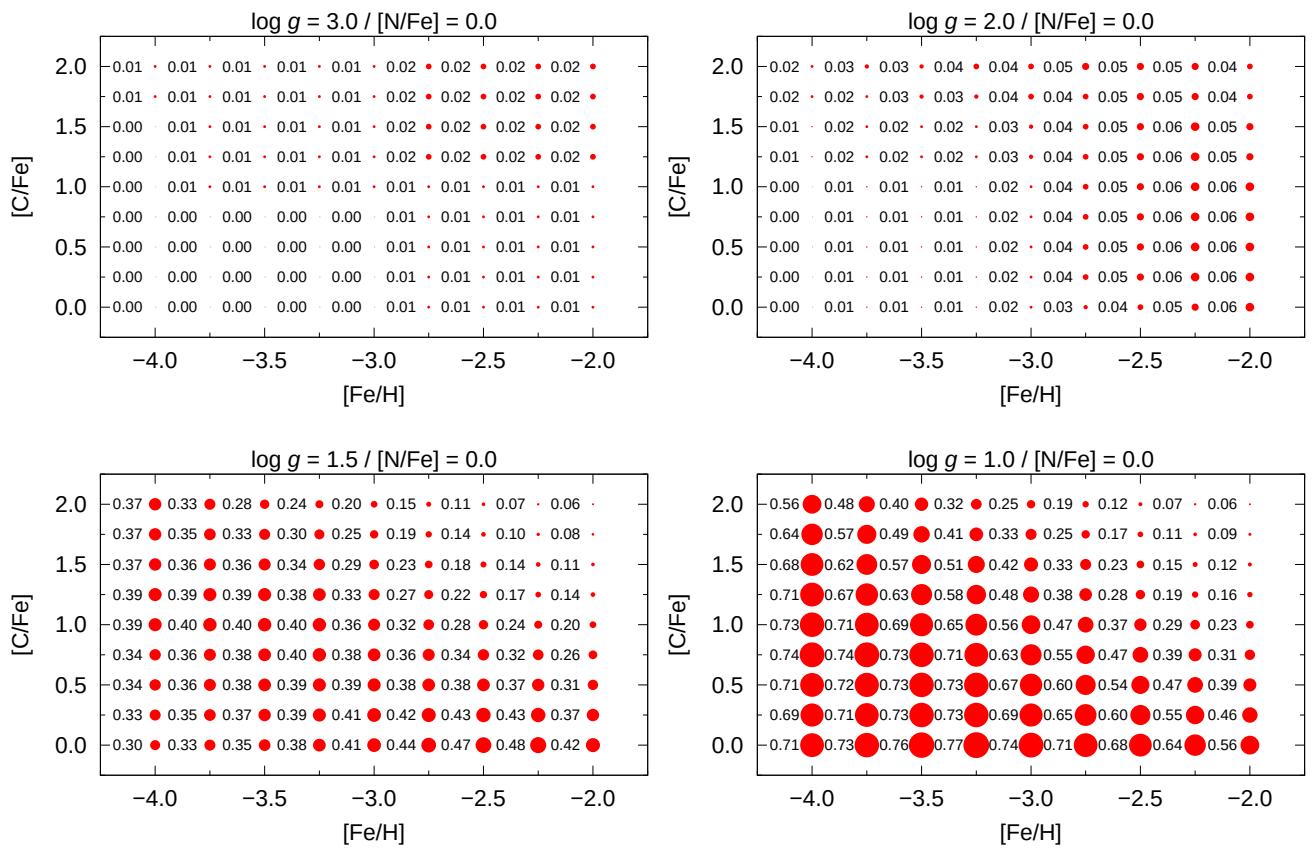

**Figure 10.** Carbon-abundance corrections for log $g$ = 3.0, 2.0, 1.5, and 1.0, assuming [N/Fe] =0.0 as the initial nitrogen abundance. The size of each point is proportional to the correction value, which is shown on the left side of each symbol. The step size is 0.25 dex in both [Fe/H] and [C/Fe].

Figure 10 shows the result of this exercise. Each panel shows the corrections for different log $g$ values, where the size of the points are proportional to the numbers shown on the left side of each point. As already mentioned, the corrections would be mostly affected by uncertainties in the measured parameters for log $g$ ≤ 1.0. For example, a star with log $g$ = 1.0, [Fe/H] = −3.0, and [C/Fe] = +1.0 has a determined correction of +0.56 dex. Assuming an uncertainty of ±0.25 dex in [C/Fe], the corrections would vary from +0.48 dex to +0.63 dex (−0.08 dex and +0.07 dex from the calculated value). For log $g$ = 1.5, the corrections would change by −0.03 dex and +0.02 dex, and for log $g$=2.0 the corrections would change by no more than ±0.01 dex. A similar exercise can be performed for [Fe/H], even though the uncertainties of its measurement based on high-resolution spectra are often on the order of ±0.10 dex or less.

Nitrogen abundances are much more challenging to determine in the optical spectra of metal-poor stars. The CN band at 3883 Å can be used if available (assuming a fixed carbon abundance - Placco et al. 2013), or better, the NH molecular feature at 3360 Å (Placco et al. 2014a). Since a large number of our literature-sample stars lack determinations of nitrogen abundances, we studied the effect of a poor choice of initial [N/Fe] on the carbon corrections. As seen in Figure 1, the corrections are higher for models with low initial carbon abundance and high initial nitrogen abundance. One possibility to assess if this would be a possible physical scenario is by looking at the distribution of [C/Fe] as a function of [N/Fe] for stars that did not evolve through the RGB. Results are shown in Figure 11.

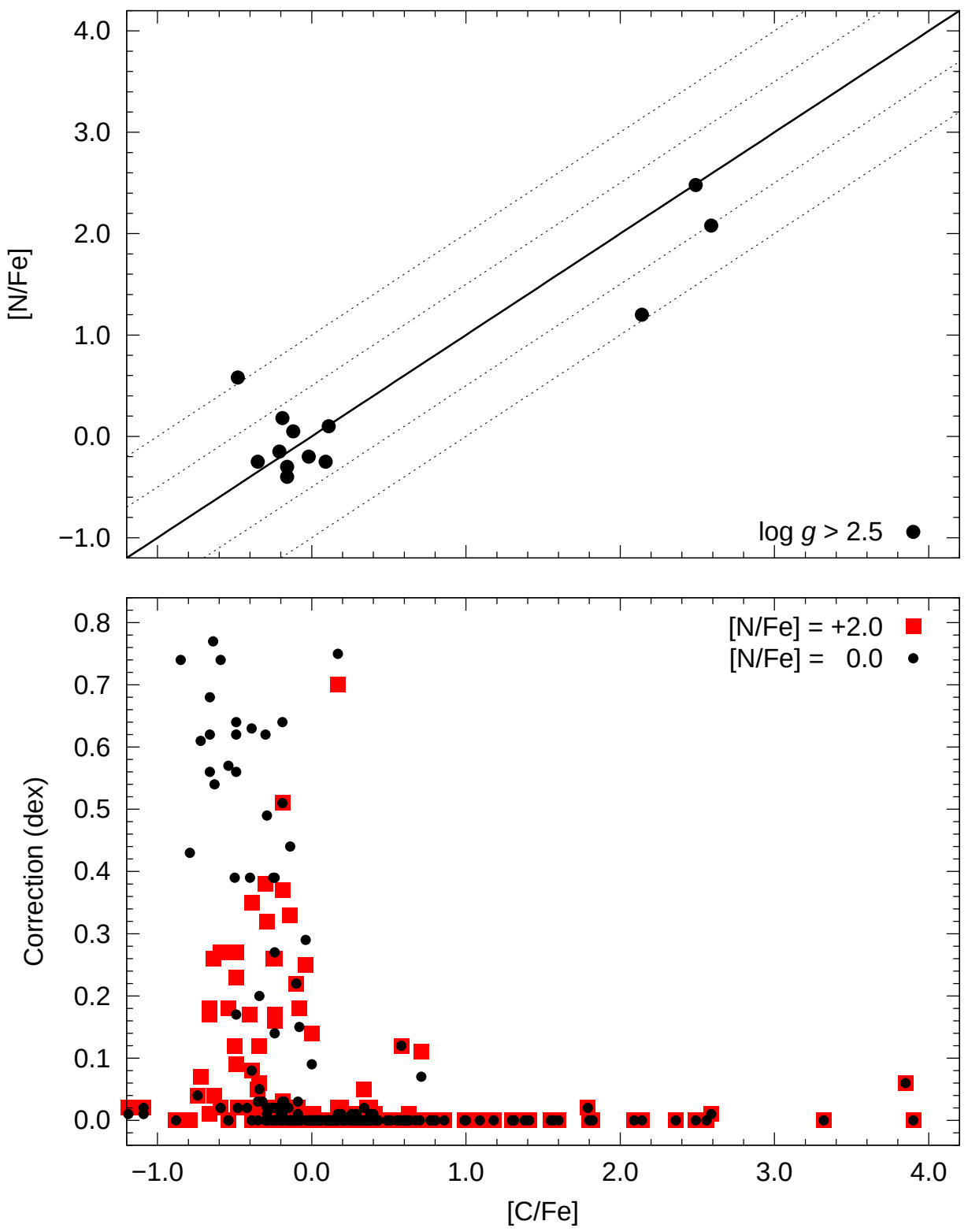

**Figure 11.** Top panel: [N/Fe] vs. [C/Fe] for the literature stars with $\log g > 2.5$ and available carbon and nitrogen abundances. The solid line is [C/N] = 0, and the dotted lines [C/N] = $-1.0/-0.5/0.5/1.0$. Bottom panel: Comparison between the carbon corrections as a function of [C/Fe] for the [N/Fe] = 0.0 (black filled circles) and [N/Fe] = +2.0 (red filled squares) models.

The upper panel of Figure 11 shows the behavior of high-resolution [N/Fe] vs. [C/Fe] measurements for stars with $\log g > 2.5$. One can see that the majority of the stars are within $\pm0.5$ dex from the [N/Fe]=[C/Fe] line. This suggests that a good approximation for the initial nitrogen abundance could be the same value as the carbon abundance. The lower panel of Figure 11 shows the carbon corrections for the [N/Fe] =0.0 and [N/Fe] =2.0 models, for stars with measured nitrogen abundances. As expected from the models, the corrections are larger when both carbon and nitrogen abundances are low, and are negligible for stars with [C/Fe] $> +2.0$. Since the choice of initial [N/Fe] seems to affect mostly the stars with lower [C/Fe], we chose, for simplicity, the [N/Fe] = 0.0 model for the determination of the CEMP-star frequencies in Section 6. Moreover, since there are no large differences in the carbon corrections between the nitrogen models for [C/Fe]$\geq +0.5$, this will not affect the CEMP-star frequencies calculations discussed below.

#### 4.3.2. *Uncertainty in* $\log g$

An additional source of uncertainty on the carbon-correction determination is the one associated with the measured surface gravity, $\log g$. This is a combination of the uncertainty in the model atmosphere, and on the ability to reliably measure Fe I and Fe II lines in the spectra, which is particularly challenging for stars with [Fe/H]$< -3.0$, even in high-resolution. For the carbon-correction determinations, this uncertainty has a greater impact for stars with $\log g \leq 2.2$, assuming a typical uncertainty of $\sim$0.3 dex. To evaluate the extent of the $\log g$ uncertainty on the [C/Fe] correction, we calculated the corrections for a small grid of [C/Fe] and $\log g$ values, assuming [Fe/H] = $-2.5$ and [N/Fe] = 0.0. Then, for each case, we then redetermined the corrections for two additional cases: $\log g$ +0.30 and $\log g$ $-0.30$.

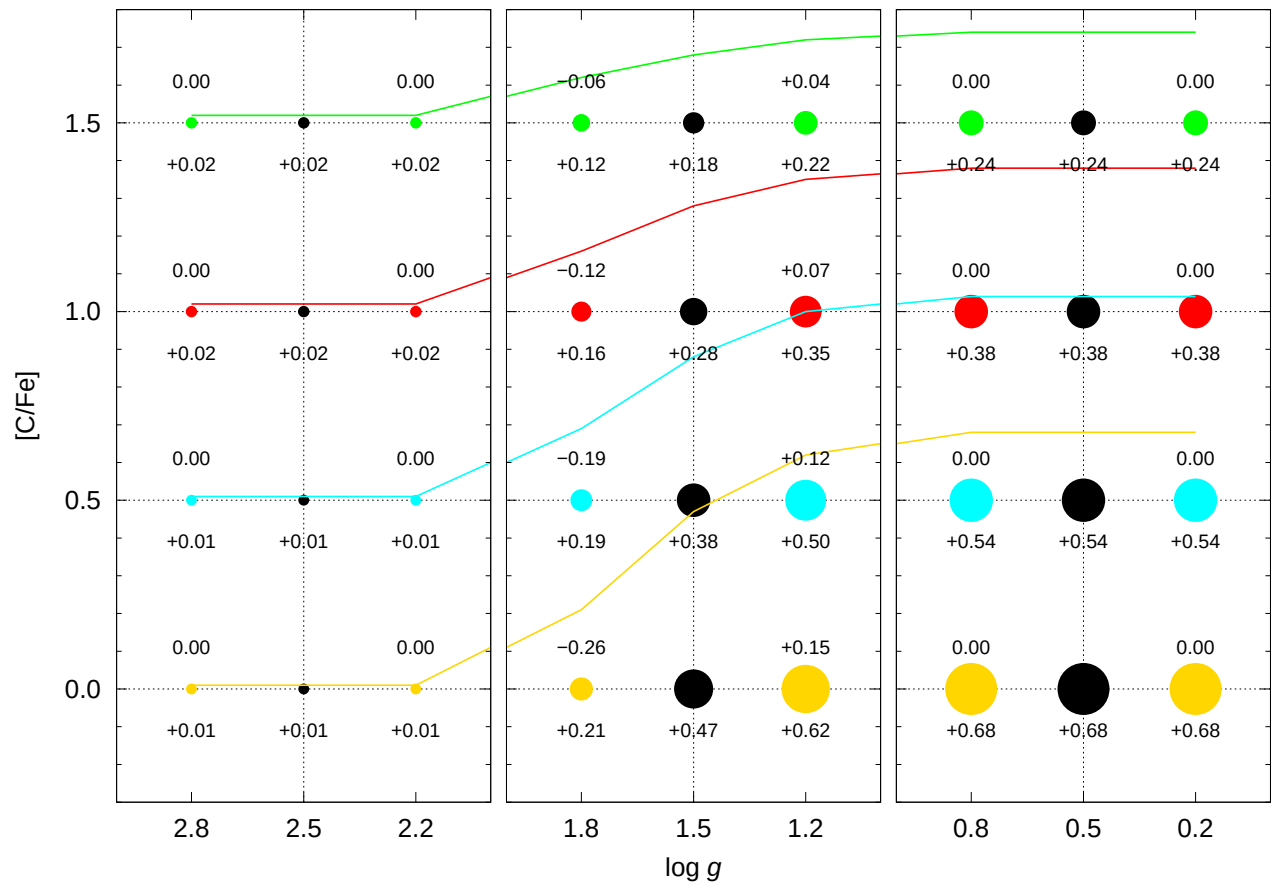

**Figure 12.** Changes in carbon-abundance corrections as a function of $\log g$. The black filled circles, crossed by the dotted lines, are the grid values, with two colored points at $\log g$ $\pm0.3$. Each color represents a different initial [C/Fe]. The values below the points are the [C/Fe] corrections using [N/Fe]=0.00, and the values above the points are difference in the carbon correction from the grid value. The colored solid lines show the corrected [C/Fe] for the circles along the horizontal dotted lines. The size of each point is proportional to the [C/Fe] correction.

Figure 12 shows how these changes in $\log g$ affect the carbon-abundance corrections. The black circles crossed by the dotted lines represent the grid points, and the filled circles at $\pm0.30$ dex in $\log g$ show the changes in the carbon corrections. The numbers below each symbol are the difference between the correction for the shifted $\log g$ value and the grid point, and the solid lines shows the corrected [C/Fe] for each point, matched by its color. For example, for a star with measured [Fe/H] = $-2.5$, [C/Fe] = +1.0 and $\log g$ = 1.5, the calculated [C/Fe] correction for [N/Fe] =0.0 is +0.28 dex. Assuming a $\pm0.3$ dex uncertainty in $\log g$, the correction value would vary between +0.16 dex and +0.35 dex. For a measured $\log g$ = 2.5, the corrections would not vary.

As expected, uncertainties of up to $\pm0.5$ dex in $\log g$ for stars in the $\log g > 3.0$ regime will not have any effect on the derived corrections. For the $\log g$ = 0.5 case, the shifts in $\log g$ produce no deviations, since $\log g$ values outside the shifted model range assume a constant cor-rection. In addition, the $\log g = 1.5$ case is where the corrections are mostly affected. Even then, the introduced shifts in $\log g$ do not produce deviations of more than $\pm 0.30$ dex in the corrections. Once again, these are within the usual $2\sigma$ uncertainties related to observed $\log g$ and [C/Fe] values.

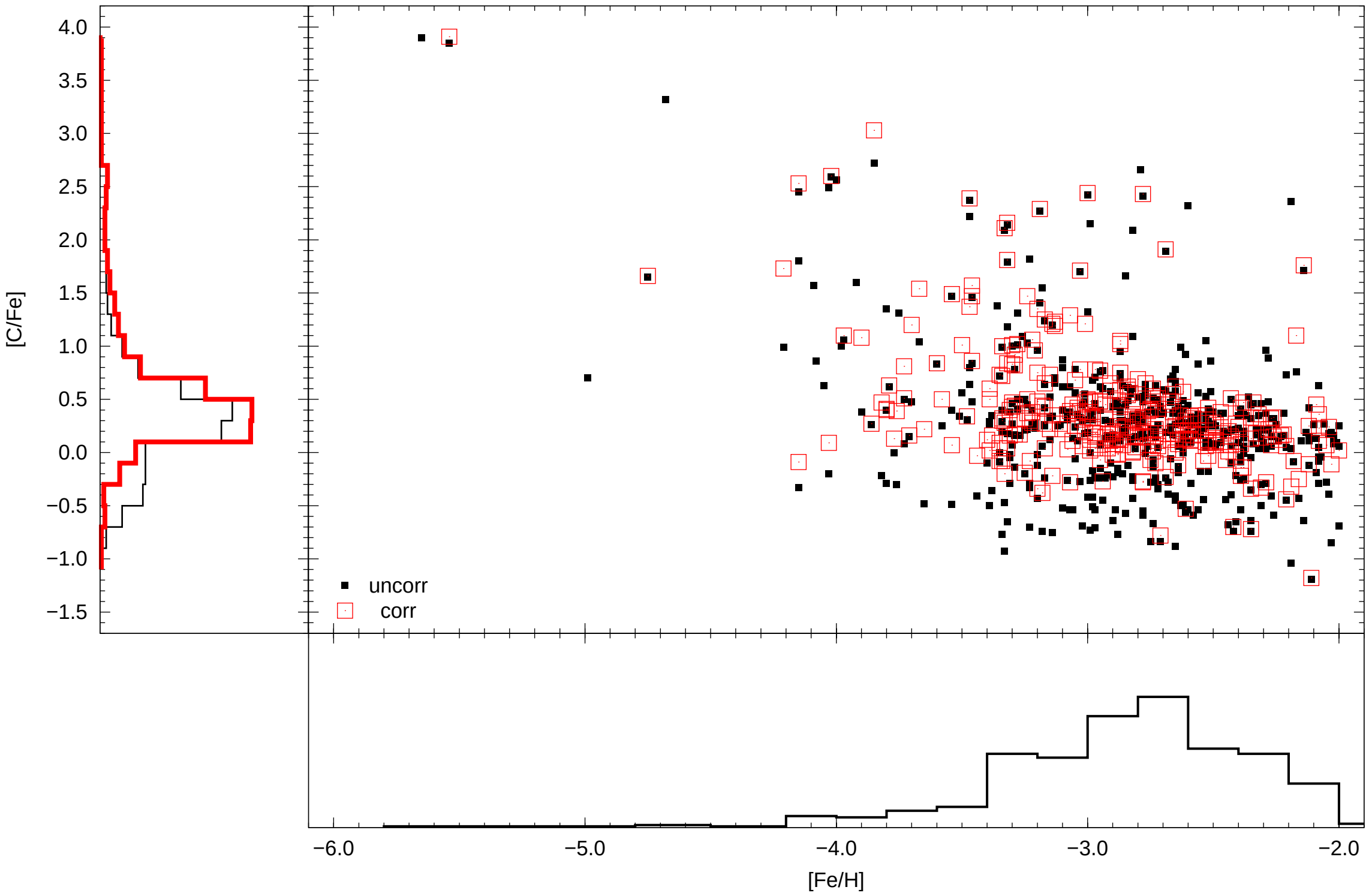

**Figure 13.** Carbonicities, [C/Fe], for the literature stars, as a function of the metallicity, [Fe/H], for the 505 stars with [Fe/H] $\leq -2.0$ selected from the literature. The black filled squares represent the measured abundances, while the red open squares show the corrected values for stars with non-zero corrections (for the [N/Fe] = 0.00 case). The marginal distributions of each variable, including the corrected values, are shown as histograms.

## 5. CARBON-ABUNDANCE CORRECTIONS

### 5.1. *Literature Sample*

The correction procedure explained above was applied to the literature data described in Section 3. Figure 13 shows the distribution of the carbonicities for the 505 selected literature stars (with [Fe/H]$\leq -2.0$), as a function of metallicity, for both uncorrected (black filled dots) and corrected (red open circles) abundances. We only plot corrections different than zero. The applied corrections are based on the [N/Fe] = 0.0 model, and the histograms on the left and bottom panels also show the change in behavior of the carbon distribution. There is no significant change for stars with [C/Fe]$> +2.0$, and the distribution shifts to higher values for [C/Fe]$< +0.5$. The bulk of the stars in the [C/Fe]$< 0.0$ region have corrected values that place them in the $0.0 <$ [C/Fe]$< +0.5$ range. This shift changes the overall behavior of the carbon abundances, and the applied corrections will affect the CEMP-star frequencies as a function of [Fe/H] (see Section 6 for further details).

Figure 14 shows the distribution of the corrected [C/Fe], as a function of luminosity (upper panel), and $\log g$ (lower panel), for the 505 stars selected from the literature. The evolutionary phases are based on the work of Gratton et al. (2000). The green solid line is the CEMP criteria from Aoki et al. (2007), and the black solid lines are the theoretical models [C/Fe] = $-0.50/+0.70/+1.50/+2.50$ (assuming [Fe/H] = $-3.3$ and [N/Fe] = 0.0) shifted by 0.5 dex (see details on Section 4). Black filled squares are the measured abundances; red open squares show the corrected values (using the [N/Fe] = 0.00 corrections). Also shown (green solid line) is the luminosity-dependent CEMP criteria from Aoki et al. (2007). The models displayed in Figure 14 serve as guidelines, and were not used to correct all of the carbon abundances (see Section 4 for details).

One can see that the criteria set by Aoki et al. (2007) under-estimates the carbon depletion when compared to the shifted models, which leads to an under-estimation of the carbon-abundance corrections, and hence decreases the CEMP-star frequencies. The decreasing [C/Fe] trend for increasing luminosities is flatter for the corrected values.

Assuming [C/Fe]$< +0.5$, the average carbon abundance for stars with $\log g < 2$ is [C/Fe]$= -0.21$ for the uncorrected abundances, and [C/Fe]$= +0.23$ for the corrected abundances, while for stars with $\log g > 3$ the average is [C/Fe]$= +0.24$. This agreement on the average carbon abundance also holds for [C/Fe]$< +0.7$ ($\log g < 2$: [C/Fe]$= -0.17$ uncorrected, [C/Fe]$= +0.27$ corrected, and $\log g > 3$: [C/Fe]$= +0.30$) and [C/Fe]$< +1.0$ ($\log g < 2$: [C/Fe]$= -0.13$ uncorrected, [C/Fe]$= +0.31$ corrected, and $\log g > 3$: [C/Fe]$= +0.35$). This demonstrates that our procedure is capable of recovering the amount of carbon depleted during the stellar evolution on the giant branch, and hence should yield more realistic values for the CEMP-star frequencies as a function of metallicity.

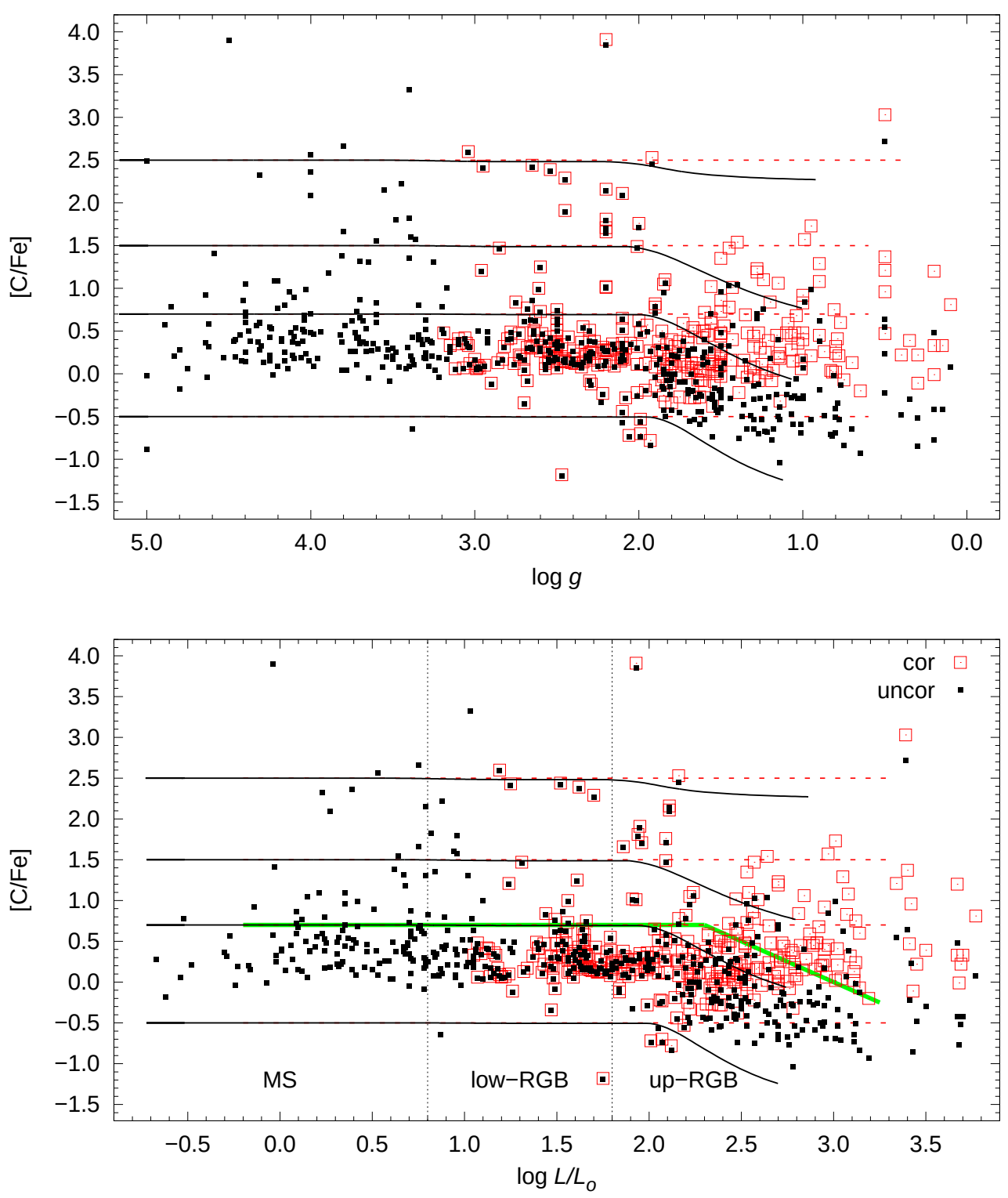

**Figure 14.** [C/Fe], as a function of $\log g$ (upper panels) and luminosity (lower panel), for the 505 stars with [Fe/H] $\leq -2.0$ selected from the literature. The evolutionary phases are based on the work of Gratton et al. (2000). The green solid line is the CEMP criteria from Aoki et al. (2007). Black solid lines are the theoretical models for [C/Fe] $= -0.50/+0.70/+1.50/+2.50$ (assuming [Fe/H] $= -3.3$ and [N/Fe] $= 0.0$). The horizontal red dashed lines show the initial carbon abundances of the models. The black filled squares represent the measured abundances, while the red open squares show the corrected values for stars with non-zero corrections (for the [N/Fe] = 0.00 case).

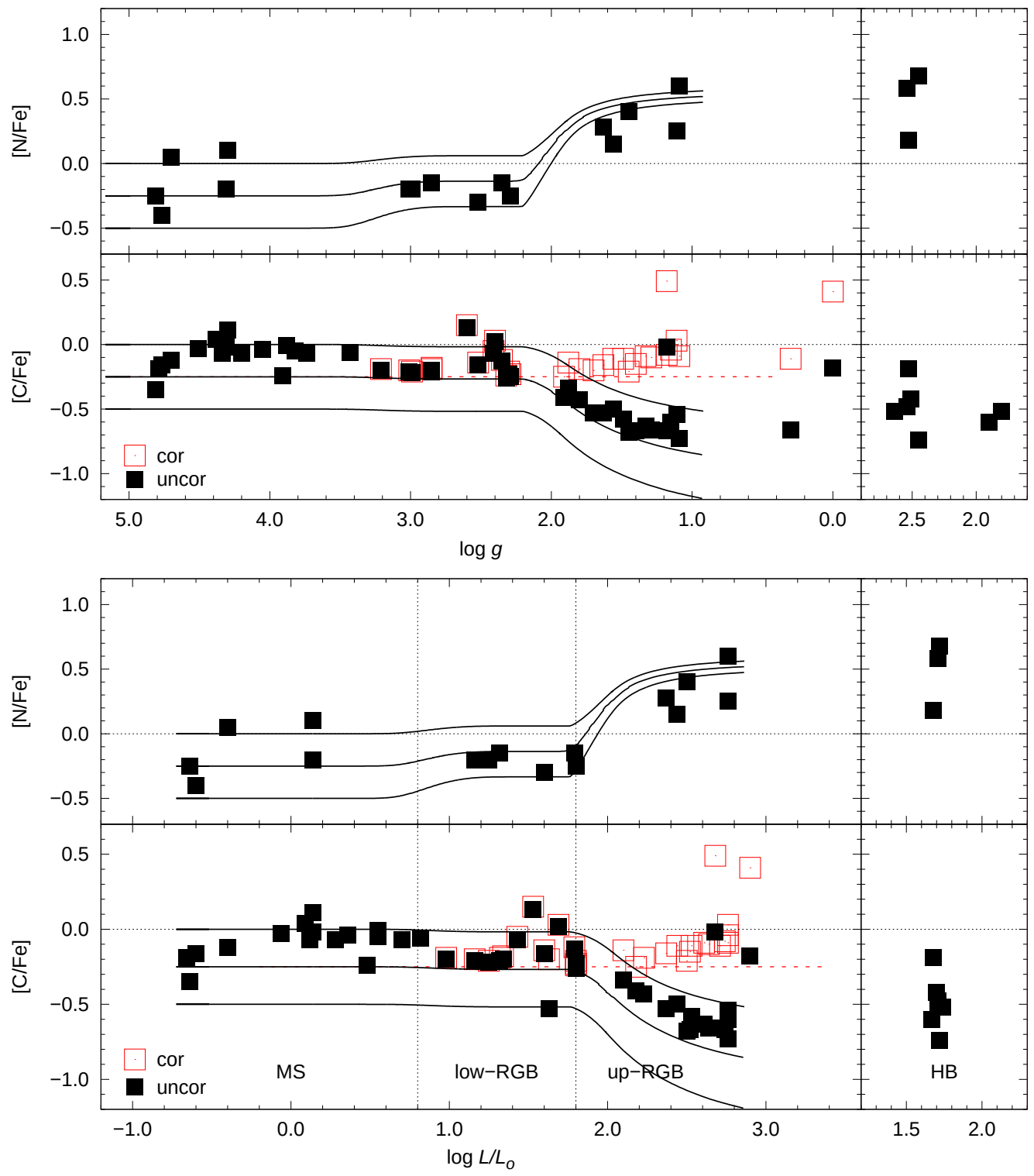

**Figure 15.** Carbon- and nitrogen-abundance ratios, as a function of $\log g$ (upper panels) and luminosity (lower panels), for the data from Gratton et al. (2000). The evolutionary phases were also taken from their Figure 7. Black solid lines are the theoretical models for [N/Fe] $= -0.50/-0.25/0.00$ (assuming [Fe/H] $= -1.3$ and [C/Fe] $= 0.0$) and [C/Fe] $= -0.50/-0.25/0.00$ (assuming [Fe/H] $= -1.3$ and [N/Fe] $= 0.0$). The horizontal red dashed lines show the initial carbon abundances of the models. The black filled squares represent the measured abundances, while the red open squares show the corrected values for stars with non-zero corrections (for the [N/Fe] = 0.00 case).

### 5.2. *The Gratton et al. Sample*

We use the new theoretical models described in this work, and the corresponding carbon-abundance corrections, to further explore the data published by Gratton et al. (2000). These authors studied the mixing along the RGB in metal-poor field stars, and mapped out the effect on the observed [C/Fe] and [N/Fe]. Figure 15 reproduces their Figure 10, where the upper panels show the behavior of [C/Fe] and [N/Fe], as a function of the $\log g$, and in the lower panels, as a function of the luminosity. The black filled squares are their published abundances, and the red open squares show the corrected carbon abundances, for case of [N/Fe] = 0.0. For comparison, we show the shifted theoretical models for [N/Fe] $= -0.50/-0.25/0.00$ (assuming [Fe/H] $= -1.3$ and [C/Fe] $= 0.0$) and [C/Fe] $= -0.50/-0.25/0.00$ (assuming [Fe/H] $= -1.3$ and [N/Fe] $=0.0$). The [C/Fe] $= -0.25$ models are a linear interpolation between the 0.0 and $-0.5$ models.

It is remarkable how well the theoretical models shown in Figure 15 reproduce the behavior of both the carbon- and nitrogen-abundance ratios. Gratton et al. find that the average carbon-abundance ratio for their unevolved stars ($\log L/L_{\odot} < 0.8$) is [C/Fe] $= -0.09$, while the average for stars on the upper RGB is [C/Fe] $= -0.58$. By recalculating the average abundance for the upper-RGB stars using the corrected carbon abundances, we find an average of [C/Fe] $= -0.08$. The fact that the corrected average matches the one for unevolved stars may be a hint that our assumption of an early mixing onset compared to the models is correct, or further processing could have occured in these objects.

## 6. THE CUMULATIVE FREQUENCIES OF CEMP STARS IN THE GALACTIC HALO AS A FUNCTION OF [FE/H]

Figure 16 shows the cumulative CEMP-star frequencies for metal-poor stars, for carbonicities [C/Fe] $\geq +0.5/+0.7/+1.0$, as a function metallicity ($-5.0 \leq$ [Fe/H] $\leq -2.0$, see discussion in Section 1), for both uncorrected and corrected [C/Fe]. The [Fe/H] step size is 0.1 dex, and the carbon corrections were taken considering initial [N/Fe]=0.0 (see discussion in Section 4.3). The solid lines represent the frequencies for uncorrected abundances, the dashed lines are the frequencies for the corrected [C/Fe], and the shaded areas highlight the differences between the distributions for a given [C/Fe] range. Also shown on the plot are the cumulative CEMP-star frequencies of Frebel et al. (2006), Carollo et al. (2012), and Lee et al. (2013).

The cumulative CEMP frequencies for selected [Fe/H] cuts are listed in Table 1. An important point to consider when comparing this new set of frequencies with previ-ous results is the fact that CEMP-$s$ and CEMP-$r/s$ stars were not specifically excluded previously, as they were based on medium-resolution (R$\sim 2,000$) spectroscopy, as opposed to this work, which uses data derived from high-resolution spectroscopy only. Since the metallicities for these stars are mostly concentrated in the [Fe/H]$> -3.0$ range, we expect over-estimated frequencies in this region. This effect can be quantified. The numbers in parenthesis in Table 1 show the difference (in %) between the frequencies calculated without any of the selection criteria presented in Section 3 and the adopted values shown in Table 1. As seen, the corrected cumulative CEMP frequencies for [Fe/H]$\leq -2.0$ and [C/Fe]$\geq +1.0$ roughly double (from 13% to 13% + 14%=27%) if one takes into account the CEMP stars enriched via an *extrinsic* formation scenario. This effect is weaker for decreasing metallicities, and becomes negligible once the [Fe/H]$\leq -3.5$ range is reached. This is an anticipated result due to the absence of CEMP-$s$ (and CEMP-$r/s$) stars at the lowest metallicities.

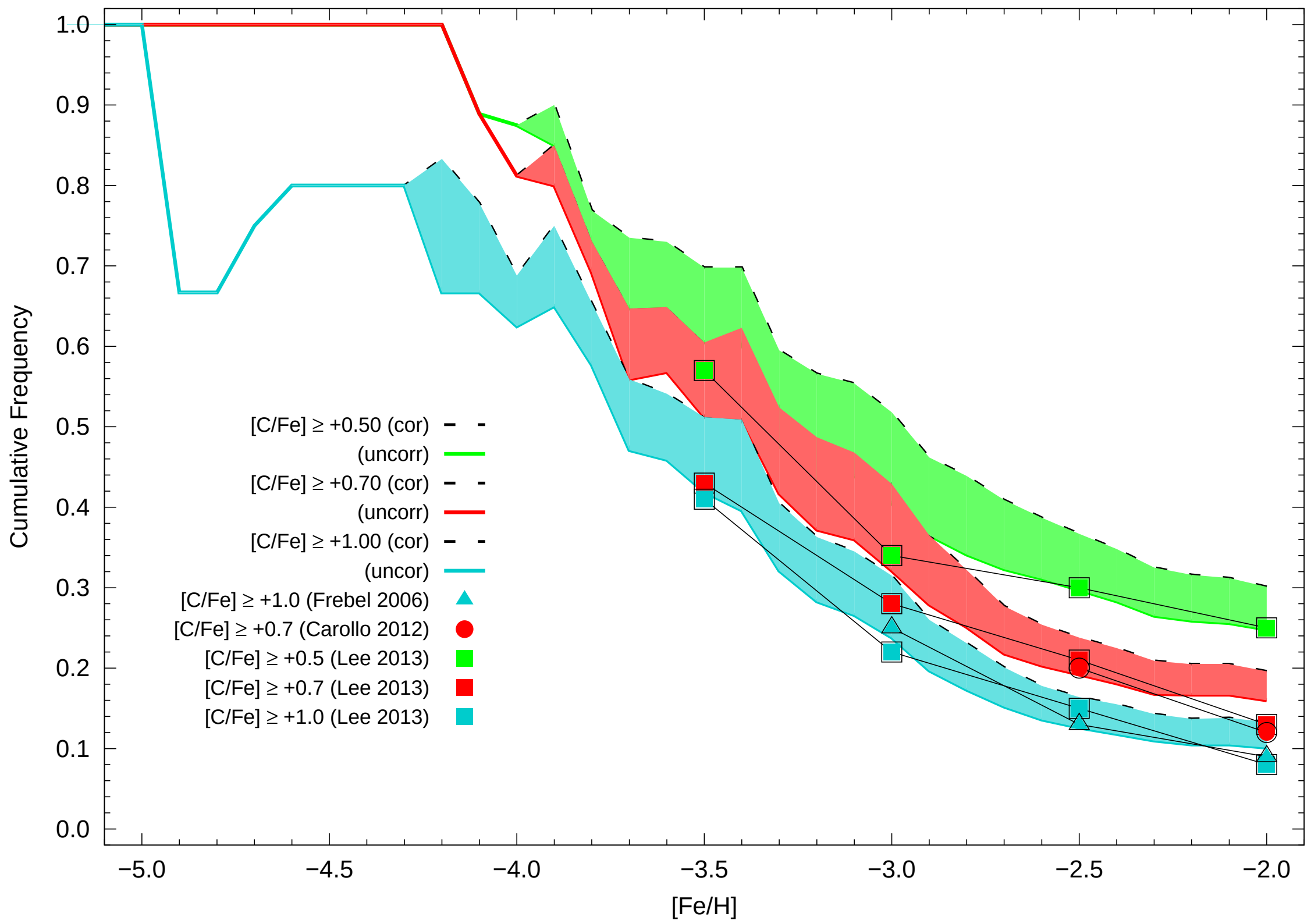

**Figure 16.** Cumulative frequencies of CEMP stars as a function of metallicity, based on the uncorrected (solid lines) and corrected (dashed lines) carbon abundances. The shaded areas highlight the differences in the frequencies for the corrected and uncorrected abundances. Note that, for the purpose of the corrections, [N/Fe] = 0 has been assumed. For comparison, we also show results from Frebel et al. (2006), Carollo et al. (2012), and Lee et al. (2013).

Inspection of Figure 16 and Table 1 reveals that, in most cases, the differences between the corrected and uncorrected values are within 10% of the entire sample; they reach up to 12% for [Fe/H] $\leq -3.0$ and [C/Fe] $\geq +0.50$ (40% uncorrected and 52% corrected), which represents a 30% increase. Our final derived cumulative CEMP-star frequencies for [C/Fe] $\geq +1.0$ (13% for [Fe/H] $\leq -2.0$, 32% for [Fe/H] $\leq -3.0$, 69% for [Fe/H] $\leq -4.0$, 100% for [Fe/H] $\leq -5.0$) are commensurate with the trend found by Frebel et al. (2006) for higher metallicities (9% for [Fe/H]$\leq -2.0$, and 25% for [Fe/H]$\leq -3.0$), even though those determinations did not take into account the evolutionary stage, nor the addition of CEMP-$s$ and CEMP-$r/s$ to the analysis (and were based on a relatively small number of stars with measured [C/Fe]). We find that the cumulative CEMP-star frequencies estimated by Carollo et al. (2012) are slightly lower for [Fe/H]$\leq -2.0$, and agree well for [Fe/H]$\leq -2.5$. For Lee et al. (2013), the frequencies are overall lower for the [C/Fe] $\geq +0.5/+0.7$ bins, even when comparing with our uncorrected cumulative frequencies. For the [C/Fe] $\geq +1.0$ regime, the cumulative frequencies for stars with [Fe/H]$\leq -3.0$ agree well with the uncorrected frequencies from this work. Even though the sample sizes are considerably smaller than ours, we find that the cumulative CEMP-star frequencies (for stars with [C/Fe]$> +1.0$ and [Fe/H]$\leq -2.0$) from Lucatello et al. (2006) (21%±2%) are over-estimated, while the results reported by Cohen et al. (2005) (14%±4%) are lower by a few percent. Note that, even though these studies used high-resolution data for their frequency calculations, CEMP-$s$ and CEMP-$r/s$ were included.

There are no differences between the cumulative CEMP-star frequencies for the uncorrected and corrected cases for [Fe/H]$< -4.5$. Among the five stars analyzed in this metallicity range, there are two giants ($\log g$=2.2), HE 0557$-$4840 and HE 0107$-$5240, with [C/Fe] corrections of +0.01 dex and +0.06 dex, respectively. In addition, the most iron-poor star found to date, SMSS J031300.36$-$670839.3 ([Fe/H] $\leq -7.1$; Keller et al. 2014), has $\log g$=2.3, and exhibits a remarkably high carbon abundance ([C/Fe]$> +4.5$), with negligible carbon correction. The sudden decrease in the frequency between $-5.0 \geq$ [Fe/H] $\geq -4.5$ and [C/Fe] $\geq +1.0$ is due to the presence of SDSS J102915 ([C/Fe]$\leq +0.7$; Caffau et al. 2011). It is important to note that this analysis is still limited by small-number statistics for stars in the [Fe/H]$\leq -4.0$ range, and further observations are required to firmly establish the frequencies.

To estimate the uncertainties on the derived frequencies from Table 1, we recalculated the corrected cumulative CEMP-star frequencies using nine different scenarios: (i) [N/Fe] $= -0.5$ model corrections; (ii) [N/Fe] $= +2.0$ model corrections; (iii) observed $\log g - 0.3$; (iv) observed $\log g + 0.3$; (v) observed $\log g$ adding a random uncertainty ranging from $-0.3$ dex to $+0.3$ dex; (vi) changing the CEMP-$s$ and CEMP-$r/s$ restriction to [Ba/Fe]$> +0.8$ (see Section 3 for further details); (vii) using carbon-abundance corrections determined by models only, without any shifts on $\log g$; (viii) introducing a $+0.3$ dex shift on the model $\log g$, and; (ix) taking into account 3D effects on the carbon abundances (Asplund 2005). The differences (in %) between the cumulative frequencies in each of these cases and the corrected values in Table 1 are shown in Table 2. Comparing the changes in the frequencies due to the choice of initial [N/Fe] shows that these changes are always less than $\pm 1$%. This is expected, since the largest differences in the corrections for [N/Fe] $= -0.5$ and [N/Fe] $= +2.0$ were from stars with measured [C/Fe] $<0.0$. Concerning the changes in $\log g$, one can see that the differences are spread between $-6$% and $+8$% of the adopted values when all the stars are subject to the same shift in $\log g$. However, when adding a random uncertainty to the distribution, the absolute changes in the frequencies are less than $\pm 1$%.

The changes in the CEMP-star frequencies for cases (vii) and (viii) are between $+1$% and $-7$%. This is within expectations, since the absence of or a $+0.3$ dex shift in $\log g$ under-estimate the carbon corrections for the stars in the upper RGB. We also considered the influence of possible 3D effects on the carbon-abundance determinations would have on our derived frequencies. It has been suggested (e.g., Collet et al. 2007) that these effects can lead to an over-estimate of [C/Fe]$=+0.5$ to $+0.9$ for the CH feature for red giants at [Fe/H]$\sim -3.0$. For the sample stars with $\log g \leq 3.0$, we applied [C/Fe] offsets of $-0.3$ for $-2.5<$[Fe/H]$\leq -2.0$, $-0.5$ for $-3.0<$[Fe/H]$\leq -2.5$, and $-0.7$ for [Fe/H]$\leq -3.0$. The frequencies decreased between 5-14% for [Fe/H]$> -2.5$; the most affected cut was at [Fe/H]$\leq -3.5$, with a decrease ranging from 13-26%. The cumulative CEMP-star frequencies for stars with [Fe/H]$< -4.5$ were only affected for the case of [C/Fe]$\geq +1.0$. However, recent studies (Placco et al. 2014b) have found that the differences between carbon abundances determined from near-ultraviolet C I lines and the CH band at 4300Å are within $\sim 0.2$ dex for the [Fe/H]$= -3.8$ subgiant BD+44°493, so the 3D effect on the carbon abundance must be carefully evaluated.

## 7. CONCLUSIONS

CEMP-star frequencies are important inputs for Galactic chemical-evolution models (e.g., Kobayashi & Nakasato 2011), since they constrain the IMF in the early stages. In this work we present improved cumulative CEMP-star frequencies, taking into account the evolutionary status of metal-poor field stars. The amount of carbon depleted during the evolution on the RGB was quantified by matching the observed carbon abundance to yields from stellar-evolution models. The offset added to the models to account for an early carbon depletion onset may indicate that the extent of thermohaline mixing is under-estimated, or that additional process(es), such as rotation, internal gravity waves, and magnetic fields (Maeder et al. 2013), must be considered.

Our final derived cumulative CEMP-star frequencies are: (i) [C/Fe] $\geq +0.7$ – 20% for [Fe/H] $\leq -2.0$, 43% for [Fe/H] $\leq -3.0$, 81% for [Fe/H] $\leq -4.0$, and 100% for [Fe/H] $\leq -5.0$; (ii) [C/Fe] $\geq +1.0$ – 13% for [Fe/H] $\leq -2.0$, 32% for [Fe/H] $\leq -3.0$, 69% for [Fe/H] $\leq -4.0$, and 100% for [Fe/H] $\leq -5.0$. For this exercise we used 505 stars with [Fe/H] $\leq -2.0$ from the literature, with atmospheric parameters and abundances determined from high-resolution spectroscopy.

This is the largest high-resolution sample yet considered for such an analysis. These values exclude the recognized CEMP-$s$ and CEMP-$r/s$ stars from the calculations, since their observed carbon abundance is "contaminated" from its evolved AGB companion in a binary system. We also developed an online tool that provides the carbon corrections for a given set of $\log g$, [Fe/H], and [C/Fe]. Further work may include correction of carbon abundances for stars observed with medium-resolution spectroscopy (Frebel et al. 2006; Placco et al. 2010, 2011; Carollo et al. 2012; Lee et al. 2013; Kordopatis et al. 2013; Beers et al. 2014). However, reliable $\log g$ information must be provided, to avoid large uncertainties on the carbon abundance corrections.

Table 3 lists the input data used for the determination of the CEMP-star frequencies, as well as the correction for the [N/Fe]=0.0 case. Also listed, for completeness, are the CEMP-$s$ and CEMP-$r/s$ stars excluded from the calculations. Even though these stars are not suitable for the CEMP-star frequency determinations, they also experience carbon depletion during their evolution, and our machinary allows its measurement. The luminosities are derived from the Aoki et al. (2007) prescription, using M=0.8 $M_{\odot}$. It is interesting to note from Table 3 that 16 stars are CEMP stars (based on corrected [C/Fe]$\geq +0.7$), but are not sub-classified further, due to lack of information on their [Ba/Fe] ratios. Of these, 11 stars have [Fe/H]$< -3.0$, which is the range where CEMP-no stars are most commom. In addition, there are 53 stars which can be considered carbon-enriched, with $+0.5 \leq$[C/Fe]$\leq +0.7$, but that do not satisfy our criterion for classication as CEMP stars. Of these, 46 have measured [Ba/Fe], with 41 exhibiting [Ba/Fe]$< 0.0$. These can be classified as likely CEMP-no stars, and further observations of such stars should help resolve their proper classifications.

Even though the cumulative frequencies presented in this work only changed by modest amounts when compared to the uncorrected results, our approach of taking into account the evolutionary status of the stars and also excluding CEMP-$s$ and CEMP-$r/s$ stars from the estimates provides the currently most reliable estimate of the CEMP frequencies. This effort would clearly benefit from additional [N/Fe] measurements for a large number of stars in our sample, as well as measurements for [Ba/Fe] (and [Eu/Fe]) in order to enable the identification of additional CEMP-no stars. As additional obser-vations of evolved CEMP stars near the tip of the RGB become available, the impact of our corrections on the derived cumulative CEMP-star frequencies will increase.

Furthermore, the ability to describe the true CEMP-star frequencies allows the quantification and assessment of the proper formation channels of the two distinctive metal-poor stellar populations in the Galactic halo at [Fe/H]$< -3.0$: carbon-normal and carbon-rich (Norris et al. 2013b). These two populations are thought to be formed by gas clouds that were influenced by at least two different primary cooling channels (Frebel et al. 2007; Schneider et al. 2012; Ji et al. 2014). Once the number and type of progenitors are quantified, it will be possible to build a more reliable model of the stellar populations of the early Milky Way, and by extension, for other galaxies. Besides that, by having a more reliable characterization of these populations, it will become possible to compare the occurance rate of CEMP-no stars with the recently discovered carbon-enhanced damped Ly$\alpha$ systems (Cooke et al. 2011, 2012) which carry abundance patterns that resemble those from massive, carbon-producing first stars.

V.M.P. is a Gemini Science Fellow, and acknowledges support from the Gemini Observatory. A.F. is supported by NSF CAREER grant AST-1255160. T.C.B. acknowledges partial support for this work from grants PHY 08-22648; Physics Frontier Center/Joint Institute or Nuclear Astrophysics (JINA), and PHY 14-30152; Physics Frontier Center/JINA Center for the Evolution of the Elements (JINA-CEE), awarded by the US National Science Foundation. R.J.S. is the recipient of a Sofja Kovalevskaja Award from the Alexander von Humboldt Foundation.

## REFERENCES

Akerman, C. J., Carigi, L., Nissen, P. E., Pettini, M., & Asplund, M. 2004, A&A, 414, 931
Allen, D. M., Ryan, S. G., Rossi, S., Beers, T. C., & Tsangarides, S. A. 2012, A&A, 548, A34
Angelou, G. C., Church, R. P., Stancliffe, R. J., Lattanzio, J. C., & Smith, G. H. 2011, ApJ, 728, 79
Angelou, G. C., Stancliffe, R. J., Church, R. P., Lattanzio, J. C., & Smith, G. H. 2012, ApJ, 749, 128
Aoki, W., Beers, T. C., Christlieb, N., et al. 2007, ApJ, 655, 492
Aoki, W., Norris, J. E., Ryan, S. G., Beers, T. C., & Ando, H. 2002, PASJ, 54, 933
Aoki, W., Suda, T., Boyd, R. N., Kajino, T., & Famiano, M. A. 2013, ApJ, 766, L13
Aoki, W., Honda, S., Beers, T. C., et al. 2005, ApJ, 632, 611
Aoki, W., Frebel, A., Christlieb, N., et al. 2006, ApJ, 639, 897
Aoki, W., Beers, T. C., Sivarani, T., et al. 2008, ApJ, 678, 1351
Asplund, M. 2005, ARA&A, 43, 481
Asplund, M., Grevesse, N., Sauval, A. J., & Scott, P. 2009, ARA&A, 47, 481
Barklem, P. S., Christlieb, N., Beers, T. C., et al. 2005, A&A, 439, 129
Beers, T. C., & Christlieb, N. 2005, ARA&A, 43, 531
Beers, T. C., Norris, J. E., Placco, V. M., et al. 2014, ApJ, 794, 58
Bisterzo, S., Gallino, R., Straniero, O., Cristallo, S., & Käppeler, F. 2011, MNRAS, 418, 284
Caffau, E., Bonifacio, P., François, P., et al. 2011, Nature, 477, 67
Carollo, D., Freeman, K., Beers, T. C., et al. 2014, ApJ, 788, 180
Carollo, D., Beers, T. C., Bovy, J., et al. 2012, ApJ, 744, 195
Charbonnel, C., & Zahn, J.-P. 2007, A&A, 467, L15
Christlieb, N., Bessell, M. S., Beers, T. C., et al. 2002, Nature, 419, 904
Christlieb, N., Beers, T. C., Barklem, P. S., et al. 2004, A&A, 428, 1027
Cohen, J. G., Christlieb, N., McWilliam, A., et al. 2008, ApJ, 672, 320
Cohen, J. G., Christlieb, N., Thompson, I., et al. 2013, ApJ, 778, 56
Cohen, J. G., Shectman, S., Thompson, I., et al. 2005, ApJ, 633, L109
Collet, R., Asplund, M., & Trampedach, R. 2007, A&A, 469, 687
Cooke, R., Pettini, M., & Murphy, M. T. 2012, MNRAS, 3437
Cooke, R., Pettini, M., Steidel, C. C., Rudie, G. C., & Nissen, P. E. 2011, MNRAS, 417, 1534
Cooke, R. J., & Madau, P. 2014, ApJ, 791, 116
Cui, W. Y., Sivarani, T., & Christlieb, N. 2013, A&A, 558, A36
Demarque, P., Woo, J.-H., Kim, Y.-C., & Yi, S. K. 2004, ApJS, 155, 667
Denissenkov, P. A., & Merryfield, W. J. 2011, ApJ, 727, L8
Eggleton, P. P. 1971, MNRAS, 151, 351
Frebel, A., Johnson, J. L., & Bromm, V. 2007, MNRAS, 380, L40
Frebel, A., Kirby, E. N., & Simon, J. D. 2010, Nature, 464, 72
Frebel, A., & Norris, J. E. 2013, Metal-Poor Stars and the Chemical Enrichment of the Universe (Published), 55
Frebel, A., Aoki, W., Christlieb, N., et al. 2005, Nature, 434, 871
Frebel, A., Christlieb, N., Norris, J. E., et al. 2006, ApJ, 652, 1585
Goswami, A., Aoki, W., Beers, T. C., et al. 2006, MNRAS, 372, 343
Gratton, R. G., Sneden, C., Carretta, E., & Bragaglia, A. 2000, A&A, 354, 169
Hansen, T., Andersen, J., Nordström, B., Buchhave, L. A., & Beers, T. C. 2011, ApJ, 743, L1
Hansen, T., Hansen, C. J., Christlieb, N., et al. 2014, ApJ, 787, 162
Herwig, F. 2005, ARA&A, 43, 435
Hollek, J. K., Frebel, A., Placco, V. M., et al. 2014, ApJ submitted
Hollek, J. K., Frebel, A., Roederer, I. U., et al. 2011, ApJ, 742, 54
Ito, H., Aoki, W., Beers, T. C., et al. 2013, ApJ, 773, 33
Ito, H., Aoki, W., Honda, S., & Beers, T. C. 2009, ApJ, 698, L37
Ji, A. P., Frebel, A., & Bromm, V. 2014, ApJ, 782, 95
Johnson, J. A., Herwig, F., Beers, T. C., & Christlieb, N. 2007, ApJ, 658, 1203
Jonsell, K., Barklem, P. S., Gustafsson, B., et al. 2006, A&A, 451, 651
Keller, S. C., Bessell, M. S., Frebel, A., et al. 2014, Nature, 506, 463
Kippenhahn, R., Ruschenplatt, G., & Thomas, H.-C. 1980, A&A, 91, 175
Kobayashi, C., & Nakasato, N. 2011, ApJ, 729, 16
Kordopatis, G., Gilmore, G., Steinmetz, M., et al. 2013, AJ, 146, 134
Lai, D. K., Bolte, M., Johnson, J. A., et al. 2008, ApJ, 681, 1524
Lai, D. K., Johnson, J. A., Bolte, M., & Lucatello, S. 2007, ApJ, 667, 1185
Lee, Y. S., Suda, T., Beers, T. C., & Stancliffe, R. J. 2014, ApJ, 788, 131
Lee, Y. S., Beers, T. C., Masseron, T., et al. 2013, AJ, 146, 132
Lucatello, S., Beers, T. C., Christlieb, N., et al. 2006, ApJ, 652, L37
Lucatello, S., Tsangarides, S., Beers, T. C., et al. 2005, ApJ, 625, 825
Maeder, A., Meynet, G., Lagarde, N., & Charbonnel, C. 2013, A&A, 553, A1
Mashonkina, L., Ryabtsev, A., & Frebel, A. 2012, A&A, 540, A98
Masseron, T., Johnson, J. A., Lucatello, S., et al. 2012, ApJ, 751, 14
Masseron, T., Johnson, J. A., Plez, B., et al. 2010, A&A, 509, A93+
Mathis, S., Decressin, T., Eggenberger, P., & Charbonnel, C. 2013, A&A, 558, A11
McWilliam, A., Preston, G. W., Sneden, C., & Searle, L. 1995, AJ, 109, 2757
Meléndez, J., & Barbuy, B. 2002, ApJ, 575, 474
Meynet, G., Ekström, S., & Maeder, A. 2006, A&A, 447, 623
Meynet, G., Hirschi, R., Ekstrom, S., et al. 2010, A&A, 521, A30
Nomoto, K., Tominaga, N., Umeda, H., Kobayashi, C., & Maeda, K. 2006, Nuclear Physics A, 777, 424
Norris, J. E., Christlieb, N., Korn, A. J., et al. 2007, ApJ, 670, 774

Norris, J. E., Bessell, M. S., Yong, D., et al. 2013a, ApJ, 762, 25
Norris, J. E., Yong, D., Bessell, M. S., et al. 2013b, ApJ, 762, 28
Placco, V. M., Frebel, A., Beers, T. C., et al. 2014a, ApJ, 781, 40
—. 2013, ApJ, 770, 104
Placco, V. M., Kennedy, C. R., Rossi, S., et al. 2010, AJ, 139, 1051
Placco, V. M., Kennedy, C. R., Beers, T. C., et al. 2011, AJ, 142, 188
Placco, V. M., Beers, T. C., Roederer, I. U., et al. 2014b, ApJ, 790, 34
Pols, O. R., Izzard, R. G., Stancliffe, R. J., & Glebbeek, E. 2012, A&A, 547, A76
Preston, G. W., & Sneden, C. 2001, AJ, 122, 1545
Preston, G. W., Sneden, C., Thompson, I. B., Shectman, S. A., & Burley, G. S. 2006, AJ, 132, 85
Roederer, I. U., Lawler, J. E., Sneden, C., et al. 2008a, ApJ, 675, 723
Roederer, I. U., Preston, G. W., Thompson, I. B., Shectman, S. A., & Sneden, C. 2014, ApJ, 784, 158
Roederer, I. U., Sneden, C., Thompson, I. B., Preston, G. W., & Shectman, S. A. 2010, ApJ, 711, 573
Roederer, I. U., Frebel, A., Shetrone, M. D., et al. 2008b, ApJ, 679, 1549
Schneider, R., Omukai, K., Bianchi, S., & Valiante, R. 2012, MNRAS, 419, 1566
Simmerer, J., Sneden, C., Cowan, J. J., et al. 2004, ApJ, 617, 1091
Sivarani, T., Beers, T. C., Bonifacio, P., et al. 2006, A&A, 459, 125
Sneden, C., Cowan, J. J., Lawler, J. E., et al. 2003, ApJ, 591, 936
Spite, M., Caffau, E., Bonifacio, P., et al. 2013, A&A, 552, A107
Spite, M., Cayrel, R., Hill, V., et al. 2006, A&A, 455, 291
Stancliffe, R. J., Church, R. P., Angelou, G. C., & Lattanzio, J. C. 2009, MNRAS, 396, 2313
Stancliffe, R. J., & Eldridge, J. J. 2009, MNRAS, 396, 1699
Stancliffe, R. J., & Lattanzio, J. C. 2011, Why Galaxies Care about AGB Stars II: Shining Examples and Common Inhabitants, 445, 29
Suda, T., Katsuta, Y., Yamada, S., et al. 2008, PASJ, 60, 1159
Thompson, I. B., Ivans, I. I., Bisterzo, S., et al. 2008, ApJ, 677, 556
Tominaga, N., Umeda, H., & Nomoto, K. 2007, ApJ, 660, 516
Ulrich, R. K. 1972, ApJ, 172, 165
Umeda, H., & Nomoto, K. 2005, ApJ, 619, 427
Viallet, M., Meakin, C., Arnett, D., & Mocák, M. 2013, ApJ, 769, 1
Woosley, S. E., & Weaver, T. A. 1995, ApJS, 101, 181
Yong, D., Norris, J. E., Bessell, M. S., et al. 2013, ApJ, 762, 26
Zhang, L., Karlsson, T., Christlieb, N., et al. 2011, A&A, 528, A92

**Table 1**
Cumulative CEMP-star Frequencies

| | % [C/Fe] (+ΔCEMP-$s$/-$rs$) | | | | | |
|---|---|---|---|---|---|---|
| | ≥ +0.50 | | ≥ +0.70 | | ≥ +1.00 | |
| [Fe/H]≤ | uncor | cor | uncor | cor | uncor | cor |
| −2.0 | 25 (+12) | 30 (+12) | 16 (+14) | 20 (+13) | 10 (+14) | 13 (+14) |
| −2.5 | 30 (+09) | 37 (+08) | 19 (+11) | 24 (+10) | 13 (+11) | 16 (+12) |
| −3.0 | 40 (+07) | 52 (+05) | 32 (+07) | 43 (+05) | 24 (+07) | 32 (+06) |
| −3.5 | 60 (+00) | 70 (+00) | 51 (+00) | 60 (+00) | 42 (+00) | 51 (+00) |
| −4.0 | 88 (+00) | 88 (+00) | 81 (+00) | 81 (+00) | 62 (+00) | 69 (+00) |
| −4.5 | 100 (+00) | 100 (+00) | 100 (+00) | 100 (+00) | 80 (+00) | 80 (+00) |
| −5.0 | 100 (+00) | 100 (+00) | 100 (+00) | 100 (+00) | 100 (+00) | 100 (+00) |

**Note**. — The values in parenthesis show the increase (in %) in the cumulative CEMP-star frequencies due to the addition of CEMP-$s$ and CEMP-$r/s$ stars.

**Table 2**
Uncertainties in Derived Cumulative CEMP-star Frequencies

| [Fe/H]≤ | (i) [N/Fe] −0.5 | (ii) [N/Fe] +2.0 | (iii) Δlog $g$ +0.3 | (iv) Δlog $g$ −0.3 | (v) Δlog $g$ random | (vi) [Ba/Fe] < +0.8 | (vii) Model only | (viii) Model +0.3 | (ix) Δ[C/Fe] 3D |
|---|---|---|---|---|---|---|---|---|---|
| | | | | Δ% ([C/Fe] ≥ +0.50) | | | | | |
| −2.0 | 1 | 0 | −3 | −2 | 1 | 0 | −1 | 1 | −10 |
| −2.5 | 0 | 0 | −4 | −2 | 0 | 1 | −4 | −2 | −14 |
| −3.0 | 1 | 0 | −6 | −4 | 0 | 0 | −7 | −3 | −20 |
| −3.5 | 0 | 0 | −3 | 0 | 0 | 0 | −3 | −3 | −26 |
| −4.0 | 0 | 0 | 0 | 0 | 0 | 0 | 0 | 0 | −13 |
| −4.5 | 0 | 0 | 0 | 0 | 0 | 0 | 0 | 0 | 0 |
| −5.0 | 0 | 0 | 0 | 0 | 0 | 0 | 0 | 0 | 0 |
| | | | | Δ% ([C/Fe] ≥ +0.70) | | | | | |
| −2.0 | 0 | 0 | −1 | 5 | −1 | 0 | −1 | 0 | −7 |
| −2.5 | 0 | 0 | −1 | 6 | 0 | 1 | −3 | −2 | −9 |
| −3.0 | 0 | 0 | −2 | 8 | −1 | 1 | −7 | −4 | −18 |
| −3.5 | 0 | 0 | −2 | 0 | 0 | 0 | −4 | −2 | −16 |
| −4.0 | 0 | 0 | 0 | 0 | 0 | 0 | 0 | 0 | −6 |
| −4.5 | 0 | 0 | 0 | 0 | 0 | 0 | 0 | 0 | 0 |
| −5.0 | 0 | 0 | 0 | 0 | 0 | 0 | 0 | 0 | 0 |
| | | | | Δ% ([C/Fe] ≥ +1.00) | | | | | |
| −2.0 | −1 | 0 | 4 | 1 | 0 | 1 | 0 | 1 | −5 |
| −2.5 | 0 | 0 | 5 | 1 | 0 | 0 | −2 | 0 | −6 |
| −3.0 | −1 | 0 | 6 | 1 | 0 | 0 | −5 | −2 | −14 |
| −3.5 | 0 | 0 | 3 | 0 | 0 | 0 | −4 | −2 | −18 |
| −4.0 | 0 | 0 | 0 | 0 | 0 | 0 | 0 | 0 | −13 |
| −4.5 | 0 | 0 | 0 | 0 | 0 | 0 | 0 | 0 | −20 |
| −5.0 | 0 | 0 | 0 | 0 | 0 | 0 | 0 | 0 | 0 |

**Table 3**
Data for Literature Stars

| Name | $T_{\rm eff}$ (K) | log $g$ (cgs) | log $L$ ($L_\odot$) | [Fe/H] | [N/Fe] | [C/Fe] | Δ[C/Fe] ([N/Fe]=0) | [C/Fe]$_c$ | [Sr/Fe] | [Ba/Fe] | Class | I/O | Ref. |
|---|---|---|---|---|---|---|---|---|---|---|---|---|---|
| BD+02:3375 | 5926 | 4.63 | −0.24 | −2.08 | ··· | −0.04 | 0.00 | −0.04 | ··· | ··· | | 1 | Lai et al. (2007) |
| BD+03:740 | 6485 | 4.31 | 0.23 | −2.70 | ≤ +0.29 | +0.59 | 0.00 | +0.59 | ··· | −0.42 | | 1 | Yong et al. (2013) |
| BD+06:0648 | 4400 | 0.90 | 2.97 | −2.09 | ··· | −0.19 | +0.64 | +0.45 | ··· | ··· | | 1 | Aoki et al. (2008) |
| BD+10:2495 | 4710 | 1.30 | 2.69 | −2.31 | ··· | −0.30 | +0.62 | +0.32 | ··· | ··· | | 1 | Roederer et al. (2010) |
| BD+17:3248 | 5240 | 2.72 | 1.45 | −2.17 | +0.65 | −0.37 | +0.01 | −0.36 | ··· | +0.69 | | 0 | Yong et al. (2013) |
| BD+23:3130 | 5262 | 2.76 | 1.42 | −2.52 | −0.54 | +0.11 | +0.01 | +0.12 | ··· | −0.56 | | 1 | Yong et al. (2013) |
| BD+24:1676 | 6241 | 3.81 | 0.67 | −2.46 | ≤ +0.21 | +0.37 | 0.00 | +0.37 | −0.05 | −0.33 | | 1 | Lai et al. (2008) |
| BD+44:493 | 5430 | 3.40 | 0.84 | −3.80 | +0.32 | +1.35 | 0.00 | +1.35 | ··· | −0.60 | CEMP-no | 1 | Ito et al. (2013) |
| BD−01:2582 | 5148 | 2.86 | 1.28 | −2.21 | ··· | +0.76 | +0.01 | +0.77 | ··· | +1.50 | CEMP-$s/rs$ | 0 | Simmerer et al. (2004) |
| BD−04:3208 | 6360 | 4.01 | 0.50 | −2.27 | ··· | +0.15 | 0.00 | +0.15 | ··· | ··· | | 1 | Akerman et al. (2004) |
| BD−18:0271 | 4245 | 0.70 | 3.11 | −2.35 | ··· | −0.64 | +0.77 | +0.13 | ··· | ··· | | 1 | Meléndez & Barbuy (2002) |
| BD−18:5550 | 4558 | 0.81 | 3.12 | −3.20 | −0.36 | −0.02 | +0.77 | +0.75 | ··· | −0.74 | CEMP-no | 1 | Yong et al. (2013) |
| BS 16023−046 | 6571 | 4.25 | 0.32 | −2.83 | ··· | +0.55 | 0.00 | +0.55 | −0.08 | ··· | | 1 | Yong et al. (2013) |
| BS 16033−008 | 5300 | 2.70 | 1.49 | −2.84 | ··· | +0.61 | +0.01 | +0.62 | −0.31 | −1.20 | | 1 | Aoki et al. (2005) |
| BS 16076−006 | 5566 | 3.32 | 0.96 | −3.51 | ··· | +0.34 | 0.00 | +0.34 | ≤ +0.10 | ≤ −1.00 | | 1 | Yong et al. (2013) |
| BS 16077−007 | 6486 | 4.31 | 0.23 | −2.77 | ≤ +0.80 | +0.60 | 0.00 | +0.60 | +0.21 | −0.23 | | 1 | Yong et al. (2013) |
| BS 16080−054 | 4902 | 1.75 | 2.31 | −2.94 | +0.75 | −0.45 | +0.18 | −0.27 | +0.38 | −0.21 | | 1 | Yong et al. (2013) |
| BS 16082−129 | 4868 | 1.67 | 2.38 | −2.84 | ··· | +0.29 | +0.25 | +0.54 | −0.69 | −0.97 | | 1 | Yong et al. (2013) |
| BS 16083−172 | 5300 | 3.10 | 1.09 | −2.53 | ··· | +0.34 | +0.01 | +0.35 | +0.19 | −0.33 | | 1 | Aoki et al. (2005) |
| BS 16084−160 | 4727 | 1.27 | 2.72 | −3.20 | +0.78 | −0.12 | +0.60 | +0.48 | −2.07 | −2.06 | | 1 | Yong et al. (2013) |
| BS 16085−050 | 4882 | 1.93 | 2.12 | −2.71 | ··· | −0.84 | +0.06 | −0.78 | −1.67 | −1.62 | | 1 | Yong et al. (2013) |
| BS 16089−013 | 4900 | 1.70 | 2.36 | −2.82 | ··· | −0.23 | +0.24 | +0.01 | +0.21 | −0.09 | | 1 | Aoki et al. (2005) |
| BS 16467−062 | 5310 | 2.80 | 1.40 | −3.80 | ≤ +0.45 | +0.40 | +0.01 | +0.41 | −1.67 | ≤ −0.57 | | 1 | Yong et al. (2013) |
| BS 16469−075 | 4919 | 1.78 | 2.28 | −3.25 | ··· | +0.21 | +0.11 | +0.32 | +0.27 | −1.12 | | 1 | Yong et al. (2013) |
| BS 16472−018 | 4946 | 2.08 | 1.99 | −2.29 | ··· | −0.29 | +0.01 | −0.28 | +0.19 | −0.14 | | 1 | Lai et al. (2007) |
| BS 16477−003 | 4879 | 1.66 | 2.39 | −3.39 | ≤ −0.26 | +0.29 | +0.21 | +0.50 | +0.11 | −0.45 | | 1 | Yong et al. (2013) |
| BS 16543−092 | 4523 | 1.14 | 2.78 | −2.19 | ··· | −1.04 | +0.72 | −0.32 | −0.36 | −0.63 | | 1 | Lai et al. (2007) |
| BS 16543−097 | 5000 | 2.10 | 1.99 | −2.52 | ··· | +0.29 | +0.02 | +0.31 | +0.12 | −0.12 | | 1 | Aoki et al. (2005) |
| BS 16547−006 | 6047 | 3.72 | 0.70 | −2.21 | ··· | +0.73 | 0.00 | +0.73 | −0.44 | −0.52 | CEMP-no | 1 | Lai et al. (2007) |
| BS 16550−087 | 4754 | 1.32 | 2.68 | −3.54 | +1.11 | −0.49 | +0.56 | +0.07 | +0.53 | −0.75 | | 1 | Yong et al. (2013) |
| BS 16928−053 | 4679 | 1.16 | 2.82 | −2.93 | +1.05 | −0.24 | +0.69 | +0.45 | −0.19 | −0.79 | | 1 | Yong et al. (2013) |
| BS 16929−005 | 5229 | 2.61 | 1.56 | −3.34 | +0.32 | +0.99 | +0.01 | +1.00 | ··· | −0.41 | CEMP-no | 1 | Yong et al. (2013) |
| BS 16934−002 | 4500 | 1.00 | 2.91 | −2.82 | +0.64 | −0.26 | +0.74 | +0.48 | −1.27 | −1.65 | | 1 | Aoki et al. (2007) |
| BS 16945−054 | 5281 | 2.99 | 1.20 | −2.76 | ··· | +0.06 | +0.01 | +0.07 | −0.04 | −0.01 | | 1 | Lai et al. (2007) |
| BS 16968−061 | 6343 | 3.82 | 0.69 | −2.85 | ··· | +0.45 | 0.00 | +0.45 | −0.47 | −0.23 | | 1 | Yong et al. (2013) |
| BS 16981−009 | 5259 | 2.92 | 1.26 | −2.71 | ··· | +0.36 | 0.00 | +0.36 | ··· | ··· | | 1 | Lai et al. (2007) |
| BS 17139−007 | 5918 | 3.66 | 0.72 | −2.25 | ··· | +0.26 | 0.00 | +0.26 | −0.55 | −0.62 | | 1 | Lai et al. (2007) |
| BS 17439−065 | 5100 | 1.60 | 2.53 | −2.39 | ··· | −0.54 | +0.40 | −0.14 | −0.22 | −0.42 | | 1 | Aoki et al. (2005) |
| BS 17569−049 | 4645 | 1.09 | 2.87 | −2.84 | +0.86 | −0.12 | +0.71 | +0.59 | +0.38 | +0.20 | | 1 | Yong et al. (2013) |
| BS 17570−063 | 6233 | 4.46 | 0.01 | −2.95 | ··· | +0.40 | 0.00 | +0.40 | +0.08 | −0.26 | | 1 | Yong et al. (2013) |
| BS 17583−100 | 5950 | 3.66 | 0.73 | −2.53 | ··· | +0.53 | 0.00 | +0.53 | +0.20 | −0.33 | | 1 | Yong et al. (2013) |
| CD−35:14849 | 6125 | 4.11 | 0.33 | −2.41 | ··· | +0.22 | 0.00 | +0.22 | ··· | ··· | | 1 | Akerman et al. (2004) |
| CD−38:245 | 4857 | 1.54 | 2.50 | −4.15 | +1.07 | −0.33 | +0.24 | −0.09 | ··· | −0.76 | | 1 | Yong et al. (2013) |
| CD−42:14278 | 5812 | 4.25 | 0.10 | −2.12 | ··· | +0.15 | 0.00 | +0.15 | ··· | ··· | | 1 | Akerman et al. (2004) |
| CS 22166−016 | 5388 | 3.26 | 0.96 | −2.30 | ··· | +0.21 | 0.00 | +0.21 | +0.30 | −0.24 | | 1 | Lai et al. (2007) |
| CS 22169−035 | 4654 | 1.10 | 2.87 | −2.95 | +1.02 | −0.24 | +0.72 | +0.48 | +0.02 | −1.19 | | 1 | Yong et al. (2013) |
| CS 22172−002 | 4893 | 1.68 | 2.37 | −3.77 | +0.24 | 0.00 | +0.13 | +0.13 | −1.21 | −1.17 | | 1 | Yong et al. (2013) |
| CS 22174−007 | 5265 | 2.78 | 1.40 | −2.39 | ≤ +1.42 | +0.41 | +0.01 | +0.42 | ··· | −0.14 | | 1 | Yong et al. (2013) |
| CS 22174−012 | 4934 | 2.06 | 2.01 | −2.35 | ··· | −0.74 | +0.02 | −0.72 | −0.45 | −0.95 | | 1 | Lai et al. (2007) |
| CS 22175−007 | 5108 | 2.46 | 1.67 | −2.81 | ··· | +0.15 | +0.01 | +0.16 | +0.27 | −0.52 | | 1 | Barklem et al. (2005) |
| CS 22177−009 | 6423 | 4.37 | 0.16 | −2.98 | ··· | +0.38 | 0.00 | +0.38 | −0.26 | ··· | | 1 | Yong et al. (2013) |
| CS 22182−033 | 5700 | 3.30 | 1.02 | −2.67 | ··· | +0.24 | 0.00 | +0.24 | +0.01 | −0.41 | | 1 | Preston et al. (2006) |
| CS 22183−015 | 5450 | 3.00 | 1.24 | −2.82 | +2.09 | +2.33 | +0.02 | +2.35 | ··· | +1.85 | CEMP-$s/rs$ | 0 | Masseron et al. (2012) |
| CS 22183−031 | 5202 | 2.54 | 1.62 | −3.17 | ··· | +0.42 | +0.01 | +0.43 | +0.14 | −0.33 | | 1 | Yong et al. (2013) |

**Table 3** — *Continued*

| Name | $T_{\rm eff}$ (K) | log $g$ (cgs) | log $L$ ($L_{\odot}$) | [Fe/H] | [N/Fe] | [C/Fe] | Δ[C/Fe] ([N/Fe]=0) | [C/Fe]$_c$ | [Sr/Fe] | [Ba/Fe] | Class | I/O | Ref. |
|---|---|---|---|---|---|---|---|---|---|---|---|---|---|
| CS 22185−007 | 5193 | 2.73 | 1.43 | −2.28 | ··· | +0.21 | +0.01 | +0.22 | −0.42 | −0.51 | | 1 | Lai et al. (2007) |
| CS 22186−023 | 5066 | 2.19 | 1.92 | −2.72 | ··· | +0.26 | +0.01 | +0.27 | −0.05 | −0.98 | | 1 | Barklem et al. (2005) |
| CS 22186−025 | 4871 | 1.66 | 2.39 | −3.07 | +0.98 | −0.54 | +0.26 | −0.28 | −0.02 | +0.02 | | 1 | Yong et al. (2013) |
| CS 22189−009 | 4944 | 1.83 | 2.24 | −3.48 | +0.27 | +0.31 | +0.03 | +0.34 | −0.85 | −1.29 | | 1 | Yong et al. (2013) |
| CS 22872−102 | 5911 | 3.60 | 0.78 | −2.94 | ≤ +0.55 | +0.60 | 0.00 | +0.60 | −0.17 | −0.62 | | 1 | Yong et al. (2013) |
| CS 22873−055 | 4551 | 0.82 | 3.11 | −2.99 | +1.07 | −0.73 | +0.75 | +0.02 | −0.39 | −0.45 | | 1 | Yong et al. (2013) |
| CS 22873−128 | 4960 | 2.10 | 1.98 | −2.86 | ··· | +0.18 | +0.01 | +0.19 | −0.43 | −0.96 | | 1 | McWilliam et al. (1995) |
| CS 22873−166 | 4516 | 0.77 | 3.14 | −2.74 | +1.05 | −0.13 | +0.73 | +0.60 | +0.18 | −0.70 | | 1 | Yong et al. (2013) |
| CS 22877−001 | 4790 | 1.45 | 2.57 | −3.24 | −0.15 | +1.03 | +0.44 | +1.47 | −0.02 | −0.58 | CEMP-no | 1 | Roederer et al. (2014) |
| CS 22877−011 | 5130 | 2.35 | 1.79 | −2.90 | ··· | +0.07 | +0.01 | +0.08 | −1.21 | −1.34 | | 1 | McWilliam et al. (1995) |
| CS 22878−027 | 6319 | 4.41 | 0.09 | −2.51 | ≤ +1.06 | +0.86 | 0.00 | +0.86 | −0.29 | ≤ −0.75 | CEMP-no | 1 | Yong et al. (2013) |
| CS 22878−101 | 4796 | 1.44 | 2.58 | −3.31 | +1.33 | −0.29 | +0.45 | +0.16 | −0.13 | −0.64 | | 1 | Yong et al. (2013) |
| CS 22878−121 | 5450 | 1.90 | 2.34 | −2.38 | ··· | −0.25 | +0.11 | −0.14 | +0.18 | −0.08 | | 1 | Preston et al. (2006) |
| CS 22879−097 | 5800 | 2.30 | 2.05 | −2.43 | ··· | +0.50 | +0.01 | +0.51 | +0.26 | −0.47 | | 1 | Preston et al. (2006) |
| CS 22879−103 | 5775 | 1.85 | 2.49 | −2.08 | ··· | +0.05 | +0.18 | +0.23 | +0.68 | +0.21 | | 1 | Preston et al. (2006) |
| CS 22880−074 | 5621 | 3.50 | 0.80 | −2.29 | −0.10 | +1.30 | 0.00 | +1.30 | +0.24 | +1.31 | CEMP-$s/rs$ | 0 | Yong et al. (2013) |
| CS 22880−086 | 5196 | 2.54 | 1.62 | −3.05 | ··· | +0.24 | +0.01 | +0.25 | −0.03 | −0.85 | | 1 | Yong et al. (2013) |
| CS 22881−036 | 6200 | 4.00 | 0.47 | −2.00 | ··· | +2.04 | 0.00 | +2.04 | +0.69 | +1.84 | CEMP-$s/rs$ | 0 | Preston & Sneden (2001) |
| CS 22882−001 | 5950 | 2.50 | 1.89 | −2.49 | ··· | +0.06 | +0.01 | +0.07 | +0.29 | +0.13 | | 1 | Preston et al. (2006) |
| CS 22885−096 | 4992 | 1.93 | 2.16 | −3.86 | +0.26 | +0.26 | +0.01 | +0.27 | −1.26 | −1.10 | | 1 | Yong et al. (2013) |
| CS 22886−042 | 4798 | 1.48 | 2.54 | −2.83 | ≤ +2.03 | +0.12 | +0.44 | +0.56 | −0.07 | −0.32 | | 1 | Yong et al. (2013) |
| CS 22888−031 | 6241 | 4.47 | 0.01 | −3.31 | ··· | +0.38 | 0.00 | +0.38 | +0.28 | ··· | | 1 | Yong et al. (2013) |
| CS 22891−171 | 5100 | 1.60 | 2.53 | −2.25 | +1.67 | +1.56 | +0.12 | +1.68 | ··· | +2.48 | CEMP-$s/rs$ | 0 | Allen et al. (2012) |
| CS 22891−184 | 5600 | 2.20 | 2.09 | −2.59 | ··· | +0.26 | +0.01 | +0.27 | +0.09 | −0.08 | | 1 | Preston et al. (2006) |
| CS 22891−200 | 4490 | 0.50 | 3.40 | −3.47 | +1.20 | +0.64 | +0.73 | +1.37 | −1.32 | −0.93 | CEMP-no | 1 | McWilliam et al. (1995) |
| CS 22891−209 | 4699 | 1.18 | 2.80 | −3.32 | +1.12 | −0.65 | +0.69 | +0.04 | +0.18 | −0.55 | | 1 | Yong et al. (2013) |
| CS 22892−052 | 4710 | 1.50 | 2.49 | −3.10 | +1.00 | +1.05 | +0.38 | +1.43 | +0.68 | +0.96 | CEMP | 0 | Masseron et al. (2012) |
| CS 22893−010 | 5150 | 2.45 | 1.69 | −2.93 | +1.55 | +0.13 | +0.01 | +0.14 | +0.01 | −1.28 | | 1 | Roederer et al. (2014) |
| CS 22896−154 | 5100 | 2.28 | 1.85 | −2.85 | −0.23 | +0.23 | +0.01 | +0.24 | +0.63 | +0.51 | | 1 | Yong et al. (2013) |
| CS 22897−008 | 4795 | 1.43 | 2.59 | −3.50 | +0.24 | +0.56 | +0.45 | +1.01 | +0.55 | −1.00 | CEMP-no | 1 | Yong et al. (2013) |
| CS 22898−027 | 6250 | 3.70 | 0.78 | −2.26 | +0.90 | +2.20 | 0.00 | +2.20 | +0.98 | +2.23 | CEMP-$s/rs$ | 0 | Masseron et al. (2012) |
| CS 22937−072 | 5450 | 2.10 | 2.14 | −2.74 | ··· | +0.51 | +0.02 | +0.53 | +0.17 | −0.18 | | 1 | Preston et al. (2006) |
| CS 22941−012 | 7200 | 4.20 | 0.53 | −2.02 | +0.09 | +0.09 | 0.00 | +0.09 | +0.29 | +0.01 | | 1 | Sneden et al. (2003) |
| CS 22942−019 | 5000 | 2.40 | 1.69 | −2.64 | ··· | +2.00 | +0.02 | +2.02 | +1.50 | +1.92 | CEMP-$s/rs$ | 0 | Aoki et al. (2002) |
| CS 22943−132 | 5850 | 3.60 | 0.76 | −2.67 | +0.49 | +0.69 | 0.00 | +0.69 | ··· | −0.05 | | 1 | Roederer et al. (2014) |
| CS 22943−201 | 5970 | 2.45 | 1.95 | −2.69 | +1.56 | +1.89 | +0.02 | +1.91 | ··· | −0.53 | CEMP-no | 1 | Roederer et al. (2014) |
| CS 22944−032 | 5293 | 2.82 | 1.37 | −2.98 | ≤ −0.44 | +0.31 | +0.01 | +0.32 | −0.24 | −0.76 | | 1 | Yong et al. (2013) |
| CS 22944−039 | 5350 | 1.50 | 2.71 | −2.41 | ··· | −0.22 | +0.49 | +0.27 | +0.47 | −0.19 | | 1 | Preston et al. (2006) |
| CS 22945−024 | 5120 | 2.35 | 1.78 | −2.58 | +0.26 | +2.31 | +0.02 | +2.33 | +0.30 | +1.44 | CEMP-$s/rs$ | 0 | Roederer et al. (2014) |
| CS 22945−028 | 5126 | 2.55 | 1.59 | −2.66 | ··· | +0.17 | +0.01 | +0.18 | +0.16 | −0.18 | | 1 | Barklem et al. (2005) |
| CS 22947−187 | 5200 | 1.50 | 2.66 | −2.51 | +1.73 | +1.07 | +0.27 | +1.34 | +0.58 | +1.24 | CEMP-$s/rs$ | 0 | Masseron et al. (2012) |
| CS 22948−006 | 5550 | 1.95 | 2.32 | −2.62 | ··· | +0.49 | +0.08 | +0.57 | −0.22 | −0.68 | | 1 | Preston et al. (2006) |
| CS 22948−027 | 4600 | 1.00 | 2.95 | −2.57 | +1.80 | +2.00 | +0.14 | +2.14 | +0.95 | +1.85 | CEMP-$s/rs$ | 0 | Masseron et al. (2012) |
| CS 22948−066 | 4830 | 1.55 | 2.48 | −3.18 | +1.05 | −0.74 | +0.36 | −0.38 | −0.35 | −1.15 | | 1 | Roederer et al. (2014) |
| CS 22948−104 | 5000 | 2.30 | 1.79 | −2.76 | ··· | +0.38 | +0.01 | +0.39 | ··· | ··· | | 1 | Masseron et al. (2012) |
| CS 22949−008a | 6300 | 3.50 | 0.99 | −2.09 | +0.41 | +1.55 | 0.00 | +1.55 | ··· | +1.32 | CEMP-$s/rs$ | 0 | Masseron et al. (2012) |
| CS 22949−008b | 5300 | 4.70 | −0.51 | −2.09 | +0.41 | +1.55 | 0.00 | +1.55 | ··· | +1.32 | CEMP-$s/rs$ | 0 | Masseron et al. (2012) |
| CS 22949−037 | 4630 | 0.95 | 3.01 | −4.21 | +2.50 | +0.99 | +0.74 | +1.73 | ··· | −0.78 | CEMP-no | 1 | Roederer et al. (2014) |
| CS 22949−048 | 4828 | 1.81 | 2.22 | −2.73 | ··· | −0.24 | +0.14 | −0.10 | −1.47 | ≤ −1.38 | | 1 | Lai et al. (2007) |
| CS 22950−046 | 4769 | 1.37 | 2.64 | −3.39 | ··· | −0.50 | +0.52 | +0.02 | −0.62 | −1.23 | | 1 | Yong et al. (2013) |
| CS 22951−077 | 5350 | 1.75 | 2.46 | −2.43 | ··· | −0.10 | +0.25 | +0.15 | −0.02 | −0.34 | | 1 | Preston et al. (2006) |
| CS 22952−015 | 4824 | 1.50 | 2.53 | −3.44 | +1.31 | −0.41 | +0.38 | −0.03 | −0.87 | −1.33 | | 1 | Yong et al. (2013) |
| CS 22953−003 | 5002 | 2.01 | 2.08 | −2.93 | +0.12 | +0.30 | +0.02 | +0.32 | +0.33 | +0.49 | | 1 | Yong et al. (2013) |
| CS 22953−037 | 6627 | 4.11 | 0.47 | −2.70 | ··· | +0.37 | 0.00 | +0.37 | ≤ −1.20 | ≤ −0.25 | | 1 | Yong et al. (2013) |
| CS 22956−050 | 4844 | 1.56 | 2.48 | −3.39 | +0.31 | +0.27 | +0.33 | +0.60 | −0.33 | −0.78 | | 1 | Yong et al. (2013) |

**Table 3** — *Continued*

| Name | $T_{\rm eff}$ (K) | log $g$ (cgs) | log $L$ ($L_\odot$) | [Fe/H] | [N/Fe] | [C/Fe] | Δ[C/Fe] ([N/Fe]=0) | [C/Fe]$_c$ | [Sr/Fe] | [Ba/Fe] | Class | I/O | Ref. |
|---|---|---|---|---|---|---|---|---|---|---|---|---|---|
| CS 22957−013 | 4904 | 1.96 | 2.10 | −2.64 | ··· | +0.06 | +0.06 | +0.12 | −0.17 | −0.60 | | 1 | Barklem et al. (2005) |
| CS 22957−022 | 5146 | 2.40 | 1.74 | −2.92 | +0.21 | +0.16 | +0.01 | +0.17 | −0.31 | −1.09 | | 1 | Yong et al. (2013) |
| CS 22957−027 | 5220 | 2.65 | 1.52 | −3.00 | +1.50 | +2.42 | +0.02 | +2.44 | ··· | −1.00 | CEMP-no | 1 | Roederer et al. (2014) |
| CS 22958−042 | 5760 | 3.55 | 0.79 | −2.99 | +2.25 | +2.15 | 0.00 | +2.15 | −0.52 | −1.02 | CEMP-no | 1 | Roederer et al. (2014) |
| CS 22958−083 | 4900 | 1.90 | 2.16 | −2.97 | ··· | +0.71 | +0.07 | +0.78 | −0.14 | ≤ −0.82 | CEMP-no | 1 | Preston et al. (2006) |
| CS 22960−010 | 5737 | 4.85 | −0.52 | −2.65 | ··· | +0.78 | 0.00 | +0.78 | ··· | ··· | CEMP | 1 | Barklem et al. (2005) |
| CS 22960−053 | 5200 | 2.10 | 2.06 | −3.14 | ··· | +2.05 | +0.02 | +2.07 | ··· | +0.86 | CEMP | 0 | Aoki et al. (2007) |
| CS 22960−064 | 5060 | 2.20 | 1.91 | −2.77 | +1.35 | +0.14 | +0.01 | +0.15 | ··· | −0.20 | | 1 | Roederer et al. (2014) |
| CS 22963−004 | 5597 | 3.34 | 0.95 | −3.54 | +0.80 | +0.40 | 0.00 | +0.40 | −0.73 | −0.51 | | 1 | Yong et al. (2013) |
| CS 22964−161a | 6050 | 3.70 | 0.72 | −2.37 | ··· | +1.58 | 0.00 | +1.58 | +0.55 | +1.36 | CEMP-$s/rs$ | 0 | Thompson et al. (2008) |
| CS 22964−161b | 5850 | 4.10 | 0.26 | −2.39 | ··· | +1.40 | 0.00 | +1.40 | +0.45 | +1.30 | CEMP-$s/rs$ | 0 | Thompson et al. (2008) |
| CS 22964−214 | 6180 | 3.75 | 0.71 | −2.95 | +1.51 | +0.62 | 0.00 | +0.62 | −0.25 | −0.54 | | 1 | Roederer et al. (2014) |
| CS 22965−016 | 4904 | 1.99 | 2.07 | −2.42 | ··· | −0.74 | +0.04 | −0.70 | +0.12 | ≤ −1.07 | | 1 | Lai et al. (2007) |
| CS 22965−029 | 5467 | 3.38 | 0.87 | −2.14 | ··· | −0.64 | 0.00 | −0.64 | +0.64 | +0.16 | | 1 | Lai et al. (2007) |
| CS 22965−054 | 6137 | 3.68 | 0.77 | −3.10 | ··· | +0.62 | 0.00 | +0.62 | +0.19 | ≤ −0.48 | | 1 | Yong et al. (2013) |
| CS 22966−011 | 6298 | 4.43 | 0.06 | −3.02 | ··· | +0.45 | 0.00 | +0.45 | +0.13 | −0.05 | | 1 | Yong et al. (2013) |
| CS 22966−057 | 5364 | 3.07 | 1.14 | −2.43 | +0.10 | +0.06 | +0.01 | +0.07 | −0.01 | −0.24 | | 1 | Yong et al. (2013) |
| CS 22968−001 | 6000 | 4.00 | 0.41 | −3.01 | ··· | +0.48 | 0.00 | +0.48 | −0.23 | −0.69 | | 1 | Preston et al. (2006) |
| CS 22968−014 | 4864 | 1.60 | 2.44 | −3.58 | +0.24 | +0.25 | +0.25 | +0.50 | −1.69 | −1.77 | | 1 | Yong et al. (2013) |
| CS 29491−053 | 4700 | 1.21 | 2.77 | −3.03 | +0.82 | −0.27 | +0.66 | +0.39 | −0.15 | −0.89 | | 1 | Yong et al. (2013) |
| CS 29491−069 | 5103 | 2.45 | 1.68 | −2.81 | ··· | +0.14 | +0.01 | +0.15 | +0.29 | +0.32 | | 1 | Barklem et al. (2005) |
| CS 29491−109 | 4736 | 1.50 | 2.50 | −2.90 | ··· | −0.23 | +0.43 | +0.20 | −0.28 | −0.97 | | 1 | Barklem et al. (2005) |
| CS 29493−090 | 4692 | 1.28 | 2.70 | −3.13 | +0.76 | +0.70 | +0.53 | +1.23 | ··· | +0.52 | CEMP | 1 | Barklem et al. (2005) |
| CS 29495−041 | 4740 | 1.34 | 2.66 | −2.74 | +0.40 | −0.07 | +0.55 | +0.48 | −0.06 | −0.65 | | 1 | Yong et al. (2013) |
| CS 29497−004 | 5013 | 2.23 | 1.87 | −2.81 | ··· | +0.18 | +0.01 | +0.19 | +0.67 | +1.16 | | 0 | Barklem et al. (2005) |
| CS 29497−030 | 6650 | 3.50 | 1.09 | −2.70 | +1.88 | +2.38 | +0.01 | +2.39 | +1.37 | +2.17 | CEMP-$s/rs$ | 0 | Masseron et al. (2012) |
| CS 29497−034 | 4983 | 1.96 | 2.13 | −3.00 | +2.63 | +2.72 | +0.04 | +2.76 | +1.05 | +2.28 | CEMP-$s/rs$ | 0 | Yong et al. (2013) |
| CS 29497−040 | 5487 | 3.41 | 0.84 | −2.63 | ··· | +0.31 | 0.00 | +0.31 | ≤ +0.03 | ≤ −0.89 | | 1 | Lai et al. (2007) |
| CS 29498−043 | 4440 | 0.50 | 3.39 | −3.85 | +1.71 | +2.72 | +0.31 | +3.03 | −0.30 | −0.51 | CEMP-no | 1 | Roederer et al. (2014) |
| CS 29499−060 | 6595 | 4.05 | 0.52 | −2.50 | ··· | +0.38 | 0.00 | +0.38 | −0.40 | ≤ −0.51 | | 1 | Yong et al. (2013) |
| CS 29502−042 | 5039 | 2.09 | 2.02 | −3.27 | −0.43 | +0.16 | +0.01 | +0.17 | −1.88 | −1.69 | | 1 | Yong et al. (2013) |
| CS 29502−092 | 4820 | 1.50 | 2.53 | −3.20 | +1.00 | +0.96 | +0.39 | +1.35 | −0.30 | −1.46 | CEMP-no | 1 | Roederer et al. (2014) |
| CS 29506−007 | 6522 | 3.94 | 0.61 | −2.73 | ≤ +0.93 | +0.49 | 0.00 | +0.49 | +0.18 | +0.09 | | 1 | Yong et al. (2013) |
| CS 29506−090 | 6603 | 4.18 | 0.39 | −2.62 | ··· | +0.41 | 0.00 | +0.41 | +0.46 | −0.35 | | 1 | Yong et al. (2013) |
| CS 29509−027 | 6850 | 3.50 | 1.14 | −2.42 | +2.44 | +1.74 | 0.00 | +1.74 | +1.03 | +1.50 | CEMP-$s/rs$ | 0 | Roederer et al. (2014) |
| CS 29510−058 | 5192 | 2.73 | 1.43 | −2.34 | ··· | +0.46 | +0.01 | +0.47 | +0.27 | −0.14 | | 1 | Lai et al. (2007) |
| CS 29512−073 | 5600 | 3.40 | 0.89 | −2.04 | +0.56 | +1.20 | 0.00 | +1.20 | ··· | +1.12 | CEMP-$s/rs$ | 0 | Masseron et al. (2012) |
| CS 29513−032 | 5810 | 3.30 | 1.05 | −2.08 | ··· | +0.63 | 0.00 | +0.63 | ··· | ··· | | 1 | Roederer et al. (2010) |
| CS 29516−024 | 4637 | 1.04 | 2.92 | −3.05 | −0.76 | −0.06 | +0.74 | +0.68 | −0.48 | −0.90 | | 1 | Yong et al. (2013) |
| CS 29517−025 | 5647 | 3.53 | 0.77 | −2.02 | ··· | −0.39 | 0.00 | −0.39 | +0.77 | +0.71 | | 0 | Lai et al. (2007) |
| CS 29518−051 | 5100 | 2.29 | 1.84 | −2.64 | +0.82 | −0.13 | +0.01 | −0.12 | +0.15 | −0.45 | | 1 | Yong et al. (2013) |
| CS 29522−046 | 5974 | 3.72 | 0.68 | −2.12 | −0.33 | +0.42 | 0.00 | +0.42 | +0.16 | −0.13 | | 1 | Yong et al. (2013) |
| CS 29526−110 | 6800 | 4.10 | 0.53 | −2.38 | +1.49 | +2.29 | 0.00 | +2.29 | +0.93 | +2.15 | CEMP-$s/rs$ | 0 | Aoki et al. (2008) |
| CS 29527−015 | 6577 | 3.89 | 0.68 | −3.32 | ··· | +1.18 | 0.00 | +1.18 | +0.44 | ··· | CEMP | 1 | Yong et al. (2013) |
| CS 29528−028 | 6800 | 4.00 | 0.63 | −2.86 | ··· | +2.77 | 0.00 | +2.77 | ··· | +3.27 | CEMP-$s/rs$ | 0 | Aoki et al. (2007) |
| CS 29528−041 | 6150 | 4.00 | 0.45 | −3.30 | +3.04 | +1.57 | 0.00 | +1.57 | −0.10 | +0.89 | CEMP | 0 | Sivarani et al. (2006) |
| CS 30301−015 | 4889 | 1.73 | 2.32 | −2.73 | +1.70 | +1.60 | +0.14 | +1.74 | +0.37 | +1.45 | CEMP-$s/rs$ | 0 | Yong et al. (2013) |
| CS 30301−024 | 6584 | 4.03 | 0.54 | −2.54 | ··· | +0.23 | 0.00 | +0.23 | ··· | −0.28 | | 1 | Yong et al. (2013) |
| CS 30306−082 | 5598 | 3.50 | 0.79 | −2.43 | ··· | +0.21 | 0.00 | +0.21 | −0.22 | −0.41 | | 1 | Lai et al. (2007) |
| CS 30306−132 | 5047 | 2.16 | 1.95 | −2.56 | ··· | +0.34 | +0.01 | +0.35 | +0.18 | +0.22 | | 1 | Yong et al. (2013) |
| CS 30308−035 | 4806 | 1.78 | 2.24 | −3.35 | ··· | 0.00 | +0.09 | +0.09 | −0.71 | ··· | | 1 | Barklem et al. (2005) |
| CS 30312−059 | 4908 | 1.75 | 2.31 | −3.22 | ≤ −0.48 | +0.27 | +0.11 | +0.38 | +0.02 | −0.14 | | 1 | Yong et al. (2013) |
| CS 30312−100 | 5117 | 2.34 | 1.79 | −2.66 | ≤ +1.85 | +0.54 | +0.01 | +0.55 | ··· | −0.73 | | 1 | Yong et al. (2013) |
| CS 30314−067 | 4320 | 0.50 | 3.34 | −3.01 | +1.18 | +0.55 | +0.66 | +1.21 | −0.27 | −0.55 | CEMP-no | 1 | Roederer et al. (2014) |
| CS 30315−001 | 4565 | 1.14 | 2.79 | −2.97 | ··· | −0.54 | +0.72 | +0.18 | +0.42 | −0.87 | | 1 | Barklem et al. (2005) |

**Table 3** — *Continued*

| Name | $T_{eff}$ (K) | log $g$ (cgs) | log $L$ ($L_\odot$) | [Fe/H] | [N/Fe] | [C/Fe] | Δ[C/Fe] ([N/Fe]=0) | [C/Fe]$_c$ | [Sr/Fe] | [Ba/Fe] | Class | I/O | Ref. |
|---|---|---|---|---|---|---|---|---|---|---|---|---|---|
| CS 30315−029 | 4541 | 1.07 | 2.85 | −3.33 | ··· | −0.47 | +0.75 | +0.28 | −0.09 | +0.36 | | 1 | Barklem et al. (2005) |
| CS 30315−093 | 5638 | 3.52 | 0.78 | −2.39 | ··· | −0.09 | 0.00 | −0.09 | −0.11 | −0.52 | | 1 | Lai et al. (2007) |
| CS 30319−020 | 5200 | 2.50 | 1.66 | −2.35 | ··· | +0.32 | +0.01 | +0.33 | −0.01 | −0.73 | | 1 | Aoki et al. (2005) |
| CS 30325−028 | 4894 | 1.73 | 2.32 | −2.87 | −0.22 | +0.38 | +0.21 | +0.59 | +0.19 | −0.49 | | 1 | Yong et al. (2013) |
| CS 30325−094 | 4948 | 1.85 | 2.22 | −3.35 | +0.18 | 0.00 | +0.04 | +0.04 | −2.14 | −1.88 | | 1 | Yong et al. (2013) |
| CS 30327−038 | 5100 | 1.70 | 2.43 | −2.64 | ··· | −0.19 | +0.27 | +0.08 | −0.59 | −1.46 | | 1 | Aoki et al. (2005) |
| CS 30329−004 | 5000 | 1.50 | 2.59 | −2.75 | ··· | −0.28 | +0.46 | +0.18 | +0.07 | −0.89 | | 1 | Aoki et al. (2005) |
| CS 30329−129 | 5467 | 3.37 | 0.88 | −2.24 | ··· | +0.11 | 0.00 | +0.11 | +0.04 | −0.60 | | 1 | Lai et al. (2007) |
| CS 30336−049 | 4827 | 1.51 | 2.52 | −4.03 | +1.00 | −0.20 | +0.29 | +0.09 | −1.53 | −1.50 | | 1 | Lai et al. (2008) |
| CS 30337−097 | 4865 | 1.81 | 2.23 | −2.74 | ··· | −0.08 | +0.15 | +0.07 | ··· | ··· | | 1 | Barklem et al. (2005) |
| CS 30338−089 | 4886 | 1.72 | 2.33 | −2.78 | +1.27 | +2.06 | +0.11 | +2.17 | ··· | +2.30 | CEMP-$s/rs$ | 0 | Yong et al. (2013) |
| CS 30339−041 | 5478 | 2.10 | 2.15 | −2.21 | ··· | −0.45 | +0.01 | −0.44 | +0.25 | +0.07 | | 1 | Barklem et al. (2005) |
| CS 30339−069 | 6326 | 3.79 | 0.71 | −3.05 | ··· | +0.56 | 0.00 | +0.56 | −0.00 | ··· | | 1 | Yong et al. (2013) |
| CS 30343−063 | 4412 | 0.83 | 3.04 | −2.97 | ··· | −0.71 | +0.75 | +0.04 | +0.05 | −0.95 | | 1 | Barklem et al. (2005) |
| CS 30492−001 | 5790 | 3.65 | 0.70 | −2.35 | +1.12 | −0.05 | 0.00 | −0.05 | ··· | −0.45 | | 1 | Roederer et al. (2014) |
| CS 31060−047 | 4749 | 1.55 | 2.45 | −2.72 | ··· | −0.30 | +0.41 | +0.11 | −0.04 | −1.07 | | 1 | Barklem et al. (2005) |
| CS 31061−032 | 6448 | 4.33 | 0.20 | −2.56 | ··· | +0.56 | 0.00 | +0.56 | +0.31 | −0.40 | | 1 | Yong et al. (2013) |
| CS 31062−012 | 6190 | 4.47 | −0.01 | −2.67 | +1.20 | +2.12 | 0.00 | +2.12 | ··· | +2.32 | CEMP-$s/rs$ | 0 | Yong et al. (2013) |
| CS 31062−041 | 4990 | 2.00 | 2.09 | −2.65 | ≤ +1.59 | +0.58 | +0.04 | +0.62 | −0.29 | −0.42 | | 1 | Yong et al. (2013) |
| CS 31062−050 | 5607 | 3.49 | 0.80 | −2.28 | +1.20 | +2.00 | 0.00 | +2.00 | ··· | +2.30 | CEMP-$s/rs$ | 0 | Yong et al. (2013) |
| CS 31069−064 | 5468 | 3.38 | 0.87 | −2.02 | ··· | +0.16 | 0.00 | +0.16 | −0.53 | −0.90 | | 1 | Lai et al. (2007) |
| CS 31070−058 | 4864 | 1.89 | 2.15 | −2.12 | ··· | +0.11 | +0.14 | +0.25 | +0.17 | −0.21 | | 1 | Lai et al. (2007) |
| CS 31072−118 | 4606 | 1.25 | 2.70 | −3.06 | ··· | −0.54 | +0.65 | +0.11 | −0.07 | −1.09 | | 1 | Barklem et al. (2005) |
| CS 31078−018 | 5100 | 2.27 | 1.86 | −2.99 | −0.38 | +0.37 | +0.01 | +0.38 | +0.26 | +0.38 | | 1 | Yong et al. (2013) |
| CS 31080−095 | 6050 | 4.50 | −0.08 | −2.85 | +0.67 | +2.67 | 0.00 | +2.67 | −0.31 | +0.72 | CEMP | 0 | Sivarani et al. (2006) |
| CS 31082−001 | 4866 | 1.66 | 2.38 | −2.75 | −0.51 | +0.16 | +0.30 | +0.46 | +0.75 | +1.02 | | 0 | Yong et al. (2013) |
| CS 31085−024 | 5778 | 4.64 | −0.30 | −2.80 | ≤ −0.24 | +0.36 | 0.00 | +0.36 | −0.24 | −0.57 | | 1 | Yong et al. (2013) |
| G 64−37 | 6318 | 4.16 | 0.34 | −3.12 | ··· | +0.25 | 0.00 | +0.25 | ··· | ··· | | 1 | Simmerer et al. (2004) |
| G 77−61 | 4000 | 5.00 | −1.30 | −4.03 | +2.48 | +2.49 | 0.00 | +2.49 | ··· | ≤ +1.00 | CEMP-no | 1 | Masseron et al. (2012) |
| HD 2796 | 4923 | 1.84 | 2.22 | −2.31 | +0.85 | −0.50 | +0.17 | −0.33 | +0.33 | −0.14 | | 1 | Yong et al. (2013) |
| HD 3008 | 4250 | 1.03 | 2.78 | −2.08 | ··· | −0.29 | +0.65 | +0.36 | ··· | +0.08 | | 1 | Simmerer et al. (2004) |
| HD 4306 | 4854 | 1.61 | 2.43 | −3.04 | ··· | +0.11 | +0.31 | +0.42 | −0.07 | −1.17 | | 1 | Yong et al. (2013) |
| HD 5223 | 4500 | 1.00 | 2.91 | −2.11 | ··· | +1.58 | +0.12 | +1.70 | +1.49 | +1.88 | CEMP-$s/rs$ | 0 | Goswami et al. (2006) |
| HD 5426 | 4910 | 1.90 | 2.16 | −2.39 | ··· | −0.04 | +0.12 | +0.08 | +0.24 | −0.14 | | 1 | McWilliam et al. (1995) |
| HD 6268 | 4696 | 1.23 | 2.75 | −2.74 | ··· | −0.67 | +0.68 | +0.01 | +0.11 | −0.04 | | 1 | Yong et al. (2013) |
| HD 13979 | 4970 | 1.15 | 2.93 | −2.63 | ··· | −0.50 | +0.70 | +0.20 | −0.91 | −0.73 | | 1 | McWilliam et al. (1995) |
| HD 23798 | 4450 | 1.06 | 2.83 | −2.26 | ··· | −0.59 | +0.74 | +0.15 | ··· | ··· | | 1 | Simmerer et al. (2004) |
| HD 26169 | 4750 | 1.35 | 2.65 | −2.80 | −0.45 | +0.15 | +0.54 | +0.69 | +0.44 | −0.50 | | 1 | Roederer et al. (2014) |
| HD 29574 | 4250 | 0.80 | 3.01 | −2.00 | ··· | −0.69 | +0.71 | +0.02 | ··· | +0.59 | | 1 | Simmerer et al. (2004) |
| HD 85773 | 4268 | 0.87 | 2.95 | −2.62 | ··· | −0.49 | +0.75 | +0.26 | ··· | −0.49 | | 1 | Simmerer et al. (2004) |
| HD 88609 | 4541 | 0.80 | 3.12 | −2.98 | ··· | −0.51 | +0.75 | +0.24 | −0.01 | −1.05 | | 1 | Yong et al. (2013) |
| HD 103036 | 4200 | 1.14 | 2.65 | −2.04 | ··· | −0.39 | +0.63 | +0.24 | ··· | ··· | | 1 | Simmerer et al. (2004) |
| HD 103545 | 4666 | 1.64 | 2.33 | −2.45 | ··· | −0.44 | +0.37 | −0.07 | ··· | −0.04 | | 1 | Simmerer et al. (2004) |
| HD 107752 | 4649 | 1.63 | 2.34 | −2.78 | ··· | −0.59 | +0.32 | −0.27 | ··· | −0.24 | | 1 | Simmerer et al. (2004) |
| HD 108317 | 5234 | 2.68 | 1.49 | −2.18 | ··· | −0.09 | +0.01 | −0.08 | ··· | +0.58 | | 1 | Simmerer et al. (2004) |
| HD 110184 | 4250 | 0.79 | 3.02 | −2.72 | ··· | −0.34 | +0.75 | +0.41 | −0.02 | −0.01 | | 1 | Simmerer et al. (2004) |
| HD 115444 | 4711 | 1.23 | 2.76 | −2.65 | ··· | −0.41 | +0.66 | +0.25 | +0.32 | +0.14 | | 1 | Yong et al. (2013) |
| HD 119516 | 5382 | 2.47 | 1.75 | −2.11 | ··· | −1.19 | +0.01 | −1.18 | ··· | ··· | | 1 | Simmerer et al. (2004) |
| HD 122563 | 4843 | 1.62 | 2.42 | −2.54 | +0.70 | −0.44 | +0.36 | −0.08 | −0.38 | −1.12 | | 1 | Yong et al. (2013) |
| HD 126587 | 4853 | 1.61 | 2.43 | −3.00 | ··· | +0.19 | +0.31 | +0.50 | −0.09 | −0.10 | | 1 | Yong et al. (2013) |
| HD 128279 | 5050 | 2.15 | 1.96 | −2.52 | −0.39 | +0.08 | +0.01 | +0.09 | −0.51 | −0.64 | | 1 | Roederer et al. (2014) |
| HD 135148 | 4183 | 1.24 | 2.54 | −2.17 | ··· | +0.76 | +0.34 | +1.10 | ··· | +0.30 | CEMP | 1 | Simmerer et al. (2004) |
| HD 140283 | 5711 | 3.53 | 0.79 | −2.56 | ··· | +0.28 | 0.00 | +0.28 | −0.05 | −0.94 | | 1 | Yong et al. (2013) |
| HD 165195 | 4237 | 0.78 | 3.02 | −2.60 | ··· | −0.54 | +0.75 | +0.21 | +0.11 | +0.23 | | 1 | Simmerer et al. (2004) |
| HD 178443 | 5180 | 1.65 | 2.50 | −2.05 | ··· | −0.28 | +0.35 | +0.07 | +0.27 | +0.01 | | 1 | McWilliam et al. (1995) |

**Table 3** — *Continued*

| Name | $T_{\rm eff}$ (K) | log $g$ (cgs) | log $L$ ($L_\odot$) | [Fe/H] | [N/Fe] | [C/Fe] | Δ[C/Fe] ([N/Fe]=0) | $[C/Fe]_c$ | [Sr/Fe] | [Ba/Fe] | Class | I/O | Ref. |
|---|---|---|---|---|---|---|---|---|---|---|---|---|---|
| HD 186478 | 4629 | 1.07 | 2.89 | −2.68 | +0.62 | −0.28 | +0.73 | +0.45 | +0.32 | −0.04 | | 1 | Yong et al. (2013) |
| HD 196944 | 5255 | 2.74 | 1.44 | −2.44 | +1.30 | +1.20 | +0.02 | +1.22 | +0.89 | +1.10 | CEMP-$s/rs$ | 0 | Yong et al. (2013) |
| HD 200654 | 5160 | 2.55 | 1.60 | −2.80 | ··· | +0.27 | +0.01 | +0.28 | −0.44 | −0.83 | | 1 | McWilliam et al. (1995) |
| HD 214925 | 4050 | 0.30 | 3.43 | −2.03 | ··· | −0.85 | +0.74 | −0.11 | ··· | ··· | | 1 | Aoki et al. (2008) |
| HD 215801 | 6005 | 3.81 | 0.60 | −2.29 | ··· | +0.04 | 0.00 | +0.04 | ··· | ··· | | 1 | Akerman et al. (2004) |
| HD 216143 | 4450 | 0.80 | 3.09 | −2.27 | ··· | −0.41 | +0.73 | +0.32 | +0.21 | −0.27 | | 1 | Aoki et al. (2008) |
| HD 237846 | 4600 | 1.00 | 2.95 | −3.29 | ··· | +0.17 | +0.75 | +0.92 | ··· | ··· | CEMP | 1 | Roederer et al. (2010) |
| HD 340279 | 6273 | 4.25 | 0.24 | −2.62 | ··· | +0.17 | 0.00 | +0.17 | ··· | ··· | | 1 | Akerman et al. (2004) |
| HE 0005−0002 | 4726 | 1.58 | 2.41 | −3.06 | ··· | +0.13 | +0.32 | +0.45 | +0.29 | −0.03 | | 1 | Barklem et al. (2005) |
| HE 0007−1832 | 6500 | 3.80 | 0.75 | −2.79 | +2.11 | +2.66 | 0.00 | +2.66 | +0.22 | +0.09 | CEMP | 1 | Cohen et al. (2013) |
| HE 0008−3842 | 4327 | 0.65 | 3.19 | −3.33 | ··· | −0.93 | +0.73 | −0.20 | −1.40 | −1.80 | | 1 | Barklem et al. (2005) |
| HE 0009−6039 | 5250 | 2.70 | 1.48 | −3.29 | ··· | +0.16 | +0.01 | +0.17 | ··· | −1.11 | | 1 | Cohen et al. (2013) |
| HE 0011−0035 | 4950 | 1.80 | 2.27 | −2.99 | ··· | 0.00 | +0.12 | +0.12 | ··· | −0.27 | | 1 | Cohen et al. (2013) |
| HE 0012−1441 | 5750 | 3.50 | 0.83 | −2.52 | +0.54 | +1.59 | 0.00 | +1.59 | ··· | +1.15 | CEMP-$s/rs$ | 0 | Cohen et al. (2013) |
| HE 0013−0257 | 4500 | 0.50 | 3.41 | −3.82 | ··· | −0.22 | +0.69 | +0.47 | ··· | −1.16 | | 1 | Hollek et al. (2011) |
| HE 0013−0522 | 4900 | 1.70 | 2.36 | −3.24 | ··· | +0.22 | +0.15 | +0.37 | ··· | −0.89 | | 1 | Hollek et al. (2011) |
| HE 0015+0048 | 4600 | 0.90 | 3.05 | −3.07 | ··· | +0.62 | +0.67 | +1.29 | ··· | −1.17 | CEMP-no | 1 | Hollek et al. (2011) |
| HE 0017−4346 | 6198 | 3.80 | 0.66 | −3.07 | ··· | +2.90 | 0.00 | +2.90 | ··· | +1.28 | CEMP-$s/rs$ | 0 | Cohen et al. (2013) |
| HE 0017−4838 | 5149 | 2.23 | 1.91 | −3.23 | ··· | −0.33 | 0.00 | −0.33 | +0.78 | −0.83 | | 1 | Barklem et al. (2005) |
| HE 0018−1349 | 5719 | 4.62 | −0.29 | −2.27 | ··· | +0.32 | 0.00 | +0.32 | +0.05 | +0.43 | | 1 | Barklem et al. (2005) |
| HE 0023−4825 | 5816 | 3.63 | 0.72 | −2.06 | ··· | +0.27 | 0.00 | +0.27 | +0.08 | −0.11 | | 1 | Barklem et al. (2005) |
| HE 0024−2523 | 6625 | 4.30 | 0.28 | −2.70 | +2.12 | +2.62 | 0.00 | +2.62 | +0.44 | +1.52 | CEMP-$s/rs$ | 0 | Masseron et al. (2012) |
| HE 0029−1839 | 5010 | 2.19 | 1.91 | −2.50 | ··· | +0.27 | +0.01 | +0.28 | +0.13 | −1.35 | | 1 | Barklem et al. (2005) |
| HE 0037−0348 | 4920 | 1.80 | 2.26 | −3.15 | +0.92 | +0.12 | +0.10 | +0.22 | ··· | −0.52 | | 1 | Cohen et al. (2013) |
| HE 0037−2657 | 4982 | 2.21 | 1.88 | −3.22 | ··· | +0.27 | +0.01 | +0.28 | −0.54 | −1.14 | | 1 | Barklem et al. (2005) |
| HE 0039−4154 | 4784 | 1.50 | 2.52 | −3.18 | ··· | +0.06 | +0.38 | +0.44 | −0.14 | +0.02 | | 1 | Cohen et al. (2013) |
| HE 0043−2845 | 5517 | 4.42 | −0.16 | −2.91 | ··· | +0.15 | 0.00 | +0.15 | −0.67 | ··· | | 1 | Barklem et al. (2005) |
| HE 0044−2459 | 5242 | 2.92 | 1.25 | −3.28 | ··· | +0.41 | 0.00 | +0.41 | −1.55 | ··· | | 1 | Barklem et al. (2005) |
| HE 0044−4023 | 5694 | 3.26 | 1.06 | −2.56 | ··· | +0.36 | 0.00 | +0.36 | +0.03 | −0.40 | | 1 | Barklem et al. (2005) |
| HE 0048−0611 | 5180 | 2.50 | 1.65 | −2.66 | +0.13 | +0.37 | +0.01 | +0.38 | ··· | −0.43 | | 1 | Cohen et al. (2013) |
| HE 0048−6408 | 4360 | 0.35 | 3.50 | −3.76 | ··· | −0.30 | +0.69 | +0.39 | −0.82 | −1.37 | | 1 | Placco et al. (2014a) |
| HE 0049−5700 | 5952 | 4.08 | 0.31 | −2.40 | ··· | +0.35 | 0.00 | +0.35 | +0.19 | −0.47 | | 1 | Barklem et al. (2005) |
| HE 0051−2304 | 4537 | 1.22 | 2.70 | −2.44 | ··· | −0.68 | +0.69 | +0.01 | +0.03 | −0.11 | | 1 | Barklem et al. (2005) |
| HE 0054−0657 | 5908 | 4.40 | −0.02 | −2.00 | ··· | +0.25 | 0.00 | +0.25 | +0.05 | −0.13 | | 1 | Barklem et al. (2005) |
| HE 0055−2314 | 6290 | 4.40 | 0.09 | −2.66 | ··· | +0.72 | 0.00 | +0.72 | ··· | −0.49 | CEMP-no | 1 | Cohen et al. (2013) |
| HE 0057−4541 | 5083 | 2.55 | 1.57 | −2.31 | ··· | +0.13 | +0.01 | +0.14 | +0.16 | −0.16 | | 1 | Barklem et al. (2005) |
| HE 0057−5959 | 5257 | 2.65 | 1.53 | −4.08 | +2.15 | +0.86 | 0.00 | +0.86 | ··· | −0.46 | CEMP-no | 1 | Yong et al. (2013) |
| HE 0058−0244 | 5620 | 3.40 | 0.89 | −2.76 | +1.78 | +1.93 | +0.01 | +1.94 | +0.42 | +1.97 | CEMP-$s/rs$ | 0 | Cohen et al. (2013) |
| HE 0102−0633 | 6012 | 3.70 | 0.71 | −3.10 | ··· | +0.87 | 0.00 | +0.87 | −0.41 | −0.54 | CEMP-no | 1 | Cohen et al. (2013) |
| HE 0102−1213 | 6100 | 3.65 | 0.79 | −3.28 | ··· | +1.31 | 0.00 | +1.31 | ··· | ≤ −0.64 | CEMP-no | 1 | Yong et al. (2013) |
| HE 0103−0357 | 5406 | 3.20 | 1.03 | −3.15 | −0.03 | +0.52 | 0.00 | +0.52 | ··· | −0.87 | | 1 | Cohen et al. (2013) |
| HE 0104−4007 | 5154 | 2.64 | 1.50 | −3.30 | ··· | +0.46 | +0.01 | +0.47 | −0.88 | −0.69 | | 1 | Barklem et al. (2005) |
| HE 0104−5300 | 4732 | 1.43 | 2.57 | −3.31 | ··· | −0.05 | +0.45 | +0.40 | −0.49 | −1.57 | | 1 | Barklem et al. (2005) |
| HE 0105−6141 | 5218 | 2.83 | 1.34 | −2.55 | ··· | +0.16 | +0.01 | +0.17 | +0.08 | +0.13 | | 1 | Barklem et al. (2005) |
| HE 0107−5240 | 5100 | 2.20 | 1.93 | −5.54 | +2.43 | +3.85 | +0.06 | +3.91 | ≤ −0.20 | ≤ +0.82 | CEMP-no | 1 | Yong et al. (2013) |
| HE 0109−0742 | 5315 | 1.85 | 2.35 | −2.52 | ··· | −0.18 | +0.15 | −0.03 | +0.07 | −0.22 | | 1 | Barklem et al. (2005) |
| HE 0111−1454 | 4632 | 1.05 | 2.91 | −2.99 | ··· | −0.26 | +0.74 | +0.48 | −0.03 | −0.75 | | 1 | Barklem et al. (2005) |
| HE 0121−2826 | 4955 | 1.99 | 2.09 | −2.97 | ··· | +0.46 | +0.03 | +0.49 | +0.37 | −0.64 | | 1 | Barklem et al. (2005) |
| HE 0122−1616 | 5215 | 2.50 | 1.66 | −2.80 | ··· | +0.31 | +0.01 | +0.32 | ··· | −1.07 | | 1 | Cohen et al. (2013) |
| HE 0130−1749 | 4820 | 1.60 | 2.43 | −3.34 | ··· | +0.40 | +0.32 | +0.72 | ··· | −0.72 | CEMP-no | 1 | Cohen et al. (2013) |
| HE 0131−2740 | 5351 | 3.03 | 1.18 | −3.08 | ··· | +0.31 | 0.00 | +0.31 | ··· | ··· | | 1 | Barklem et al. (2005) |
| HE 0131−3953 | 5928 | 3.83 | 0.56 | −2.71 | ··· | +2.36 | 0.00 | +2.36 | +0.56 | +2.12 | CEMP-$s/rs$ | 0 | Barklem et al. (2005) |
| HE 0132−2429 | 5294 | 2.75 | 1.44 | −3.60 | +1.22 | +0.83 | +0.01 | +0.84 | +0.13 | −0.85 | CEMP-no | 1 | Cohen et al. (2008) |
| HE 0132−2439 | 5249 | 2.63 | 1.55 | −3.79 | +1.07 | +0.62 | +0.01 | +0.63 | ··· | −0.88 | | 1 | Yong et al. (2013) |
| HE 0134−1519 | 5525 | 3.17 | 1.10 | −3.98 | ≤ −0.20 | +1.00 | 0.00 | +1.00 | ··· | ≤ −0.81 | CEMP-no | 1 | Hansen et al. (2014) |

**Table 3** — *Continued*

| Name | $T_{\rm eff}$ (K) | log $g$ (cgs) | log $L$ ($L_\odot$) | [Fe/H] | [N/Fe] | [C/Fe] | Δ[C/Fe] ([N/Fe]=0) | $[{\rm C/Fe}]_c$ | [Sr/Fe] | [Ba/Fe] | Class | I/O | Ref. |
|---|---|---|---|---|---|---|---|---|---|---|---|---|---|
| HE 0134−2504 | 5204 | 2.50 | 1.66 | −2.74 | ··· | +1.79 | +0.02 | +1.81 | ··· | +1.65 | CEMP-$s/rs$ | 0 | Cohen et al. (2013) |
| HE 0139−2826 | 4900 | 1.50 | 2.56 | −3.46 | +0.03 | +0.48 | +0.38 | +0.86 | −1.51 | −1.22 | CEMP-no | 1 | Placco et al. (2014a) |
| HE 0143−0441 | 6276 | 3.84 | 0.65 | −2.32 | +0.85 | +1.82 | 0.00 | +1.82 | +0.93 | +2.42 | CEMP-$s/rs$ | 0 | Yong et al. (2013) |
| HE 0143−1135 | 5629 | 4.53 | −0.23 | −2.13 | ··· | +0.19 | 0.00 | +0.19 | −0.10 | −0.13 | | 1 | Barklem et al. (2005) |
| HE 0143−4108 | 5121 | 2.41 | 1.72 | −2.62 | ··· | +0.12 | +0.01 | +0.13 | −0.84 | −1.23 | | 1 | Barklem et al. (2005) |
| HE 0143−4146 | 4799 | 1.50 | 2.52 | −2.95 | ··· | +0.07 | +0.41 | +0.48 | −0.52 | −0.84 | | 1 | Barklem et al. (2005) |
| HE 0146−1548 | 4636 | 0.99 | 2.97 | −3.46 | ··· | +0.84 | +0.73 | +1.57 | ··· | −0.71 | CEMP-no | 1 | Yong et al. (2013) |
| HE 0157−3335 | 4775 | 1.62 | 2.39 | −3.08 | ··· | −0.26 | +0.29 | +0.03 | +0.15 | −0.46 | | 1 | Barklem et al. (2005) |
| HE 0200−0955 | 5216 | 2.57 | 1.60 | −2.47 | ··· | +0.37 | +0.01 | +0.38 | −0.26 | −0.73 | | 1 | Barklem et al. (2005) |
| HE 0206−1916 | 5073 | 2.23 | 1.89 | −2.52 | +1.61 | +2.10 | +0.02 | +2.12 | ··· | +1.99 | CEMP-$s/rs$ | 0 | Yong et al. (2013) |
| HE 0207−1423 | 5023 | 2.07 | 2.03 | −2.95 | ··· | +2.38 | +0.03 | +2.41 | ··· | +1.73 | CEMP-$s/rs$ | 0 | Yong et al. (2013) |
| HE 0212−0557 | 5075 | 2.20 | 1.92 | −2.27 | +1.09 | +1.74 | +0.02 | +1.76 | +0.04 | +2.18 | CEMP-$s/rs$ | 0 | Cohen et al. (2013) |
| HE 0226+0103 | 5250 | 2.70 | 1.48 | −2.81 | +0.88 | +0.04 | +0.01 | +0.05 | ··· | +0.12 | | 1 | Cohen et al. (2013) |
| HE 0231−4016 | 5972 | 3.59 | 0.81 | −2.08 | ··· | +1.27 | 0.00 | +1.27 | +0.77 | +1.38 | CEMP-$s/rs$ | 0 | Barklem et al. (2005) |
| HE 0233−0343 | 6075 | 3.40 | 1.03 | −4.68 | ≤ +1.20 | +3.32 | 0.00 | +3.32 | ··· | ≤ +0.70 | CEMP-no | 1 | Hansen et al. (2014) |
| HE 0240−0807 | 4729 | 1.54 | 2.46 | −2.68 | ··· | −0.39 | +0.42 | +0.03 | +0.01 | +0.16 | | 1 | Barklem et al. (2005) |
| HE 0240−6105 | 4720 | 1.26 | 2.73 | −3.23 | ··· | −0.29 | +0.59 | +0.30 | +0.46 | −0.73 | | 1 | Barklem et al. (2005) |
| HE 0243−0753 | 5066 | 2.29 | 1.82 | −2.49 | ··· | +0.25 | +0.01 | +0.26 | −0.28 | −1.07 | | 1 | Barklem et al. (2005) |
| HE 0243−5238 | 5085 | 2.35 | 1.77 | −3.04 | ··· | +0.36 | +0.01 | +0.37 | +0.06 | −0.35 | | 1 | Barklem et al. (2005) |
| HE 0244−4111 | 5624 | 3.39 | 0.91 | −2.56 | ··· | +0.21 | 0.00 | +0.21 | −0.05 | −0.37 | | 1 | Barklem et al. (2005) |
| HE 0248+0039 | 5199 | 2.57 | 1.59 | −2.53 | ··· | +0.05 | +0.01 | +0.06 | −0.15 | −0.31 | | 1 | Barklem et al. (2005) |
| HE 0251−3216 | 5750 | 3.70 | 0.63 | −3.15 | ··· | +2.32 | 0.00 | +2.32 | ··· | +1.28 | CEMP-$s/rs$ | 0 | Cohen et al. (2013) |
| HE 0255−0859 | 5250 | 4.80 | −0.62 | −2.55 | ··· | −0.18 | 0.00 | −0.18 | ··· | −0.71 | | 1 | Cohen et al. (2013) |
| HE 0256−1109 | 5891 | 4.04 | 0.34 | −2.73 | ··· | +0.63 | 0.00 | +0.63 | ··· | ··· | | 1 | Barklem et al. (2005) |
| HE 0300−0751 | 5280 | 2.97 | 1.22 | −2.27 | ··· | +0.06 | +0.01 | +0.07 | +0.23 | −0.02 | | 1 | Barklem et al. (2005) |
| HE 0302−3417a | 4400 | 0.20 | 3.67 | −3.70 | ··· | +0.48 | +0.72 | +1.20 | ··· | −2.10 | CEMP-no | 1 | Hollek et al. (2011) |
| HE 0305−4520 | 4817 | 1.56 | 2.47 | −2.91 | ··· | +0.29 | +0.36 | +0.65 | −0.56 | +0.53 | | 1 | Barklem et al. (2005) |
| HE 0305−5442 | 4904 | 1.70 | 2.36 | −3.30 | ··· | +0.27 | +0.16 | +0.43 | ··· | −2.41 | | 1 | Cohen et al. (2013) |
| HE 0308−1154 | 4984 | 2.05 | 2.04 | −2.82 | ··· | +0.34 | +0.02 | +0.36 | −1.83 | ··· | | 1 | Barklem et al. (2005) |
| HE 0312−5200 | 5204 | 2.60 | 1.56 | −3.11 | ··· | +0.26 | +0.01 | +0.27 | ··· | −0.98 | | 1 | Cohen et al. (2013) |
| HE 0315+0000 | 5013 | 2.11 | 1.99 | −2.73 | ··· | +0.14 | +0.01 | +0.15 | +0.09 | +0.30 | | 1 | Barklem et al. (2005) |
| HE 0316+0214 | 4639 | 1.39 | 2.57 | −3.14 | ··· | −0.75 | +0.53 | −0.22 | −1.38 | −1.43 | | 1 | Barklem et al. (2005) |
| HE 0317−0554 | 5400 | 3.10 | 1.13 | −2.75 | ··· | +0.40 | +0.01 | +0.41 | ··· | −0.32 | | 1 | Cohen et al. (2013) |
| HE 0317−4640 | 5872 | 4.10 | 0.27 | −2.32 | ··· | +0.18 | 0.00 | +0.18 | +0.01 | −0.24 | | 1 | Barklem et al. (2005) |
| HE 0323−4529 | 5127 | 2.51 | 1.63 | −3.14 | ··· | +0.34 | +0.01 | +0.35 | +0.38 | +0.25 | | 1 | Barklem et al. (2005) |
| HE 0324+0152a | 4775 | 1.20 | 2.81 | −3.32 | ··· | +0.18 | +0.66 | +0.84 | ··· | −1.20 | CEMP-no | 1 | Hollek et al. (2011) |
| HE 0328−1047 | 5301 | 3.03 | 1.16 | −2.24 | ··· | +0.11 | +0.01 | +0.12 | +0.11 | −0.12 | | 1 | Barklem et al. (2005) |
| HE 0330−4004 | 5946 | 3.78 | 0.61 | −2.19 | ··· | +0.04 | 0.00 | +0.04 | +0.04 | −0.08 | | 1 | Barklem et al. (2005) |
| HE 0331−4939 | 5206 | 2.60 | 1.56 | −2.87 | ··· | +0.30 | +0.01 | +0.31 | +0.03 | −0.98 | | 1 | Barklem et al. (2005) |
| HE 0332−1007 | 4750 | 1.30 | 2.70 | −2.90 | ··· | −0.64 | +0.62 | −0.02 | ··· | −0.33 | | 1 | Cohen et al. (2013) |
| HE 0333−4001 | 5892 | 4.31 | 0.07 | −2.64 | ··· | +0.28 | 0.00 | +0.28 | −0.05 | −0.21 | | 1 | Barklem et al. (2005) |
| HE 0336+0113 | 5819 | 3.59 | 0.77 | −2.60 | +1.60 | +2.25 | 0.00 | +2.25 | +1.77 | +2.69 | CEMP-$s/rs$ | 0 | Yong et al. (2013) |
| HE 0336−3829 | 5740 | 3.22 | 1.11 | −2.75 | ··· | +0.19 | 0.00 | +0.19 | −0.19 | −0.13 | | 1 | Barklem et al. (2005) |
| HE 0337−5127 | 5247 | 2.86 | 1.32 | −2.61 | ··· | +0.12 | +0.01 | +0.13 | +0.09 | −0.07 | | 1 | Barklem et al. (2005) |
| HE 0338−3945 | 6160 | 4.13 | 0.32 | −2.38 | ··· | +2.07 | 0.00 | +2.07 | +0.84 | +2.31 | CEMP-$s/rs$ | 0 | Jonsell et al. (2006) |
| HE 0340−5355 | 4862 | 1.81 | 2.23 | −2.88 | ··· | −0.15 | +0.13 | −0.02 | −0.43 | −1.46 | | 1 | Barklem et al. (2005) |
| HE 0344−0243 | 5140 | 2.30 | 1.84 | −3.35 | +0.72 | −0.09 | +0.01 | −0.08 | ··· | −1.23 | | 1 | Cohen et al. (2013) |
| HE 0347−1819 | 5198 | 4.23 | −0.07 | −2.77 | ··· | −0.01 | 0.00 | −0.01 | −0.08 | −0.56 | | 1 | Barklem et al. (2005) |
| HE 0349−0045 | 5022 | 2.30 | 1.80 | −2.23 | +0.50 | +0.15 | +0.01 | +0.16 | ··· | −1.56 | | 1 | Cohen et al. (2013) |
| HE 0353−6024 | 5320 | 3.15 | 1.05 | −3.17 | ··· | +0.25 | 0.00 | +0.25 | −1.18 | ··· | | 1 | Barklem et al. (2005) |
| HE 0400−2917 | 5065 | 2.39 | 1.72 | −2.88 | ··· | +0.11 | +0.01 | +0.12 | −0.77 | −1.14 | | 1 | Barklem et al. (2005) |
| HE 0401−0138 | 4866 | 1.76 | 2.28 | −3.34 | ··· | +0.20 | +0.12 | +0.32 | +0.21 | −0.25 | | 1 | Barklem et al. (2005) |
| HE 0401−3835 | 5458 | 3.20 | 1.04 | −3.05 | ··· | +0.78 | 0.00 | +0.78 | ··· | −0.75 | CEMP-no | 1 | Cohen et al. (2013) |
| HE 0411−5725 | 5000 | 2.10 | 1.99 | −3.08 | ··· | +0.35 | +0.01 | +0.36 | ··· | −1.25 | | 1 | Cohen et al. (2013) |
| HE 0414−0343 | 4660 | 0.75 | 3.22 | −2.38 | ··· | +1.38 | +0.21 | +1.59 | ··· | +1.73 | CEMP-$s/rs$ | 0 | Hollek et al. (2014) |

**Table 3** — *Continued*

| Name | $T_{\rm eff}$ (K) | log $g$ (cgs) | log $L$ ($L_\odot$) | [Fe/H] | [N/Fe] | [C/Fe] | Δ[C/Fe] ([N/Fe]=0) | [C/Fe]$_c$ | [Sr/Fe] | [Ba/Fe] | Class | I/O | Ref. |
|---|---|---|---|---|---|---|---|---|---|---|---|---|---|
| HE 0417−0821 | 5811 | 4.83 | −0.48 | −2.33 | ··· | +0.21 | 0.00 | +0.21 | −0.06 | −0.41 | | 1 | Barklem et al. (2005) |
| HE 0420+0123a | 4800 | 1.45 | 2.57 | −3.03 | ··· | +0.33 | +0.45 | +0.78 | ··· | +0.08 | CEMP | 1 | Hollek et al. (2011) |
| HE 0430−4404 | 6214 | 4.27 | 0.20 | −2.07 | ··· | +1.35 | 0.00 | +1.35 | +0.66 | +1.53 | CEMP-$s/rs$ | 0 | Barklem et al. (2005) |
| HE 0430−4901 | 5296 | 3.12 | 1.07 | −2.72 | ··· | +0.05 | +0.01 | +0.06 | +0.10 | +0.46 | | 1 | Barklem et al. (2005) |
| HE 0432−0923 | 5131 | 2.64 | 1.50 | −3.19 | ··· | +0.20 | +0.01 | +0.21 | +0.58 | +0.67 | | 0 | Barklem et al. (2005) |
| HE 0432−1005a | 4525 | 0.50 | 3.42 | −3.21 | ··· | +0.23 | +0.73 | +0.96 | ··· | −0.90 | CEMP-no | 1 | Hollek et al. (2011) |
| HE 0436−4008 | 5431 | 3.34 | 0.90 | −2.35 | ··· | +0.45 | 0.00 | +0.45 | +0.43 | −0.31 | | 1 | Barklem et al. (2005) |
| HE 0441−0652 | 4811 | 1.52 | 2.50 | −2.77 | +0.89 | +1.38 | +0.25 | +1.63 | ··· | +1.20 | CEMP-$s/rs$ | 0 | Yong et al. (2013) |
| HE 0441−4343 | 5629 | 3.59 | 0.71 | −2.52 | ··· | +0.29 | 0.00 | +0.29 | −0.01 | −0.65 | | 1 | Barklem et al. (2005) |
| HE 0442−1234 | 4604 | 1.34 | 2.61 | −2.41 | ··· | −0.65 | +0.61 | −0.04 | −0.22 | −0.19 | | 1 | Barklem et al. (2005) |
| HE 0445−2339 | 5165 | 2.10 | 2.05 | −2.85 | ··· | −0.57 | 0.00 | −0.57 | ··· | −0.64 | | 1 | Cohen et al. (2013) |
| HE 0447−1858 | 5570 | 3.50 | 0.78 | −2.56 | −1.12 | +0.17 | 0.00 | +0.17 | ··· | −1.61 | | 1 | Cohen et al. (2013) |
| HE 0450−4705 | 5429 | 3.34 | 0.89 | −3.10 | ··· | +0.80 | 0.00 | +0.80 | −0.38 | ··· | CEMP | 1 | Barklem et al. (2005) |
| HE 0454−4758 | 5388 | 3.33 | 0.89 | −3.10 | ··· | +0.40 | 0.00 | +0.40 | −0.26 | −0.74 | | 1 | Barklem et al. (2005) |
| HE 0501−5139 | 5861 | 3.54 | 0.83 | −2.38 | ··· | +0.36 | 0.00 | +0.36 | −0.03 | ··· | | 1 | Barklem et al. (2005) |
| HE 0501−5644 | 5033 | 2.30 | 1.80 | −2.40 | ··· | +0.23 | +0.01 | +0.24 | −0.06 | −0.50 | | 1 | Barklem et al. (2005) |
| HE 0512−3835 | 4948 | 2.01 | 2.06 | −2.39 | ··· | −0.26 | +0.04 | −0.22 | +0.13 | −0.16 | | 1 | Barklem et al. (2005) |
| HE 0513−4557 | 5629 | 3.62 | 0.68 | −2.79 | ··· | +0.35 | 0.00 | +0.35 | +0.11 | ··· | | 1 | Barklem et al. (2005) |
| HE 0516−3820 | 5342 | 3.05 | 1.16 | −2.32 | ··· | +0.35 | +0.01 | +0.36 | +0.14 | −0.16 | | 1 | Barklem et al. (2005) |
| HE 0517−1952 | 5254 | 1.99 | 2.19 | −2.61 | ··· | −0.56 | +0.03 | −0.53 | +0.07 | −0.43 | | 1 | Barklem et al. (2005) |
| HE 0519−5525 | 5580 | 3.48 | 0.80 | −2.52 | ··· | +0.25 | 0.00 | +0.25 | +0.29 | −0.24 | | 1 | Barklem et al. (2005) |
| HE 0520−1748 | 5272 | 3.06 | 1.12 | −2.52 | ··· | +0.41 | +0.01 | +0.42 | −0.44 | −0.64 | | 1 | Barklem et al. (2005) |
| HE 0524−2055 | 4739 | 1.57 | 2.43 | −2.59 | ··· | −0.29 | +0.41 | +0.12 | +0.23 | +0.02 | | 1 | Barklem et al. (2005) |
| HE 0533−5340 | 4937 | 1.80 | 2.27 | −2.67 | ··· | −0.06 | +0.18 | +0.12 | ··· | −0.94 | | 1 | Cohen et al. (2013) |
| HE 0534−4615 | 5506 | 3.40 | 0.86 | −2.01 | ··· | +0.09 | 0.00 | +0.09 | +0.15 | −0.06 | | 1 | Barklem et al. (2005) |
| HE 0547−4539 | 5152 | 2.59 | 1.55 | −3.01 | ··· | +0.46 | +0.01 | +0.47 | +0.28 | −1.08 | | 1 | Barklem et al. (2005) |
| HE 0557−4840 | 4900 | 2.20 | 1.86 | −4.75 | ··· | +1.65 | +0.01 | +1.66 | ≤ −1.04 | +0.03 | CEMP-no | 1 | Masseron et al. (2012) |
| HE 0858−0016 | 4447 | 0.75 | 3.14 | −2.75 | ··· | −0.84 | +0.77 | −0.07 | −0.60 | −0.59 | | 1 | Barklem et al. (2005) |
| HE 0926−0508 | 6249 | 4.24 | 0.24 | −2.78 | ··· | +0.53 | 0.00 | +0.53 | +0.08 | −0.36 | | 1 | Barklem et al. (2005) |
| HE 0926−0546 | 5159 | 2.50 | 1.65 | −3.73 | +1.20 | +0.50 | +0.01 | +0.51 | ··· | −0.78 | | 1 | Cohen et al. (2013) |
| HE 0938+0114 | 6777 | 4.89 | −0.27 | −2.51 | ··· | +0.57 | 0.00 | +0.57 | +0.02 | −0.19 | | 1 | Barklem et al. (2005) |
| HE 0951−1152 | 5349 | 4.75 | −0.54 | −2.62 | ··· | +0.06 | 0.00 | +0.06 | −0.40 | +0.04 | | 1 | Barklem et al. (2005) |
| HE 1005−1439 | 5202 | 2.55 | 1.61 | −3.09 | +1.79 | +2.48 | +0.02 | +2.50 | ··· | +1.17 | CEMP-$s/rs$ | 0 | Yong et al. (2013) |
| HE 1006−2218 | 6638 | 4.62 | −0.04 | −2.69 | ··· | +0.58 | 0.00 | +0.58 | −0.05 | ··· | | 1 | Zhang et al. (2011) |
| HE 1012−1540 | 5745 | 3.45 | 0.88 | −3.47 | +1.25 | +2.22 | 0.00 | +2.22 | −0.46 | −0.25 | CEMP-no | 1 | Yong et al. (2013) |
| HE 1015−0027 | 6315 | 4.40 | 0.10 | −2.65 | ··· | +0.67 | 0.00 | +0.67 | +0.53 | ··· | | 1 | Zhang et al. (2011) |
| HE 1031−0020 | 5043 | 2.13 | 1.98 | −2.79 | +2.48 | +1.63 | +0.02 | +1.65 | +0.39 | +1.61 | CEMP-$s/rs$ | 0 | Yong et al. (2013) |
| HE 1044−2509 | 5227 | 2.78 | 1.39 | −2.88 | ··· | +0.48 | +0.01 | +0.49 | +0.33 | +0.18 | | 1 | Barklem et al. (2005) |
| HE 1045+0226 | 5077 | 2.20 | 1.92 | −2.20 | ··· | +0.97 | +0.02 | +0.99 | ··· | +1.24 | CEMP-$s/rs$ | 0 | Cohen et al. (2013) |
| HE 1052−2548 | 6534 | 4.52 | 0.04 | −2.29 | ··· | +0.47 | 0.00 | +0.47 | +0.35 | +0.82 | | 0 | Barklem et al. (2005) |
| HE 1054−0059 | 4601 | 1.19 | 2.76 | −3.34 | ··· | −0.77 | +0.69 | −0.08 | −0.61 | −1.26 | | 1 | Barklem et al. (2005) |
| HE 1059−0118 | 5560 | 4.30 | −0.02 | −2.80 | ··· | +0.33 | 0.00 | +0.33 | −0.41 | ··· | | 1 | Barklem et al. (2005) |
| HE 1100−0137 | 6101 | 4.25 | 0.19 | −2.90 | ··· | +0.43 | 0.00 | +0.43 | ··· | ··· | | 1 | Barklem et al. (2005) |
| HE 1105+0027 | 6132 | 3.45 | 1.00 | −2.42 | ··· | +1.91 | 0.00 | +1.91 | +0.83 | +2.36 | CEMP-$s/rs$ | 0 | Barklem et al. (2005) |
| HE 1111−3026 | 5000 | 2.00 | 2.09 | −2.00 | ··· | +0.97 | +0.06 | +1.03 | ··· | +1.31 | CEMP-$s/rs$ | 0 | Cohen et al. (2013) |
| HE 1116−0634 | 4400 | 0.10 | 3.77 | −3.73 | ··· | +0.08 | +0.73 | +0.81 | ··· | −1.81 | CEMP-no | 1 | Hollek et al. (2011) |
| HE 1120−0153 | 6191 | 4.09 | 0.37 | −2.77 | ··· | +0.54 | 0.00 | +0.54 | −0.08 | −0.22 | | 1 | Barklem et al. (2005) |
| HE 1122−1429 | 5787 | 3.29 | 1.06 | −2.65 | ··· | +0.40 | 0.00 | +0.40 | +0.04 | −0.40 | | 1 | Barklem et al. (2005) |
| HE 1124−2335 | 5226 | 2.68 | 1.49 | −2.95 | ··· | +0.76 | +0.01 | +0.77 | +0.18 | −1.15 | CEMP-no | 1 | Barklem et al. (2005) |
| HE 1126−1735 | 5689 | 3.31 | 1.01 | −2.69 | ··· | +0.19 | 0.00 | +0.19 | −0.07 | −0.51 | | 1 | Barklem et al. (2005) |
| HE 1127−1143 | 5224 | 2.64 | 1.53 | −2.72 | ··· | +0.50 | +0.01 | +0.51 | +0.34 | +0.57 | | 1 | Barklem et al. (2005) |
| HE 1128−0823 | 5909 | 3.71 | 0.67 | −2.71 | ··· | +0.43 | 0.00 | +0.43 | +0.12 | −0.21 | | 1 | Barklem et al. (2005) |
| HE 1131+0141 | 5236 | 2.98 | 1.19 | −2.49 | ··· | +0.08 | +0.01 | +0.09 | +0.25 | +0.48 | | 1 | Barklem et al. (2005) |
| HE 1132+0125 | 5732 | 3.54 | 0.79 | −2.38 | ··· | +0.20 | 0.00 | +0.20 | +0.07 | −0.21 | | 1 | Barklem et al. (2005) |
| HE 1132+0204 | 5046 | 2.25 | 1.86 | −2.53 | ··· | +0.09 | +0.01 | +0.10 | −0.12 | −0.74 | | 1 | Barklem et al. (2005) |

**Table 3** — *Continued*

| Name | $T_{\rm eff}$ (K) | log $g$ (cgs) | log $L$ ($L_\odot$) | [Fe/H] | [N/Fe] | [C/Fe] | Δ[C/Fe] ([N/Fe]=0) | $[C/Fe]_c$ | [Sr/Fe] | [Ba/Fe] | Class | I/O | Ref. |
|---|---|---|---|---|---|---|---|---|---|---|---|---|---|
| HE 1133−1641 | 6500 | 4.20 | 0.35 | −2.82 | ··· | +1.09 | 0.00 | +1.09 | ··· | −0.55 | CEMP-no | 1 | Cohen et al. (2013) |
| HE 1135+0139 | 5487 | 1.80 | 2.45 | −2.33 | ··· | +1.11 | +0.15 | +1.26 | +0.76 | +1.04 | CEMP-$s/rs$ | 0 | Barklem et al. (2005) |
| HE 1135−0344 | 6154 | 4.03 | 0.42 | −2.63 | ··· | +0.99 | 0.00 | +0.99 | +0.48 | ··· | CEMP | 1 | Barklem et al. (2005) |
| HE 1141−0610 | 6750 | 4.10 | 0.51 | −2.28 | ··· | +0.89 | 0.00 | +0.89 | +0.09 | −0.04 | CEMP-no | 1 | Cohen et al. (2013) |
| HE 1144−1349 | 6500 | 4.16 | 0.39 | −2.61 | ··· | +0.15 | 0.00 | +0.15 | ··· | ··· | | 1 | Akerman et al. (2004) |
| HE 1148−0037 | 5964 | 4.16 | 0.24 | −3.47 | ··· | +0.80 | 0.00 | +0.80 | −0.32 | ··· | CEMP | 1 | Barklem et al. (2005) |
| HE 1150−0428 | 5208 | 2.54 | 1.62 | −3.47 | +2.52 | +2.37 | +0.02 | +2.39 | −0.32 | −0.48 | CEMP-no | 1 | Yong et al. (2013) |
| HE 1157−0518 | 4900 | 2.00 | 2.06 | −2.34 | ··· | +2.15 | +0.05 | +2.20 | ··· | +2.14 | CEMP-$s/rs$ | 0 | Aoki et al. (2007) |
| HE 1159−0525 | 4838 | 1.50 | 2.53 | −2.91 | ··· | +1.82 | +0.22 | +2.04 | ··· | +1.53 | CEMP-$s/rs$ | 0 | Cohen et al. (2013) |
| HE 1201−1512 | 5725 | 3.39 | 0.94 | −3.92 | ≤ +1.29 | +1.60 | 0.00 | +1.60 | ··· | ≤ −0.34 | CEMP-no | 1 | Yong et al. (2013) |
| HE 1207−2031 | 6281 | 4.40 | 0.09 | −2.82 | ··· | +0.60 | 0.00 | +0.60 | −0.06 | ··· | | 1 | Barklem et al. (2005) |
| HE 1208−0040 | 5995 | 4.29 | 0.12 | −2.09 | ··· | +0.13 | 0.00 | +0.13 | ··· | ··· | | 1 | Akerman et al. (2004) |
| HE 1210+0048 | 6028 | 3.73 | 0.69 | −2.28 | ··· | +0.48 | 0.00 | +0.48 | +0.03 | −0.14 | | 1 | Barklem et al. (2005) |
| HE 1210−1956 | 5790 | 3.30 | 1.05 | −2.57 | ··· | +0.18 | 0.00 | +0.18 | +0.14 | −0.47 | | 1 | Barklem et al. (2005) |
| HE 1212−0127 | 4915 | 1.85 | 2.21 | −2.16 | ··· | −0.43 | +0.18 | −0.25 | −0.09 | −0.36 | | 1 | Barklem et al. (2005) |
| HE 1214−1819 | 4916 | 1.88 | 2.18 | −3.00 | ··· | +0.31 | +0.07 | +0.38 | +0.36 | −0.10 | | 1 | Barklem et al. (2005) |
| HE 1215+0149 | 5098 | 2.37 | 1.76 | −2.90 | ··· | +0.11 | +0.01 | +0.12 | +0.39 | −0.81 | | 1 | Barklem et al. (2005) |
| HE 1217−0540 | 5700 | 4.20 | 0.12 | −2.94 | ··· | +0.77 | 0.00 | +0.77 | +0.23 | ··· | CEMP | 1 | Barklem et al. (2005) |
| HE 1219−0312 | 5140 | 2.40 | 1.74 | −2.80 | ··· | −0.12 | +0.01 | −0.11 | +0.41 | +0.65 | | 0 | Barklem et al. (2005) |
| HE 1221−0522 | 5789 | 4.04 | 0.31 | −2.84 | ··· | +0.49 | 0.00 | +0.49 | −0.35 | ··· | | 1 | Barklem et al. (2005) |
| HE 1221−1948 | 6083 | 3.81 | 0.62 | −3.36 | ··· | +1.38 | 0.00 | +1.38 | ··· | ··· | CEMP | 1 | Barklem et al. (2005) |
| HE 1222−0200 | 5288 | 2.72 | 1.47 | −2.44 | ··· | +0.19 | +0.01 | +0.20 | −0.03 | ··· | | 1 | Barklem et al. (2005) |
| HE 1222−0336 | 6052 | 3.96 | 0.46 | −2.04 | ··· | +0.18 | 0.00 | +0.18 | −0.23 | −0.25 | | 1 | Barklem et al. (2005) |
| HE 1225+0155 | 4842 | 1.80 | 2.24 | −2.75 | ··· | +0.22 | +0.18 | +0.40 | +0.18 | −0.50 | | 1 | Barklem et al. (2005) |
| HE 1226−1149 | 5120 | 2.30 | 1.83 | −2.91 | +1.43 | +0.42 | +0.01 | +0.43 | ··· | +0.90 | | 0 | Cohen et al. (2013) |
| HE 1230−1724 | 5952 | 3.98 | 0.41 | −2.30 | ··· | +0.09 | 0.00 | +0.09 | +0.04 | −0.22 | | 1 | Barklem et al. (2005) |
| HE 1237−3103 | 4851 | 1.72 | 2.32 | −2.91 | ··· | −0.10 | +0.21 | +0.11 | −0.09 | −0.65 | | 1 | Barklem et al. (2005) |
| HE 1243−1425 | 5507 | 3.19 | 1.07 | −2.67 | ··· | +0.47 | +0.01 | +0.48 | +0.34 | −0.45 | | 1 | Barklem et al. (2005) |
| HE 1245−0215 | 6500 | 3.80 | 0.75 | −2.85 | ··· | +1.66 | 0.00 | +1.66 | ··· | +0.29 | CEMP | 1 | Cohen et al. (2013) |
| HE 1245−1616 | 6191 | 4.04 | 0.42 | −2.98 | ··· | +0.68 | 0.00 | +0.68 | +0.36 | +0.19 | | 1 | Barklem et al. (2005) |
| HE 1246−1344 | 4853 | 1.65 | 2.39 | −3.40 | ··· | −0.10 | +0.22 | +0.12 | −1.15 | ··· | | 1 | Barklem et al. (2005) |
| HE 1247−2114 | 5012 | 2.08 | 2.02 | −2.61 | ··· | +0.28 | +0.02 | +0.30 | +0.46 | −0.48 | | 1 | Barklem et al. (2005) |
| HE 1248−1800 | 5288 | 2.92 | 1.27 | −2.89 | ··· | +0.49 | 0.00 | +0.49 | −0.16 | −0.18 | | 1 | Barklem et al. (2005) |
| HE 1249−2932 | 4718 | 1.52 | 2.47 | −2.65 | ··· | −0.45 | +0.45 | 0.00 | −0.78 | −1.26 | | 1 | Barklem et al. (2005) |
| HE 1249−3121 | 5373 | 3.40 | 0.82 | −3.23 | ··· | +1.82 | 0.00 | +1.82 | −0.88 | ··· | CEMP | 1 | Barklem et al. (2005) |
| HE 1251−0104 | 5084 | 2.32 | 1.80 | −2.73 | ··· | +0.21 | +0.01 | +0.22 | −0.12 | −0.42 | | 1 | Barklem et al. (2005) |
| HE 1252+0044 | 5296 | 2.98 | 1.21 | −3.28 | ··· | +0.50 | 0.00 | +0.50 | +0.80 | −0.62 | | 1 | Barklem et al. (2005) |
| HE 1252−0117 | 4847 | 1.67 | 2.37 | −2.86 | ··· | −0.20 | +0.27 | +0.07 | −0.89 | −1.70 | | 1 | Barklem et al. (2005) |
| HE 1254+0009 | 4865 | 1.86 | 2.18 | −2.95 | ··· | −0.15 | +0.08 | −0.07 | −0.24 | −0.90 | | 1 | Barklem et al. (2005) |
| HE 1256−0228 | 4860 | 1.83 | 2.21 | −2.08 | ··· | −0.08 | +0.19 | +0.11 | +0.10 | +0.07 | | 1 | Barklem et al. (2005) |
| HE 1256−0651 | 6137 | 4.05 | 0.40 | −2.36 | ··· | +0.52 | 0.00 | +0.52 | −0.25 | −0.36 | | 1 | Barklem et al. (2005) |
| HE 1258−1317 | 5175 | 2.50 | 1.65 | −2.90 | ··· | +0.17 | +0.01 | +0.18 | ··· | −0.99 | | 1 | Cohen et al. (2013) |
| HE 1259−0621 | 5787 | 3.68 | 0.67 | −2.64 | ··· | +0.37 | 0.00 | +0.37 | +0.17 | −0.02 | | 1 | Barklem et al. (2005) |
| HE 1300+0157 | 5529 | 3.25 | 1.02 | −3.75 | ≤ +0.71 | +1.31 | 0.00 | +1.31 | −1.73 | ≤ −0.85 | CEMP-no | 1 | Yong et al. (2013) |
| HE 1300−0641 | 5308 | 2.96 | 1.24 | −3.14 | ··· | +1.20 | +0.01 | +1.21 | −0.50 | −0.86 | CEMP-no | 1 | Barklem et al. (2005) |
| HE 1300−0642 | 5173 | 2.68 | 1.47 | −3.02 | ··· | +0.30 | +0.01 | +0.31 | −0.12 | −0.43 | | 1 | Barklem et al. (2005) |
| HE 1300−2201 | 6332 | 4.64 | −0.14 | −2.61 | ··· | +0.92 | 0.00 | +0.92 | +0.38 | −0.13 | CEMP-no | 1 | Barklem et al. (2005) |
| HE 1300−2431 | 5029 | 1.96 | 2.14 | −3.25 | ··· | −0.20 | +0.01 | −0.19 | −0.29 | −0.50 | | 1 | Barklem et al. (2005) |
| HE 1305+0007 | 4750 | 2.00 | 2.00 | −2.08 | ··· | +1.85 | +0.05 | +1.90 | +0.96 | +2.40 | CEMP-$s/rs$ | 0 | Goswami et al. (2006) |
| HE 1305−0331 | 6081 | 4.22 | 0.21 | −3.26 | ··· | +1.09 | 0.00 | +1.09 | −0.07 | ··· | CEMP | 1 | Barklem et al. (2005) |
| HE 1310−0536 | 4975 | 1.92 | 2.16 | −4.15 | ≤ +1.20 | +2.45 | +0.08 | +2.53 | ··· | −0.49 | CEMP-no | 1 | Hansen et al. (2014) |
| HE 1311−0131 | 4825 | 1.50 | 2.53 | −3.15 | ··· | +0.33 | +0.40 | +0.73 | ··· | −0.62 | CEMP-no | 1 | Hollek et al. (2011) |
| HE 1311−1412 | 4796 | 1.50 | 2.52 | −2.88 | ··· | −0.19 | +0.44 | +0.25 | +0.04 | −0.11 | | 1 | Barklem et al. (2005) |
| HE 1314−3036 | 4757 | 1.54 | 2.47 | −2.98 | ··· | −0.17 | +0.39 | +0.22 | −0.02 | −0.10 | | 1 | Barklem et al. (2005) |
| HE 1317−0407 | 4725 | 0.30 | 3.69 | −3.10 | ··· | −0.52 | +0.74 | +0.22 | ··· | −0.35 | | 1 | Hollek et al. (2011) |

**Table 3** — *Continued*

| Name | $T_{eff}$ (K) | log $g$ (cgs) | log $L$ ($L_\odot$) | [Fe/H] | [N/Fe] | [C/Fe] | Δ[C/Fe] ([N/Fe]=0) | [C/Fe]$_c$ | [Sr/Fe] | [Ba/Fe] | Class | I/O | Ref. |
|---|---|---|---|---|---|---|---|---|---|---|---|---|---|
| HE 1319−1935 | 4691 | 1.27 | 2.71 | −2.22 | +0.46 | +1.45 | +0.15 | +1.60 | ··· | +1.68 | CEMP-*s*/*rs* | 0 | Yong et al. (2013) |
| HE 1320+0139 | 6500 | 3.70 | 0.85 | −3.38 | ··· | +0.35 | 0.00 | +0.35 | ··· | −1.32 | | 1 | Cohen et al. (2013) |
| HE 1320−1339 | 4935 | 1.69 | 2.38 | −2.78 | ··· | −0.55 | +0.27 | −0.28 | +0.33 | −0.48 | | 1 | Barklem et al. (2005) |
| HE 1327−2326 | 6180 | 4.50 | −0.04 | −5.65 | +4.23 | +3.90 | 0.00 | +3.90 | +1.27 | ≤ +1.42 | CEMP-no | 1 | Masseron et al. (2012) |
| HE 1330−0354 | 6257 | 4.13 | 0.35 | −2.29 | ··· | +0.96 | 0.00 | +0.96 | +0.11 | −0.56 | CEMP-no | 1 | Barklem et al. (2005) |
| HE 1330−0607 | 5094 | 2.30 | 1.82 | −2.33 | ··· | +0.17 | +0.01 | +0.18 | −0.35 | −0.81 | | 1 | Barklem et al. (2005) |
| HE 1332−0309 | 5125 | 2.40 | 1.73 | −2.46 | ··· | +0.17 | +0.01 | +0.18 | +0.40 | ··· | | 1 | Barklem et al. (2005) |
| HE 1333−0340 | 6053 | 4.18 | 0.24 | −2.64 | ··· | +0.38 | 0.00 | +0.38 | +0.37 | ··· | | 1 | Zhang et al. (2011) |
| HE 1335+0135 | 5246 | 2.94 | 1.24 | −2.47 | ··· | +0.09 | +0.01 | +0.10 | +0.14 | −0.82 | | 1 | Barklem et al. (2005) |
| HE 1337+0012 | 6390 | 4.38 | 0.14 | −3.25 | +1.57 | +0.50 | 0.00 | +0.50 | +0.28 | −0.25 | | 1 | Aoki et al. (2006) |
| HE 1337−0453 | 5938 | 3.56 | 0.83 | −2.34 | ··· | +0.08 | 0.00 | +0.08 | +0.25 | −0.29 | | 1 | Barklem et al. (2005) |
| HE 1338−0052 | 5856 | 3.70 | 0.67 | −3.00 | ··· | +1.32 | 0.00 | +1.32 | ··· | −0.02 | CEMP-no | 1 | Cohen et al. (2013) |
| HE 1345−0206 | 5006 | 2.22 | 1.87 | −2.82 | ··· | +0.30 | +0.01 | +0.31 | −0.33 | −1.15 | | 1 | Barklem et al. (2005) |
| HE 1347−1025 | 5206 | 2.52 | 1.64 | −3.71 | ··· | +0.15 | +0.01 | +0.16 | −1.05 | −0.50 | | 1 | Yong et al. (2013) |
| HE 1351−1049 | 5204 | 2.85 | 1.31 | −3.46 | ··· | +1.46 | +0.01 | +1.47 | +0.13 | +0.05 | CEMP | 1 | Barklem et al. (2005) |
| HE 1357−0123 | 4600 | 1.00 | 2.95 | −3.80 | ··· | −0.29 | +0.69 | +0.40 | ··· | −1.69 | | 1 | Cohen et al. (2013) |
| HE 1405−0822 | 5220 | 1.70 | 2.47 | −2.40 | +1.34 | +1.97 | +0.08 | +2.05 | ··· | +1.95 | CEMP-*s*/*rs* | 0 | Cui et al. (2013) |
| HE 1405−2512 | 5602 | 3.30 | 0.99 | −2.80 | ··· | +0.52 | 0.00 | +0.52 | ··· | −1.15 | | 1 | Cohen et al. (2013) |
| HE 1410+0213 | 5000 | 2.00 | 2.09 | −2.14 | +1.76 | +1.71 | +0.05 | +1.76 | ··· | −0.26 | CEMP-no | 1 | Cohen et al. (2013) |
| HE 1410−0004 | 4985 | 2.00 | 2.09 | −3.09 | ··· | +2.09 | +0.04 | +2.13 | +0.26 | +1.13 | CEMP-*s*/*rs* | 0 | Masseron et al. (2012) |
| HE 1413−1954 | 6533 | 4.59 | −0.03 | −3.19 | ··· | +1.41 | 0.00 | +1.41 | −0.37 | ··· | CEMP | 1 | Barklem et al. (2005) |
| HE 1416−1032 | 5000 | 1.80 | 2.29 | −3.20 | +0.37 | −0.43 | +0.09 | −0.34 | ··· | −1.24 | | 1 | Cohen et al. (2013) |
| HE 1419−1759 | 4809 | 1.63 | 2.39 | −3.17 | ··· | −0.24 | +0.28 | +0.04 | −0.17 | −0.86 | | 1 | Barklem et al. (2005) |
| HE 1421−2006 | 5687 | 3.73 | 0.59 | −2.65 | ··· | +0.26 | 0.00 | +0.26 | 0.00 | −0.18 | | 1 | Barklem et al. (2005) |
| HE 1422−0818 | 4816 | 1.30 | 2.73 | −3.29 | ··· | −0.14 | +0.58 | +0.44 | ··· | −1.90 | | 1 | Cohen et al. (2013) |
| HE 1424−0241 | 5260 | 2.66 | 1.52 | −4.05 | ≤ +1.13 | +0.63 | 0.00 | +0.63 | ≤ −1.61 | −0.75 | | 1 | Yong et al. (2013) |
| HE 1429−0551 | 4757 | 1.39 | 2.62 | −2.60 | +1.39 | +2.28 | +0.10 | +2.38 | ··· | +1.47 | CEMP-*s*/*rs* | 0 | Yong et al. (2013) |
| HE 1430−0026 | 5855 | 4.12 | 0.25 | −2.78 | ··· | +0.48 | 0.00 | +0.48 | −0.12 | −0.31 | | 1 | Barklem et al. (2005) |
| HE 1430−1123 | 5915 | 3.75 | 0.63 | −2.71 | ··· | +1.75 | 0.00 | +1.75 | +0.34 | +1.73 | CEMP-*s*/*rs* | 0 | Barklem et al. (2005) |
| HE 1431−2142 | 6137 | 4.10 | 0.35 | −2.60 | ··· | +0.44 | 0.00 | +0.44 | +0.07 | +0.08 | | 1 | Barklem et al. (2005) |
| HE 1432−1819 | 5975 | 3.60 | 0.80 | −2.56 | ··· | +0.83 | 0.00 | +0.83 | ··· | −0.67 | CEMP-no | 1 | Cohen et al. (2013) |
| HE 1434−1442 | 5420 | 3.20 | 1.03 | −2.39 | +1.70 | +1.95 | +0.01 | +1.96 | ··· | +1.23 | CEMP-*s*/*rs* | 0 | Cohen et al. (2013) |
| HE 1439−0218 | 6150 | 3.70 | 0.75 | −2.65 | ··· | +0.66 | 0.00 | +0.66 | ··· | −0.15 | | 1 | Cohen et al. (2013) |
| HE 1439−1420 | 6056 | 3.80 | 0.62 | −2.97 | +1.09 | +1.84 | 0.00 | +1.84 | ··· | +1.31 | CEMP-*s*/*rs* | 0 | Cohen et al. (2013) |
| HE 1443+0113 | 5000 | 2.00 | 2.09 | −2.11 | ··· | +1.52 | +0.06 | +1.58 | ··· | +1.79 | CEMP-*s*/*rs* | 0 | Cohen et al. (2013) |
| HE 1446−0140 | 6090 | 4.40 | 0.03 | −2.38 | ··· | +0.14 | 0.00 | +0.14 | ··· | −0.16 | | 1 | Cohen et al. (2013) |
| HE 1447+0102 | 5100 | 1.70 | 2.43 | −2.47 | ··· | +2.48 | +0.06 | +2.54 | ··· | +2.70 | CEMP-*s*/*rs* | 0 | Aoki et al. (2007) |
| HE 1456+0230 | 5664 | 2.20 | 2.11 | −3.32 | +2.84 | +2.14 | +0.02 | +2.16 | ··· | −0.19 | CEMP-no | 1 | Cohen et al. (2013) |
| HE 1500−1628 | 4807 | 1.50 | 2.52 | −2.38 | ··· | −0.01 | +0.48 | +0.47 | −1.66 | −1.72 | | 1 | Cohen et al. (2013) |
| HE 1506−0113 | 5016 | 2.01 | 2.09 | −3.54 | +0.61 | +1.47 | +0.02 | +1.49 | ··· | −0.80 | CEMP-no | 1 | Yong et al. (2013) |
| HE 1509−0806 | 5185 | 2.50 | 1.65 | −2.91 | +2.23 | +1.98 | +0.02 | +2.00 | +1.20 | +1.93 | CEMP-*s*/*rs* | 0 | Cohen et al. (2013) |
| HE 1523−1155 | 4800 | 1.60 | 2.42 | −2.15 | ··· | +1.86 | +0.07 | +1.93 | ··· | +1.72 | CEMP-*s*/*rs* | 0 | Aoki et al. (2007) |
| HE 1528−0409 | 5000 | 1.80 | 2.29 | −2.60 | ··· | +2.41 | +0.06 | +2.47 | ··· | +2.30 | CEMP-*s*/*rs* | 0 | Aoki et al. (2007) |
| HE 2122−4707 | 5147 | 2.50 | 1.64 | −2.42 | ··· | +1.60 | +0.02 | +1.62 | ··· | +2.03 | CEMP-*s*/*rs* | 0 | Cohen et al. (2013) |
| HE 2123−0329 | 4725 | 1.15 | 2.84 | −3.22 | ··· | +0.40 | +0.66 | +1.06 | ··· | −0.85 | CEMP-no | 1 | Hollek et al. (2011) |
| HE 2133−1426 | 6300 | 4.10 | 0.39 | −3.32 | +1.62 | +1.79 | 0.00 | +1.79 | −0.15 | +2.34 | CEMP-*s*/*rs* | 0 | Cohen et al. (2013) |
| HE 2133−1432 | 5716 | 3.46 | 0.86 | −2.02 | ··· | +0.08 | 0.00 | +0.08 | +0.21 | −0.10 | | 1 | Barklem et al. (2005) |
| HE 2134+0001 | 5257 | 3.00 | 1.18 | −2.22 | ··· | +0.16 | +0.01 | +0.17 | +0.02 | −0.47 | | 1 | Barklem et al. (2005) |
| HE 2138−0314 | 5015 | 1.90 | 2.20 | −3.29 | ··· | +0.78 | +0.04 | +0.82 | ··· | −0.85 | CEMP-no | 1 | Hollek et al. (2011) |
| HE 2138−3336 | 5850 | 3.60 | 0.76 | −2.79 | +1.66 | +2.43 | 0.00 | +2.43 | +0.27 | +1.91 | CEMP-*s*/*rs* | 0 | Placco et al. (2013) |
| HE 2139−1851 | 4925 | 1.86 | 2.21 | −3.24 | ··· | +0.45 | +0.05 | +0.50 | +0.39 | −0.95 | | 1 | Barklem et al. (2005) |
| HE 2139−5432 | 5416 | 3.04 | 1.19 | −4.02 | +2.08 | +2.59 | +0.01 | +2.60 | ··· | ≤ −0.33 | CEMP-no | 1 | Yong et al. (2013) |
| HE 2141−3741 | 4945 | 1.00 | 3.07 | −3.30 | +1.37 | +0.07 | +0.76 | +0.83 | −0.57 | −1.28 | CEMP-no | 1 | Placco et al. (2014a) |
| HE 2142−5656 | 4939 | 1.85 | 2.22 | −2.87 | +0.54 | +0.95 | +0.10 | +1.05 | ··· | −0.63 | CEMP-no | 1 | Yong et al. (2013) |
| HE 2143+0030 | 4688 | 1.50 | 2.48 | −2.43 | ··· | −0.40 | +0.50 | +0.10 | +0.01 | −0.43 | | 1 | Barklem et al. (2005) |

**Table 3** — *Continued*

| Name | $T_{\rm eff}$ (K) | $\log g$ (cgs) | $\log L$ ($L_{\odot}$) | [Fe/H] | [N/Fe] | [C/Fe] | $\Delta$[C/Fe] ([N/Fe]=0) | $[{\rm C/Fe}]_c$ | [Sr/Fe] | [Ba/Fe] | Class | I/O | Ref. |
|---|---|---|---|---|---|---|---|---|---|---|---|---|---|
| HE 2145−3025 | 5608 | 2.22 | 2.07 | −2.69 | ··· | −0.24 | +0.01 | −0.23 | +0.56 | −0.36 | | 1 | Barklem et al. (2005) |
| HE 2148−1105a | 4400 | 0.20 | 3.67 | −2.98 | ··· | −0.42 | +0.75 | +0.33 | ··· | −1.09 | | 1 | Hollek et al. (2011) |
| HE 2148−1247 | 6380 | 3.90 | 0.62 | −2.30 | ··· | +1.70 | 0.00 | +1.70 | +0.84 | +1.92 | CEMP-$s/rs$ | 0 | Cohen et al. (2013) |
| HE 2151−2858 | 5598 | 4.14 | 0.15 | −2.38 | ··· | +0.06 | 0.00 | +0.06 | −0.18 | −0.30 | | 1 | Barklem et al. (2005) |
| HE 2153−2719 | 4898 | 2.01 | 2.05 | −2.49 | ··· | +0.08 | +0.03 | +0.11 | +0.10 | −0.74 | | 1 | Barklem et al. (2005) |
| HE 2155+0136 | 5331 | 3.22 | 0.98 | −2.07 | ··· | −0.04 | 0.00 | −0.04 | −0.02 | +0.09 | | 1 | Barklem et al. (2005) |
| HE 2155−3750 | 5000 | 2.30 | 1.79 | −2.59 | +1.48 | +1.78 | +0.02 | +1.80 | ··· | +1.73 | CEMP-$s/rs$ | 0 | Cohen et al. (2013) |
| HE 2156−3130 | 4692 | 1.28 | 2.70 | −3.13 | ··· | +0.65 | +0.54 | +1.19 | −0.80 | +0.43 | CEMP | 1 | Barklem et al. (2005) |
| HE 2157−3404 | 4990 | 1.30 | 2.79 | −2.89 | +0.66 | −0.54 | +0.62 | +0.08 | −0.48 | −1.39 | | 1 | Placco et al. (2014a) |
| HE 2158−0348 | 5150 | 2.44 | 1.70 | −2.57 | +1.52 | +1.87 | +0.02 | +1.89 | +0.60 | +1.75 | CEMP-$s/rs$ | 0 | Yong et al. (2013) |
| HE 2200−1946 | 5000 | 2.50 | 1.59 | −2.63 | ··· | +0.34 | +0.01 | +0.35 | ··· | −1.43 | | 1 | Cohen et al. (2013) |
| HE 2200−2030 | 5870 | 3.42 | 0.95 | −2.00 | ··· | +0.06 | 0.00 | +0.06 | +0.03 | −0.09 | | 1 | Barklem et al. (2005) |
| HE 2201−0637 | 4976 | 2.21 | 1.87 | −2.59 | ··· | +0.10 | +0.01 | +0.11 | 0.00 | −0.72 | | 1 | Barklem et al. (2005) |
| HE 2202−4831 | 5331 | 2.95 | 1.25 | −2.78 | ··· | +2.41 | +0.02 | +2.43 | ··· | −1.28 | CEMP-no | 1 | Yong et al. (2013) |
| HE 2203−3723 | 5118 | 2.20 | 1.93 | −3.30 | ··· | +1.00 | +0.01 | +1.01 | ··· | −0.24 | CEMP-no | 1 | Cohen et al. (2013) |
| HE 2204−1703 | 4932 | 1.87 | 2.20 | −2.79 | ··· | +0.17 | +0.11 | +0.28 | −0.43 | −0.80 | | 1 | Barklem et al. (2005) |
| HE 2206−2245 | 5100 | 2.45 | 1.68 | −2.73 | ··· | +0.17 | +0.01 | +0.18 | +0.05 | −0.16 | | 1 | Barklem et al. (2005) |
| HE 2215−2548 | 5035 | 2.10 | 2.00 | −3.01 | ··· | +0.18 | +0.01 | +0.19 | ··· | −0.95 | | 1 | Cohen et al. (2013) |
| HE 2216−0621 | 4671 | 1.27 | 2.70 | −3.23 | ··· | −0.70 | +0.62 | −0.08 | −0.39 | −1.31 | | 1 | Barklem et al. (2005) |
| HE 2217−0706 | 4497 | 1.01 | 2.90 | −2.58 | ··· | −0.59 | +0.75 | +0.16 | −0.09 | −0.61 | | 1 | Barklem et al. (2005) |
| HE 2217−1523 | 4847 | 1.82 | 2.22 | −2.62 | ··· | 0.00 | +0.17 | +0.17 | −0.05 | −0.57 | | 1 | Barklem et al. (2005) |
| HE 2217−4053 | 5148 | 2.40 | 1.74 | −3.37 | +0.65 | +1.14 | +0.01 | +1.15 | ··· | +1.30 | CEMP-$s/rs$ | 0 | Cohen et al. (2013) |
| HE 2219−0713 | 4789 | 1.68 | 2.34 | −2.92 | ··· | −0.21 | +0.25 | +0.04 | −0.22 | −1.41 | | 1 | Barklem et al. (2005) |
| HE 2221−0453 | 4430 | 0.73 | 3.15 | −2.00 | +0.84 | +1.83 | +0.08 | +1.91 | ··· | +1.76 | CEMP-$s/rs$ | 0 | Yong et al. (2013) |
| HE 2221−4150 | 5887 | 4.06 | 0.32 | −2.03 | ··· | +0.19 | 0.00 | +0.19 | +0.12 | +0.04 | | 1 | Barklem et al. (2005) |
| HE 2222−4156 | 5537 | 3.27 | 1.00 | −2.73 | ··· | +0.38 | 0.00 | +0.38 | +0.16 | −0.21 | | 1 | Barklem et al. (2005) |
| HE 2224+0143 | 5198 | 2.66 | 1.50 | −2.58 | ··· | +0.31 | +0.01 | +0.32 | +0.33 | +0.53 | | 1 | Barklem et al. (2005) |
| HE 2224−4103 | 5074 | 2.32 | 1.80 | −2.60 | ··· | +0.19 | +0.01 | +0.20 | −0.00 | −0.62 | | 1 | Barklem et al. (2005) |
| HE 2226−4102 | 5140 | 2.43 | 1.71 | −2.87 | ··· | +0.42 | +0.01 | +0.43 | +0.05 | −1.05 | | 1 | Barklem et al. (2005) |
| HE 2227−4044 | 5811 | 3.85 | 0.50 | −2.32 | ··· | +1.59 | 0.00 | +1.59 | +0.51 | +1.29 | CEMP-$s/rs$ | 0 | Barklem et al. (2005) |
| HE 2228−0706 | 5003 | 2.02 | 2.07 | −2.78 | +1.13 | +2.32 | +0.04 | +2.36 | ··· | +2.46 | CEMP-$s/rs$ | 0 | Yong et al. (2013) |
| HE 2228−3806 | 5175 | 2.62 | 1.53 | −3.07 | ··· | +0.38 | +0.01 | +0.39 | −0.09 | −0.62 | | 1 | Barklem et al. (2005) |
| HE 2229−4153 | 5138 | 2.47 | 1.67 | −2.59 | ··· | +0.33 | +0.01 | +0.34 | +0.19 | −0.33 | | 1 | Barklem et al. (2005) |
| HE 2231−0622 | 5211 | 2.90 | 1.26 | −2.12 | ··· | −0.12 | +0.01 | −0.11 | −0.40 | −0.51 | | 1 | Barklem et al. (2005) |
| HE 2233−4724 | 4360 | 0.40 | 3.45 | −3.65 | ··· | −0.48 | +0.70 | +0.22 | −0.77 | −1.13 | | 1 | Placco et al. (2014a) |
| HE 2234−0521 | 5332 | 3.15 | 1.05 | −2.78 | ··· | +0.32 | +0.01 | +0.33 | −0.20 | −0.89 | | 1 | Barklem et al. (2005) |
| HE 2236+0204 | 5036 | 2.10 | 2.00 | −2.86 | ··· | +0.22 | +0.01 | +0.23 | ··· | −0.05 | | 1 | Cohen et al. (2013) |
| HE 2238−0131 | 4350 | 0.15 | 3.70 | −3.00 | ··· | −0.42 | +0.75 | +0.33 | ··· | −0.46 | | 1 | Hollek et al. (2011) |
| HE 2238−2152 | 5427 | 3.28 | 0.95 | −2.40 | ··· | +0.09 | 0.00 | +0.09 | −0.19 | −0.42 | | 1 | Barklem et al. (2005) |
| HE 2239−5019 | 6125 | 3.48 | 0.96 | −4.15 | ≤ +1.20 | +1.80 | 0.00 | +1.80 | ··· | ≤ 0.00 | CEMP-no | 1 | Hansen et al. (2014) |
| HE 2240−0412 | 5852 | 4.33 | 0.04 | −2.20 | ··· | +1.26 | 0.00 | +1.26 | +0.34 | +1.28 | CEMP-$s/rs$ | 0 | Barklem et al. (2005) |
| HE 2242−1930 | 5281 | 2.99 | 1.20 | −2.21 | ··· | +0.05 | +0.01 | +0.06 | +0.06 | −0.15 | | 1 | Barklem et al. (2005) |
| HE 2244−1503 | 5122 | 2.57 | 1.56 | −2.87 | ··· | +0.11 | +0.01 | +0.12 | −0.10 | +0.40 | | 1 | Barklem et al. (2005) |
| HE 2244−2116 | 5230 | 2.70 | 1.47 | −2.35 | ··· | −0.35 | +0.01 | −0.34 | ··· | −1.92 | | 1 | Cohen et al. (2013) |
| HE 2247−3705 | 5366 | 3.04 | 1.17 | −2.26 | ··· | +0.32 | +0.01 | +0.33 | −0.23 | −0.88 | | 1 | Barklem et al. (2005) |
| HE 2247−7400 | 4829 | 1.56 | 2.47 | −2.87 | ··· | +0.70 | +0.32 | +1.02 | ··· | −0.94 | CEMP-no | 1 | Yong et al. (2013) |
| HE 2248−3345 | 5011 | 2.11 | 1.99 | −2.73 | ··· | +0.17 | +0.01 | +0.18 | −1.89 | −1.42 | | 1 | Barklem et al. (2005) |
| HE 2249−1704 | 4592 | 1.10 | 2.84 | −2.95 | ··· | +1.76 | +0.31 | +2.07 | ··· | +1.75 | CEMP-$s/rs$ | 0 | Cohen et al. (2013) |
| HE 2250−2132 | 5705 | 3.69 | 0.63 | −2.22 | ··· | +0.37 | 0.00 | +0.37 | +0.10 | −0.36 | | 1 | Barklem et al. (2005) |
| HE 2251−0821 | 5160 | 2.50 | 1.65 | −2.91 | ··· | +0.57 | +0.01 | +0.58 | ··· | −0.49 | | 1 | Cohen et al. (2013) |
| HE 2252−4225 | 4708 | 1.53 | 2.46 | −2.82 | ··· | −0.43 | +0.43 | 0.00 | +0.16 | +0.40 | | 1 | Barklem et al. (2005) |
| HE 2253−0849 | 4425 | 0.20 | 3.68 | −2.88 | ··· | −0.77 | +0.76 | −0.01 | ··· | −0.38 | | 1 | Hollek et al. (2011) |
| HE 2258−3456 | 4988 | 1.55 | 2.54 | −2.97 | ··· | −0.24 | +0.39 | +0.15 | −0.93 | −1.20 | | 1 | Barklem et al. (2005) |
| HE 2258−6358 | 4900 | 1.60 | 2.46 | −2.67 | +1.44 | +2.42 | +0.08 | +2.50 | +0.80 | +2.23 | CEMP-$s/rs$ | 0 | Placco et al. (2013) |
| HE 2259−3407 | 6266 | 4.32 | 0.16 | −2.29 | ··· | +0.37 | 0.00 | +0.37 | −0.08 | −0.57 | | 1 | Barklem et al. (2005) |

**Table 3** — *Continued*

| Name | $T_{\rm eff}$ (K) | log $g$ (cgs) | log $L$ ($L_\odot$) | [Fe/H] | [N/Fe] | [C/Fe] | Δ[C/Fe] ([N/Fe]=0) | [C/Fe]$_c$ | [Sr/Fe] | [Ba/Fe] | Class | I/O | Ref. |
|---|---|---|---|---|---|---|---|---|---|---|---|---|---|
| HE 2301−4024 | 5743 | 3.76 | 0.57 | −2.10 | ··· | +0.26 | 0.00 | +0.26 | +0.10 | +0.44 | | 1 | Barklem et al. (2005) |
| HE 2301−4126 | 5841 | 3.66 | 0.70 | −2.37 | ··· | +0.35 | 0.00 | +0.35 | +0.03 | +0.06 | | 1 | Barklem et al. (2005) |
| HE 2302−2154a | 4675 | 0.90 | 3.08 | −3.90 | ··· | +0.38 | +0.70 | +1.08 | ··· | −1.50 | CEMP-no | 1 | Hollek et al. (2011) |
| HE 2304−4153 | 4663 | 1.01 | 2.96 | −3.02 | ··· | −0.69 | +0.75 | +0.06 | −1.53 | −1.66 | | 1 | Barklem et al. (2005) |
| HE 2311+0129 | 5188 | 2.65 | 1.51 | −2.78 | ··· | +0.29 | +0.01 | +0.30 | +0.20 | −0.24 | | 1 | Barklem et al. (2005) |
| HE 2314−1554 | 5050 | 2.20 | 1.91 | −3.28 | +2.20 | +1.01 | +0.01 | +1.02 | +0.16 | −0.77 | CEMP-no | 1 | Cohen et al. (2013) |
| HE 2318−1621 | 4846 | 1.40 | 2.64 | −3.67 | +1.24 | +1.04 | +0.50 | +1.54 | −1.00 | −1.61 | CEMP-no | 1 | Placco et al. (2014a) |
| HE 2319−0852 | 4724 | 1.36 | 2.63 | −3.38 | ··· | −0.36 | +0.52 | +0.16 | −0.43 | −1.04 | | 1 | Barklem et al. (2005) |
| HE 2323−0256 | 4958 | 1.84 | 2.24 | −3.97 | +2.16 | +1.06 | +0.04 | +1.10 | ··· | −0.52 | CEMP-no | 1 | Yong et al. (2013) |
| HE 2323−6549 | 5215 | 2.60 | 1.56 | −3.35 | +0.42 | +0.72 | +0.01 | +0.73 | +0.03 | −0.58 | CEMP-no | 1 | Placco et al. (2014a) |
| HE 2325−0755 | 5665 | 3.17 | 1.14 | −2.85 | ··· | +0.17 | 0.00 | +0.17 | −0.14 | −0.62 | | 1 | Barklem et al. (2005) |
| HE 2326+0038 | 5145 | 2.51 | 1.63 | −2.78 | ··· | +0.19 | +0.01 | +0.20 | −0.35 | −0.71 | | 1 | Barklem et al. (2005) |
| HE 2327−5642 | 5050 | 2.34 | 1.77 | −2.78 | −0.20 | +0.13 | +0.01 | +0.14 | +0.40 | +0.31 | | 1 | Masseron et al. (2010) |
| HE 2329−3702 | 6060 | 3.72 | 0.71 | −2.15 | ··· | +0.11 | 0.00 | +0.11 | +0.24 | −0.19 | | 1 | Barklem et al. (2005) |
| HE 2330−0555 | 4867 | 1.65 | 2.40 | −2.98 | +1.00 | +2.09 | +0.15 | +2.24 | ··· | +1.17 | CEMP-$s/rs$ | 0 | Yong et al. (2013) |
| HE 2333−1358 | 5020 | 2.15 | 1.95 | −3.34 | ··· | +0.29 | +0.01 | +0.30 | −1.77 | ··· | | 1 | Barklem et al. (2005) |
| HE 2335−5958 | 5765 | 3.24 | 1.10 | −2.32 | ··· | +0.04 | 0.00 | +0.04 | +0.05 | −0.37 | | 1 | Barklem et al. (2005) |
| HE 2338−1311 | 5582 | 3.50 | 0.78 | −2.86 | ··· | +0.30 | 0.00 | +0.30 | ··· | ··· | | 1 | Barklem et al. (2005) |
| HE 2338−1618 | 5515 | 3.38 | 0.88 | −2.64 | ··· | +0.43 | 0.00 | +0.43 | −0.12 | ··· | | 1 | Barklem et al. (2005) |
| HE 2341−2029 | 6500 | 4.40 | 0.15 | −2.53 | +1.10 | +1.05 | 0.00 | +1.05 | −0.01 | −0.42 | CEMP-no | 1 | Cohen et al. (2013) |
| HE 2345−1919 | 5617 | 4.46 | −0.17 | −2.46 | ··· | +0.20 | 0.00 | +0.20 | −0.18 | −0.51 | | 1 | Barklem et al. (2005) |
| HE 2347−1334 | 4453 | 0.95 | 2.94 | −2.56 | ··· | −0.54 | +0.75 | +0.21 | −0.11 | −0.83 | | 1 | Barklem et al. (2005) |
| HE 2347−1448 | 6162 | 3.98 | 0.47 | −2.31 | ··· | +0.46 | 0.00 | +0.46 | +0.18 | −0.21 | | 1 | Barklem et al. (2005) |
| HE 2356−0410 | 5170 | 2.45 | 1.70 | −3.19 | +1.75 | +2.27 | +0.02 | +2.29 | −0.90 | −0.80 | CEMP-no | 1 | Yong et al. (2013) |
| HE 2357−0701 | 5142 | 2.40 | 1.74 | −3.17 | +0.87 | +0.26 | +0.01 | +0.27 | ··· | −1.34 | | 1 | Cohen et al. (2013) |
| HK17435−00532 | 5200 | 2.15 | 2.01 | −2.23 | ··· | +0.77 | +0.01 | +0.78 | +0.41 | +0.86 | CEMP | 0 | Roederer et al. (2008b) |
| LP625−44 | 5500 | 2.80 | 1.46 | −2.71 | +1.00 | +2.10 | +0.02 | +2.12 | +1.22 | +2.74 | CEMP-$s/rs$ | 0 | Masseron et al. (2012) |
| LP 815−43 | 6533 | 4.25 | 0.31 | −2.67 | ··· | +0.22 | 0.00 | +0.22 | −0.01 | ··· | | 1 | Akerman et al. (2004) |
| SDSS J0002+29 | 6150 | 4.00 | 0.45 | −3.26 | ··· | +2.63 | 0.00 | +2.63 | ··· | +1.84 | CEMP-$s/rs$ | 0 | Aoki et al. (2013) |
| SDSS J0018−09 | 4600 | 5.00 | −1.05 | −2.65 | ··· | −0.88 | 0.00 | −0.88 | ··· | ··· | | 1 | Aoki et al. (2013) |
| SDSS J0036−10 | 6479 | 4.31 | 0.23 | −2.60 | ··· | +2.32 | 0.00 | +2.32 | ··· | +0.40 | CEMP | 1 | Yong et al. (2013) |
| SDSS J0126+06 | 6900 | 4.00 | 0.65 | −3.01 | ··· | +3.08 | 0.00 | +3.08 | ··· | +3.20 | CEMP-$s/rs$ | 0 | Aoki et al. (2013) |
| SDSS J0140+23 | 5703 | 3.36 | 0.96 | −4.09 | ··· | +1.57 | 0.00 | +1.57 | ··· | ≤ −0.04 | CEMP-no | 1 | Yong et al. (2013) |
| SDSS J0259+00 | 4550 | 5.00 | −1.07 | −3.31 | ··· | −0.02 | 0.00 | −0.02 | ··· | ··· | | 1 | Aoki et al. (2013) |
| SDSS J0308+05 | 5950 | 4.00 | 0.39 | −2.19 | ··· | +2.36 | 0.00 | +2.36 | ··· | ··· | CEMP | 1 | Aoki et al. (2013) |
| SDSS J0351+10 | 5450 | 3.60 | 0.64 | −3.18 | ··· | +1.55 | 0.00 | +1.55 | ··· | ··· | CEMP-no | 1 | Aoki et al. (2013) |
| SDSS J0629+83 | 5550 | 4.00 | 0.27 | −2.82 | ··· | +2.09 | 0.00 | +2.09 | ··· | ··· | CEMP | 1 | Aoki et al. (2013) |
| SDSS J0711+67 | 5350 | 3.00 | 1.21 | −2.91 | ··· | +1.98 | +0.01 | +1.99 | ··· | +0.82 | CEMP | 0 | Aoki et al. (2013) |
| SDSS J0723+36 | 5150 | 2.20 | 1.94 | −3.32 | ··· | +1.79 | +0.02 | +1.81 | ··· | ··· | CEMP-no | 1 | Aoki et al. (2013) |
| SDSS J0741+67 | 5200 | 2.50 | 1.66 | −2.87 | ··· | +0.74 | +0.01 | +0.75 | ··· | +0.26 | CEMP | 1 | Aoki et al. (2013) |
| SDSS J0817+26 | 6300 | 4.00 | 0.49 | −3.16 | ··· | +2.41 | 0.00 | +2.41 | ··· | +0.77 | CEMP | 0 | Aoki et al. (2008) |
| SDSS J0858+35 | 5200 | 2.50 | 1.66 | −2.53 | ··· | +0.30 | +0.01 | +0.31 | ··· | −0.02 | | 1 | Aoki et al. (2013) |
| SDSS J0912+02 | 6500 | 4.50 | 0.05 | −2.50 | +1.75 | +2.17 | 0.00 | +2.17 | ··· | +1.49 | CEMP-$s/rs$ | 0 | Masseron et al. (2012) |
| SDSS J0918+37 | 6463 | 4.34 | 0.20 | −2.98 | ··· | +2.82 | 0.00 | +2.82 | ··· | +1.70 | CEMP-$s/rs$ | 0 | Yong et al. (2013) |
| SDSS J0924+40 | 6196 | 3.77 | 0.69 | −2.68 | ··· | +2.72 | 0.00 | +2.72 | ··· | +1.73 | CEMP-$s/rs$ | 0 | Yong et al. (2013) |
| SDSS J1029+17 | 5811 | 4.00 | 0.35 | −4.99 | ≤ +0.20 | +0.70 | 0.00 | +0.70 | ··· | ··· | CEMP | 1 | Caffau et al. (2011) |
| SDSS J1036+12 | 6000 | 4.00 | 0.41 | −3.20 | +1.29 | +1.47 | 0.00 | +1.47 | ··· | +1.17 | CEMP-$s/rs$ | 0 | Masseron et al. (2012) |
| SDSS J1114+18 | 6200 | 4.00 | 0.47 | −3.35 | ··· | +3.25 | 0.00 | +3.25 | ··· | +1.62 | CEMP-$s/rs$ | 0 | Spite et al. (2013) |
| SDSS J1143+20 | 6240 | 4.00 | 0.48 | −3.15 | ··· | +2.75 | 0.00 | +2.75 | ··· | +1.84 | CEMP-$s/rs$ | 0 | Spite et al. (2013) |
| SDSS J1241−08 | 5150 | 2.50 | 1.64 | −2.73 | ··· | +0.50 | +0.01 | +0.51 | ··· | −0.45 | | 1 | Aoki et al. (2013) |
| SDSS J1242−03 | 5150 | 2.50 | 1.64 | −2.77 | ··· | +0.64 | +0.01 | +0.65 | ··· | −0.11 | | 1 | Aoki et al. (2013) |
| SDSS J1245−07 | 6100 | 4.00 | 0.44 | −3.17 | ··· | +2.53 | 0.00 | +2.53 | ··· | +2.09 | CEMP-$s/rs$ | 0 | Aoki et al. (2013) |
| SDSS J1349−02 | 6200 | 4.00 | 0.47 | −3.00 | +1.60 | +2.82 | 0.00 | +2.82 | ··· | +2.17 | CEMP-$s/rs$ | 0 | Masseron et al. (2012) |
| SDSS J1422+00 | 5200 | 2.20 | 1.96 | −3.03 | ··· | +1.70 | +0.01 | +1.71 | ··· | −1.18 | CEMP-no | 1 | Aoki et al. (2013) |
| SDSS J1612+04 | 5350 | 3.30 | 0.91 | −2.86 | ··· | +0.63 | 0.00 | +0.63 | ··· | +0.05 | | 1 | Aoki et al. (2013) |

**Table 3** — *Continued*

| Name | $T_{\rm eff}$ (K) | log $g$ (cgs) | log $L$ ($L_\odot$) | [Fe/H] | [N/Fe] | [C/Fe] | Δ[C/Fe] ([N/Fe]=0) | $[C/Fe]_c$ | [Sr/Fe] | [Ba/Fe] | Class | I/O | Ref. |
|---|---|---|---|---|---|---|---|---|---|---|---|---|---|
| SDSS J1613+53 | 5350 | 2.10 | 2.11 | −3.33 | ··· | +2.09 | +0.02 | +2.11 | ··· | +0.03 | CEMP | 1 | Aoki et al. (2013) |
| SDSS J1626+14 | 6400 | 4.00 | 0.52 | −2.99 | ··· | +2.86 | 0.00 | +2.86 | ··· | +1.69 | CEMP-*s*/*rs* | 0 | Aoki et al. (2013) |
| SDSS J1646+28 | 6100 | 4.00 | 0.44 | −3.05 | ··· | +2.52 | 0.00 | +2.52 | ··· | +1.78 | CEMP-*s*/*rs* | 0 | Aoki et al. (2013) |
| SDSS J1703+28 | 5100 | 4.80 | −0.67 | −3.21 | ··· | +0.28 | 0.00 | +0.28 | ··· | ··· | | 1 | Aoki et al. (2013) |
| SDSS J1707+58 | 6700 | 4.20 | 0.40 | −2.52 | ··· | +2.14 | 0.00 | +2.14 | ··· | +3.40 | CEMP-*s*/*rs* | 0 | Aoki et al. (2008) |
| SDSS J1734+43 | 5200 | 2.70 | 1.46 | −2.51 | ··· | +1.78 | +0.02 | +1.80 | ··· | +1.61 | CEMP-*s*/*rs* | 0 | Aoki et al. (2013) |
| SDSS J1746+24 | 5350 | 2.60 | 1.61 | −3.17 | ··· | +1.24 | +0.01 | +1.25 | ··· | +0.26 | CEMP | 1 | Aoki et al. (2013) |
| SDSS J1836+63 | 5350 | 3.00 | 1.21 | −2.85 | ··· | +2.02 | +0.01 | +2.03 | ··· | +2.37 | CEMP-*s*/*rs* | 0 | Aoki et al. (2013) |
| SDSS J2047+00 | 6383 | 4.36 | 0.16 | −2.36 | ··· | +2.00 | 0.00 | +2.00 | ··· | +1.70 | CEMP-*s*/*rs* | 0 | Yong et al. (2013) |
| SDSS J2206−09 | 5100 | 2.10 | 2.03 | −3.17 | ··· | +0.64 | +0.01 | +0.65 | ··· | −0.85 | | 1 | Aoki et al. (2013) |
| SDSS J2209−00 | 6440 | 4.00 | 0.53 | −4.00 | ··· | +2.56 | 0.00 | +2.56 | ··· | ≤ +1.05 | CEMP-no | 1 | Spite et al. (2013) |
| SDSS J2357−00 | 5200 | 4.80 | −0.64 | −3.20 | ··· | +0.57 | 0.00 | +0.57 | ··· | +1.33 | | 0 | Aoki et al. (2013) |

**Note**. — Class column:
CEMP: [C/Fe] ≥ +0.7
CEMP-no: [C/Fe] ≥ +0.7 and [Ba/Fe] ≤ 0.0, or a published light-element enrichment pattern that clearly suggests CEMP-no status.
CEMP-s/rs: [C/Fe] ≥ +0.7 and [Ba/Fe] ≥ 1.0
I/O rejection flag:
"1" for accepted stars ([Ba/Fe]< +0.6, [Ba/Sr]<0.0, and upper limits for [Ba/Fe])
"0" for stars rejected on the CEMP-star frequency calculations ([Ba/Fe]> +1.0 and [Ba/Sr]>0.0)